\documentclass[12pt,letterpaper]{article} 
\usepackage{setspace,wrapfig}

\usepackage{mathptmx} 

\usepackage{mdframed}
\usepackage{natbib}
\def\lax    {${_<\atop^{\sim}}$}
\def\gax    {${_>\atop^{\sim}}$}

\usepackage[english]{babel} 

\usepackage{microtype} 

\usepackage{amsmath,amsfonts,amsthm} 

\usepackage[svgnames]{xcolor} 

\usepackage[hang, small, labelfont=bf, up, textfont=it]{caption} 

\usepackage{booktabs} 

\usepackage{lastpage} 

\usepackage{graphicx} 

\usepackage{enumitem} 
\setlist{noitemsep} 

\usepackage{sectsty} 
\allsectionsfont{\usefont{OT1}{phv}{b}{n}} 

\usepackage{geometry} 

\usepackage[T1]{fontenc} 
\usepackage[utf8]{inputenc} 

\usepackage{XCharter} 

\usepackage{fancyhdr} 
\fancypagestyle{firstpage}{ 
	\fancyhf{}
}

\newcommand{\authorstyle}[1]{{\usefont{OT1}{phv}{b}{n}\color{MidnightBlue}#1}} 

\newcommand{\institution}[1]{{\small\usefont{OT1}{phv}{m}{sl}\color{Black}#1}} 

\usepackage{titling} 

\newcommand{\HorRule}{\color{DarkGoldenrod}\rule{\linewidth}{1pt}} 

\pretitle{
	\vspace{80pt} 
	\HorRule\vspace{-30pt} 
	\fontsize{22}{28}\usefont{OT1}{phv}{b}{n}\selectfont 
	\color{MidnightBlue} 
\begin{flushleft}
}

\posttitle{\par\end{flushleft}\vskip 10pt} 

\preauthor{\begin{flushleft}} 

\postauthor{ 
\end{flushleft}
	\vspace{10pt} 
	\par\HorRule 
	\vspace{10pt} 
}

\usepackage{lettrine} 
\usepackage{fix-cm}	

\newcommand{\initial}[1]{ 
	\lettrine[lines=3,findent=4pt,nindent=0pt]{
		\color{DarkGoldenrod}
		{#1}
	}{}%
}

\usepackage{xstring} 

\newcommand{\lettrineabstract}[1]{
	\StrLeft{#1}{1}[\firstletter] 
	\initial{\firstletter}\textbf{\StrGobbleLeft{#1}{1}} 
}

\usepackage{natbib}

\usepackage{aas_macros}

\usepackage[pages=some]{background}

\title{X-rays from Stars and Planetary Systems}

\author{\authorstyle{
\noindent Jeremy J. Drake} 
\newline\newline 
\institution{Smithsonian Astrophysical Observatory,\\ 
Center for Astrophysics $\vert$ Harvard \& Smithsonian,\\ 60 Garden Street,\\
Cambridge, MA 02138, USA}, 
}

\date{A primer on stellar and planetary astrophysics enabled by the {\it Chandra} X-ray Observatory\\  ~~~ \\  July 2019} 

\begin{document}
\backgroundsetup{contents=\includegraphics{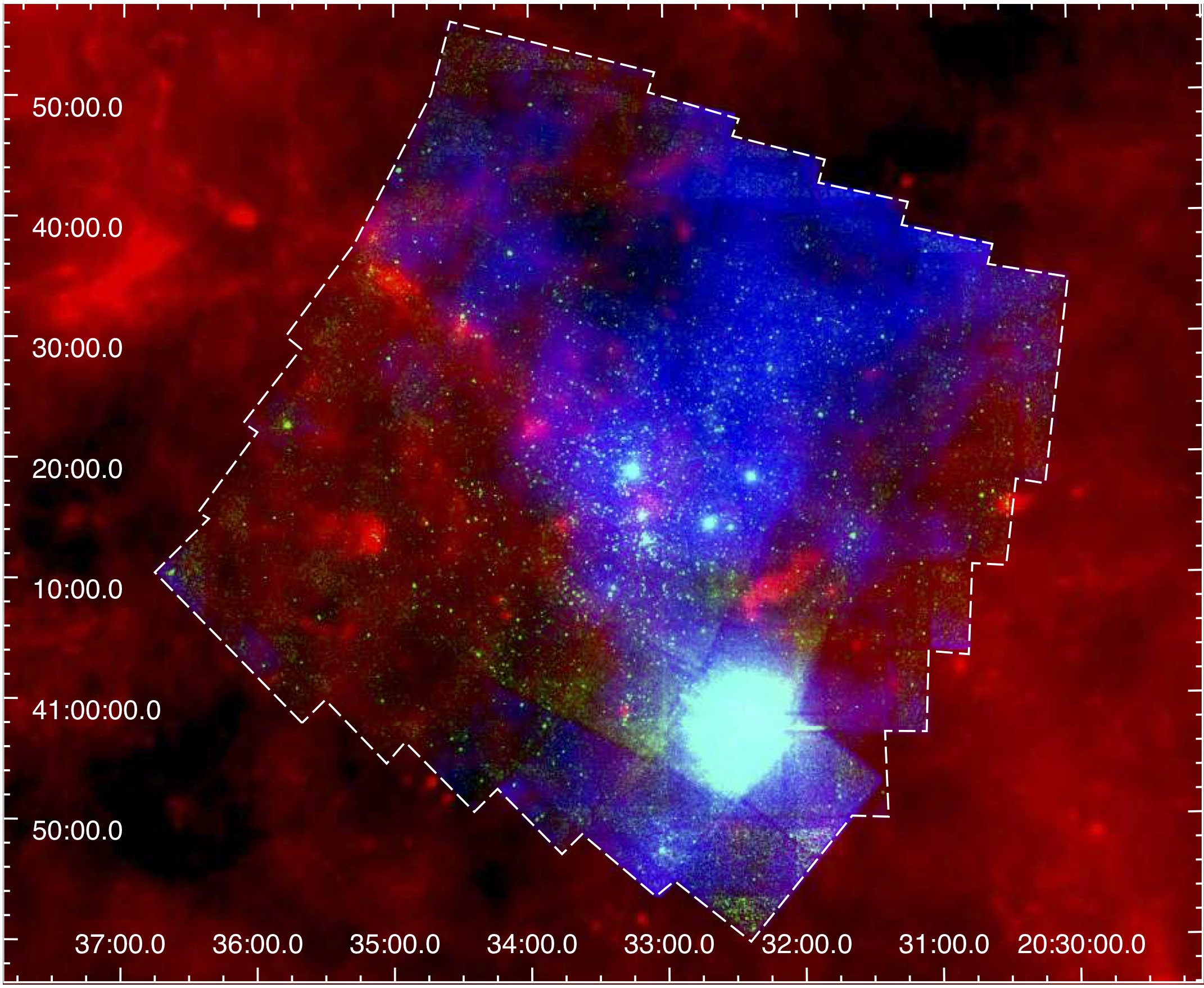},scale=2.0,opacity=0.2,vshift=0pt,angle=0}
\BgThispage

\maketitle 

\thispagestyle{firstpage} 


\newpage
\backgroundsetup{contents=\includegraphics{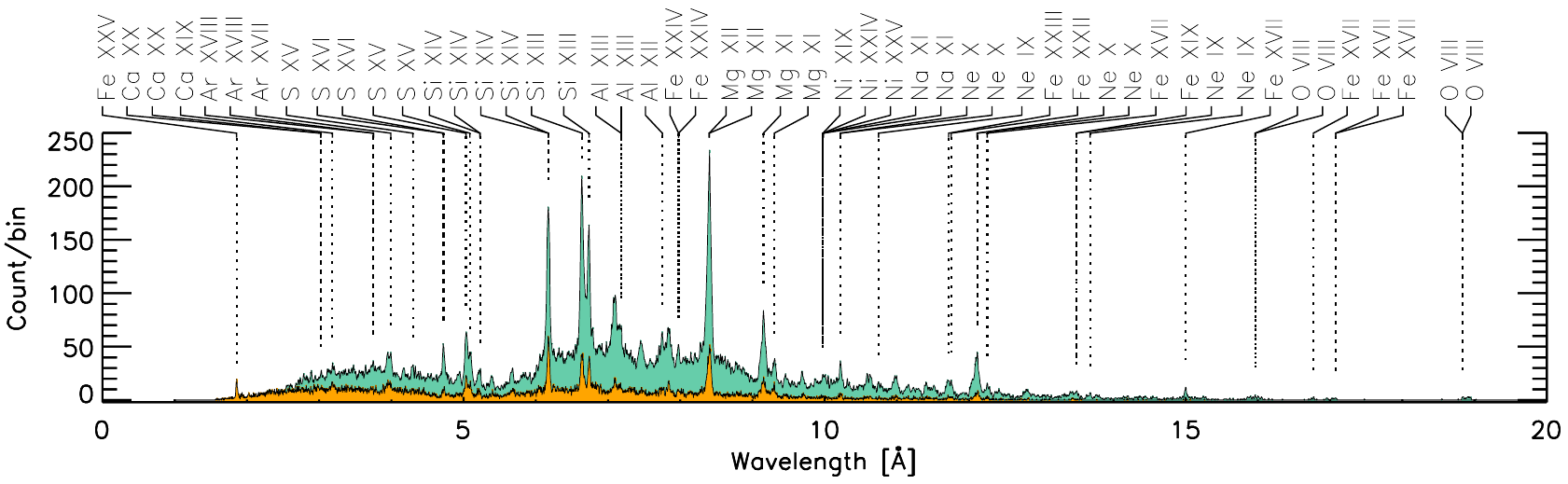},scale=2.4,opacity=0.2,vshift=-65pt,angle=0}
\BgThispage

\lettrineabstract{The Chandra X-ray Observatory has completed a remarkable twenty years in orbit.  A large part of the science program of Chandra during this time has involved the study of stars and their planetary systems.  This primer aims to give the reader a taste of the enormous range of stellar and planetary astrophysics that Chandra has enabled.  Beginning with a tour of the X-ray solar system zoo, including the stunning pulsating X-ray aurorae of Jupiter, we then move on to the hot million degree outer atmospheres of stars like our own Sun, whose X-ray emission is driven by an internal magnetic dynamo. The same emission processes are also vigorously present in the youngest stars, and we highlight some Chandra observations and results on nascent stellar and planetary systems.   Chandra surveys and high resolution spectroscopy of massive stars have provided a new window on the means by which they scavenge X-ray emission from their radiatively-driven winds, sometimes modulating this output by strong underlying stellar magnetic fields.  We touch upon the evanescent X-radiation from intermediate mass stars before arriving at the inevitable evolutionary end points of all but massive stars, first in energized X-ray emitting planetary nebulae, then in the slowly cooling, soft-X-ray emitting photospheres of white dwarfs.  We conclude with white dwarfs in close binary systems, rejuvenated by interaction with a companion and where accretion gives play to a new range of energetic behavior even more spectacular and cataclysmic than the coruscant  astrophysical road down which they have travelled.}

\newpage

\tableofcontents
\backgroundsetup{contents=\includegraphics{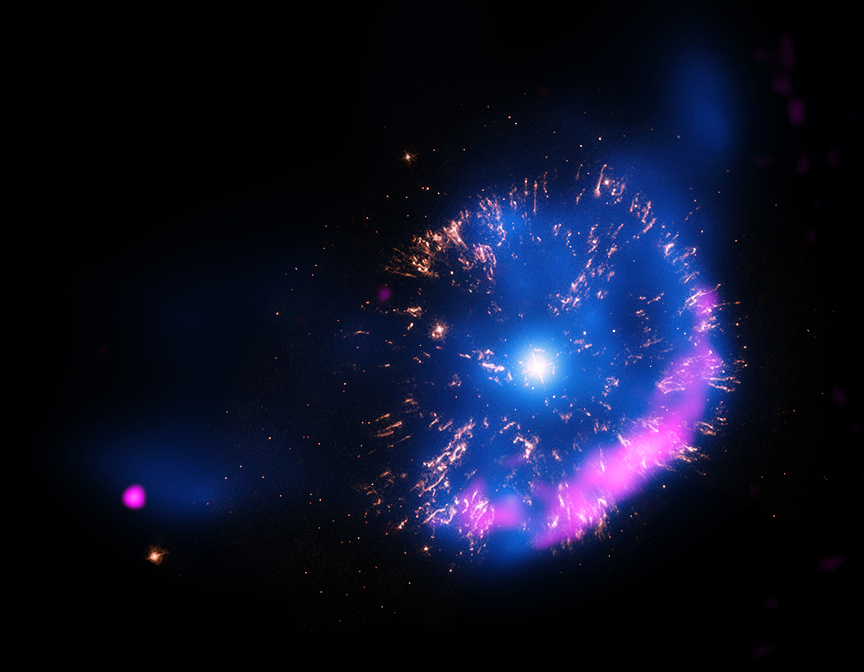},scale=0.45,opacity=1.0,vshift=-350pt,angle=0}
\BgThispage

\newpage

\section{Introduction}

This {\it primer} forms the material that makes up Chapter 4 {\it X-rays from Stars and Planetary Systems} of a forthcoming IoP e-book {\it 20 Years of Chandra Science}, written in celebration of the past 20 years of successful {\it Chandra} operations.  It is not intended as a thorough review---there are far too many omissions of interesting and important papers and results for that.  Rather, it is meant to give the reader a background on the types of studies and achievements {\it Chandra} has made in the field of X-ray studies of stars and planetary systems.

The phrase ``stars and planetary systems" covers a vast array of astrophysical phenomena, from the relatively small scales of planetary science to the hundred parsec scales of the most massive clusters of stars. This primer attempts to summarize {\it Chandra's} major contributions to understanding energetic processes in these environments through their X-ray emission.
\footnotetext[1]{A thorough description of the instrumentation and performance of {\em Chandra} can be found on http://cxc.harvard.edu/proposer/POG/ }

\begin{wrapfigure}{R}{10cm}
\vspace{-0.8cm}
\begin{mdframed}[backgroundcolor=MidnightBlue!20] 
\vspace{-0.22in}
\section*{The {\em Chandra} X-ray Observatory\footnotemark[1]} 
\vspace{-0.13in}
 \sf  NASA's flagship X-ray space telescope, launched on space shuttle mission STS-93 1999 July 23 \\
$\bullet$~~Arcsecond spatial resolution \\
$\bullet$~~High elliptical orbit with 65 hour period\\
$\bullet$~~Advanced CCD Imaging Spectrometer (ACIS) and High Resolution Camera (HRC) microchannel plate detectors\\
$\bullet$~~High Energy Transmission Grating (HETG) spectrometer, 1.2--31~\AA\ with resolving power $R\leq 1000$\\
$\bullet$~~Low Energy Transmission Grating (LETG) spectrometer, 1.2--175~\AA\ with resolving power $R\leq 2000$
\end{mdframed}
\vspace{-0.7cm}
\end{wrapfigure}

X-rays in the cosmos are primarily produced by matter heated to temperatures exceeding a million degrees, or else from matter accelerated to relativistic energies.  It might then come as a surprise that many comparatively cool objects in the solar system are now known to shine in X-rays.  

As we shall see, X-ray emission from stars themselves naturally divides them into the categories of ``high'' mass and ``low'' mass, whose energetic radiation originates from fundamentally different astrophysical processes.  The division occurs at spectral types where stars begin to develop substantial outer convection zones, which on the main-sequence corresponds to type F and later, or a mass of about $1.5 M_\odot$.  The resulting convective flows maintain rotational shear which sustains magnetic dynamo activity whose energy is eventually dissipated at the stellar surface in the form of chromospheric and coronal UV--X-ray emission and the driving of a magnetized wind.  High-mass stars instead drive winds by radiation pressure and their X-rays originate in plasma heated by shocks resulting from instabilities in this process, or by collisions within winds induced by magnetic channeling or with a wind-driving massive binary companion. 

In between high- and low-mass stars lies an intriguing but quite narrow range of stellar masses that are X-ray dark, but which might sustain X-ray activity driven by natal differential rotation for a very brief period after the pre-main sequence phase. 

Supernovae and the neutron star or back hole remnants of high-mass stars are not discussed here---they of course represent an entire sub-field in high energy astrophysics and to do any justice to these topics would require a doubling of the length of this piece.  Here, we briefly touch on the X-ray emission from the hot white dwarf remnants of lower mass stars and from white dwarfs in close, mass-transferring binary systems that form cataclysmic variables and lead to nova explosions.

\section{X-rays from Solar System Bodies}
\label{s:xraysolsys}

X-rays from solar system bodies tend to be driven ultimately by the Sun, although there is emerging evidence that planetary systems can generate X-rays themselves \citep{Dunn.etal:16}.  The story of X-rays in the solar system forms an important segment in the historical narrative of X-ray astronomy itself.  It was a proposed experiment to detect X-rays from the Moon that instead serendipitously made the first detection of a cosmic X-ray source---the low-mass X-ray binary Scorpius X-1 \citep{Giacconi.etal:62}---and ushered in the age of X-rays as a new window on the Universe. 

 In the mid-1990's, only the Earth, the Moon, and Jupiter had been detected in X-rays, along with the Sun which was ostensibly first photographed at X-ray wavelengths by a rocket-born camera in 1949 \citep{Burnight:49}---we return to this type of X-ray source in \S\ref{s:coronae} below. The Moon was studied using a proportional counter instrument from lunar orbit by the Apollo 15 and 16 missions \citep{Adler.etal:73},
while Jupiter was first observed by the {\it Einstein} satellite in 1979 \citep{Metzger.etal:83}.  {\it ROSAT} made the first detections of X-rays from comets (Comet Levy/1990c)
and the Earth's geocorona, although those detections were not recognized
until later (see Sections~\ref{s:earth}, \ref{s:moon}, \ref{s:comets}).   

{\it Chandra} has played a major role in the X-ray exploration of the solar system.
The tally of solar system objects detected in X-rays now includes asteroids, in addition to 
Mercury, Venus, Earth, Mars, Jupiter and its  moons Io and Europa, the Io plasma torus, Saturn and its rings, and Pluto.  {\it Chandra's} unique contribution to this field, as in the many other aspects of astrophysics discussed in this book, stems largely from its high spatial resolution capability combined with imaging spectroscopy.  The reader is also referred to reviews of the general topic of X-rays from solar system objects presented by \citet{Bhardwaj.etal:07}, \citet{Dennerl:14}, and \citet{Branduardi-Raymont:17}.

\subsection{X-ray Emission Mechanisms in Solar System Bodies}
\label{s:mechanisms}

Setting aside the Sun for now, the common thread for X-ray production by solar system objects is that they are essentially cold ($< 1000$~K) and so the energy to generate X-rays has to come from elsewhere.  X-ray emission from solar system bodies falls into the general categories of charge exchange emission, elastic scattering of solar X-ray photons, X-ray fluorescence following inner-shell ionization by solar X-rays, and bremsstrahlung and collisionally-excited line emission caused by impact of energetic electrons and ions.  The energy sources in these processes are electric potential energy, incident photon energy and particle kinetic energy, respectively.  Several solar system objects display X-rays produced by more than one of these mechanisms. 
We take these processes in turn below.

\begin{figure}[tbp]
\begin{center}
\includegraphics[angle=0,width=0.7\textwidth]{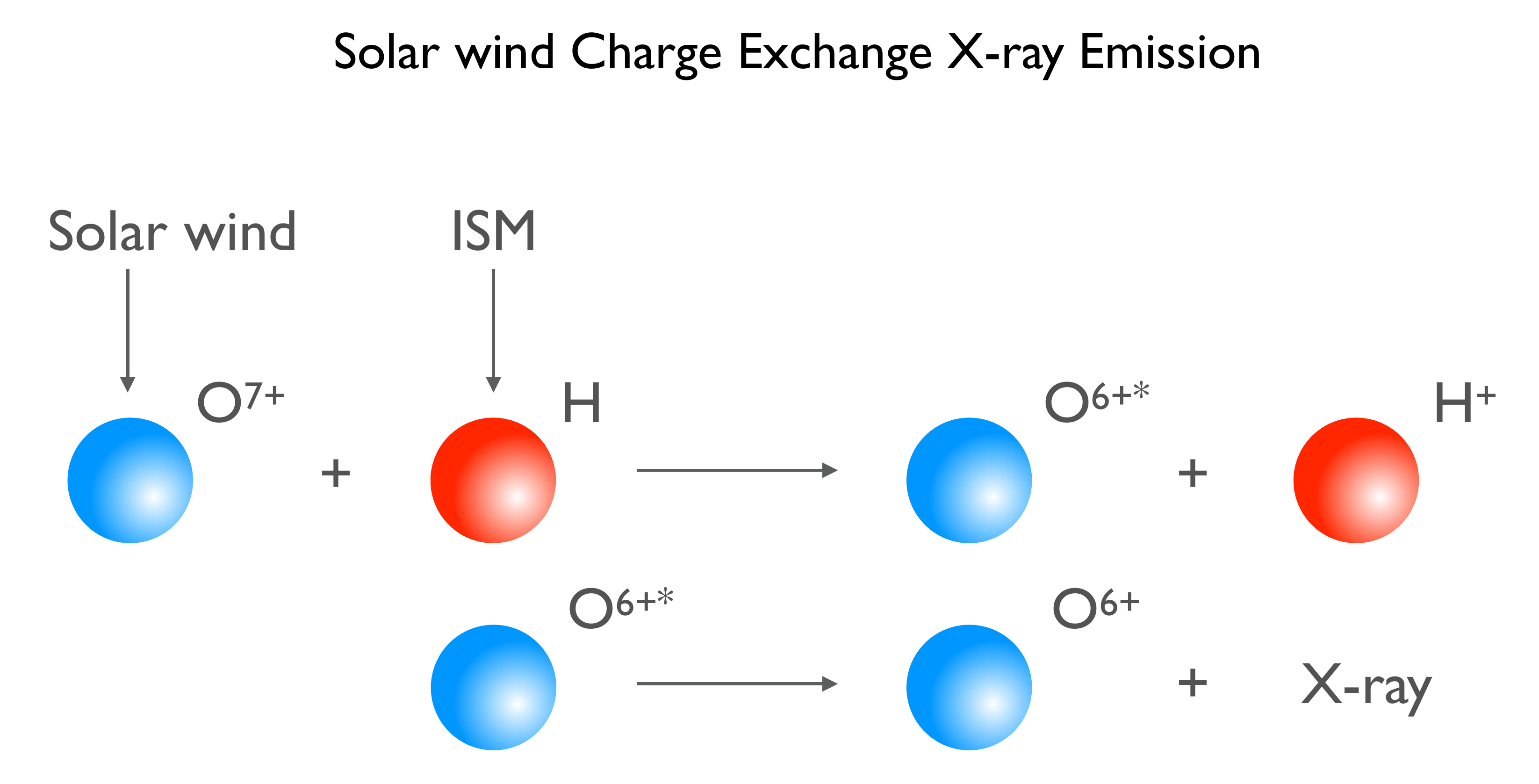}
\end{center}
\caption{Illustration of the charge exchange X-ray emission mechanism in which a highly charged ion interacts with a neutral and captures an electron into an excited state.  In this case, an O$7+$ solar wind ion interacts with a neutral H atom and the decay of the resulting O$^{6+*}$ excited state produces a soft X-ray photon.}
\label{f:cx}
\end{figure}

Charge exchange (commonly abbreviated CX) X-ray emission occurs when a highly-ionized, heavy ion encounters neutral material. Neutrals act as electron donors, and the electrons then combine with the ion in upper levels to leave the ion in an excited state.  It is the decay from this excited state that leads to the emission of an X-ray photon.   In the solar system, the Sun provides a copious supply of highly-charged ions in the form of the solar wind.  A schematic illustration of the process involving O$^{7+}$ ions interacting with neutral hydrogen atoms is shown in Figure~\ref{f:cx}.  Solar wind CX emission (SWCX) was first identified as the origin of the X-rays detected from Comet Hyatuke \citep{Cravens:97}, and has since been observed from several other solar system objects, including a number of more recent comets and the exospheres of Venus, Earth and Mars.

Coherent scattering of solar X-rays and fluorescence---essentially an inelastic scattering process---occur to some extent on all solar system bodies illuminated by the Sun, although these might not be the dominant source of X-ray emission. In the soft X-ray spectral region that {\it Chandra} observes in,  absorption cross-sections are much larger than scattering cross sections, and the majority of the incident X-ray flux is consequently absorbed. Scattered X-rays have nevertheless been observed from Earth, the Moon, Jupiter and Saturn.  We will return to these objects below.  

Fluorescence occurs when an atom is ionized in its inner shell, by either an X-ray photon or energetic electrons, protons or ions. The ``hole" created by the ionization event leaves the ion in an excited state that subsequently decays via the transition of an electron from a higher shell.  
If ionized in the $n=1$ shell (the ``K shell"), 
this transition can be either radiative, corresponding to the emission of a characteristic "K$\alpha$" photon, or non-radiative, corresponding to an Auger transition in which one or more Auger electrons are emitted. In the case of ionization of $n>1$  inner shells  (L, M etc shells), the decay channels get more complex, with transitions from higher subshells of the same shell also being possible---so called ``Coster-Kronig" transitions.  The probability of a radiative transition is very low for low $Z$ elements, and increases strongly with increasing atomic number; the fluorescence yields for O "K$\alpha$" and Fe "K$\alpha$" emission are $8.3\times 10^{-3}$ and 0.34, for example.  Consequently, 
the fluorescence process for the abundant solar system elements C, N and O is very inefficient.

Collisional excitation by electrons and ions, leading to line emission
or bremsstrahlung, can be driven by precipitation of solar wind particles
or when particles are accelerated by magnetospheric processes. Inner-shell ionization by particle impact produces characteristic X-ray line emission in the same way as fluorescence due to photoionization.  We shall see that these processes are relevant for producing auroral X-ray emission. 

\subsection{Terrestrial Planets}
\label{s:terrestrial}

\subsubsection{Venus}
\label{s:venus}

The terrestrial planets subtend sufficiently large angular diameters---up to 66'' for Venus and 25'' for Mars---to be easily resolved by {\it Chandra}. While {\it Chandra} is able to observe as close as $45.5\deg$ to the Sun---closer than any other X-ray telescope past or present---elusive Mercury lives up to its name and, while subtending up to 13'', always resides too close to the Sun for {\it Chandra} to observe.   

Venus generally lies much closer to the Sun than $45.5\deg$ and frustratingly out of reach of {\it Chandra}.  However, its most extreme elongations extend to $48\deg$ and {\it Chandra} was able to capitalize on these short windows of visibility by making the first X-ray observation of Venus in January 2001 (\citealt{Dennerl.etal:02}; see Figure~\ref{f:terrestrial}).  The X-ray signal, obtained with the ACIS-I detector, was found to result from fluorescent scattering of solar X-rays in the Venusian thermosphere, as correctly predicted by \citet{Cravens.Maurellis:01}.  Delightful confirmation of the fluorescence lines was provided by a high resolution spectrum that was also obtained by the LETG and ACIS-S detector.

Later observations of Venus in March 2006 and October 2007, when the Sun was close to minimum activity and X-ray flux, succeeded in detecting the expected much weaker SWCX emission from the interaction of the solar wind with the Venusian exosphere \citep{Dennerl:08}.

\subsubsection{Mars}
\label{s:mars}

\begin{figure}[tbp]
\begin{center}
\includegraphics[angle=0,width=0.321\textwidth]{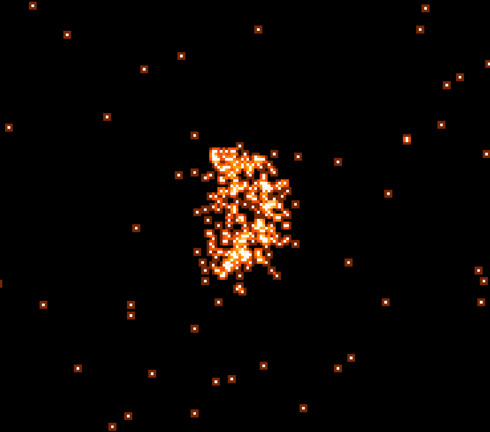}
\includegraphics[angle=0,width=0.285\textwidth]{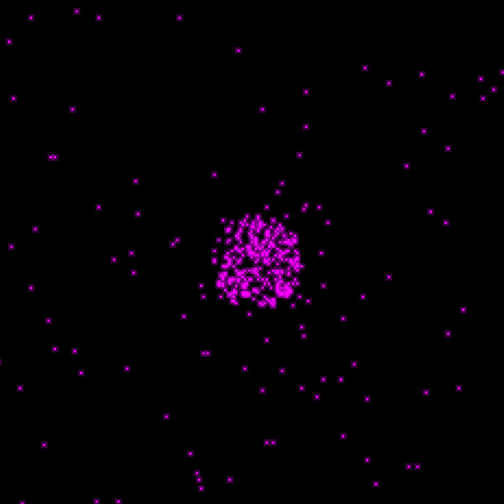}
\includegraphics[angle=0,width=0.38\textwidth]{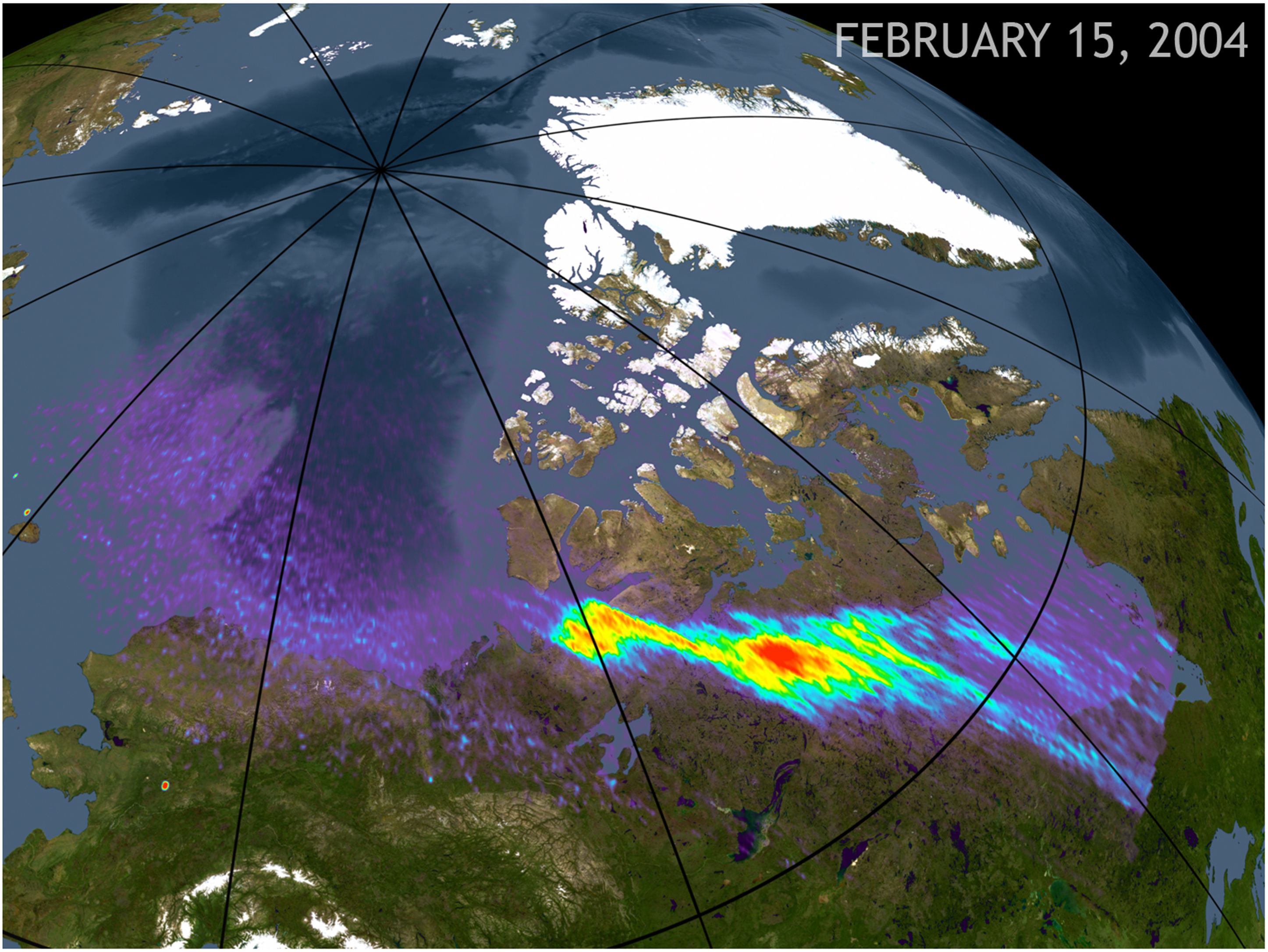}
\end{center}
\caption{{\it Chandra} images of the terrestrial planets.  From left to right: Venus, Mars, and a bright spot and X-ray emitting arc in Earth's aurora superimposed over a visible light representation. From http://chandra.harvard.edu/photo/category/solarsystem.html} 
\label{f:terrestrial}
\end{figure}

Mars was detected in X-rays for the first time in 2001 July 4 in a {\it Chandra} ACIS-I observation, appearing ``as an almost fully illuminated disk, with an indication of limb brightening at the sunward side, accompanied by some fading on the opposite side'' \citep[][Figure~\ref{f:terrestrial}, center]{Dennerl:02}.  The observation was timed when Mars was only 70 million kilometers from Earth, and also near the point in its orbit when it is closest to the Sun.  The ACIS spectrum of the disk of Mars was dominated by the O~K$\alpha$ fluorescence line excited by solar X-rays absorbed at a height of approximately 120 km above the planetary surface.  \citet{Dennerl:02} also traced a faint X-ray emitting halo out to three Mars radii consistent with a thermal bremsstrahlung spectrum with a characteristic temperature of 0.2 keV.   This spectral component was traced to SWCX, between highly charged heavy ions in the solar wind and exospheric hydrogen and oxygen around Mars.  A global dust storm was intensifying while {\it Chandra} observed, but no sign of fluorescence lines from refractory elements such as Mg, Si and Fe, that might be expected from atmospheric dust at very high altitude was detected.

The detection of Mars by {\it Chandra} is testament to its remarkable sensitivity.  The X-ray power emitted from the Martian atmosphere is very small indeed in comparison to the solar heating at Mars, amounting to only 4 megawatts.  In a more terrestrial context, this corresponds to the X-ray power of about ten thousand medical X-ray machines.  

\subsubsection{Earth}
\label{s:earth}

Earth differs from Venus and Mars in being a magnetized planet with a well-developed magnetosphere.  Such a physical system displays a rich array of phenomena associated with the interaction of the solar wind and magnetosphere, the response of the upper atmosphere to ionization by solar extreme ultraviolet and X-ray irradiation, and the complex behavior of atmospheric and precipitated solar wind ions and electrons within this dynamic system.

An X-ray emitting aurora on Earth has been known since balloon and rocket observations beginning in the late 1950s \citep[e.g.][]{Anderson:58,Winckler.etal:59} and observations by spacecraft since the 1970s.   The X-ray aurora on Earth is generated by energetic electron bremsstrahlung \citep[e.g.,][]{Berger.Seltzer:72,Bhardwaj.etal:07}.  

{\it Chandra} observed northern auroral regions of Earth using the HRC-I in 11 different 20 minute observations obtained between 2003 mid-December and 2004 mid-April \citep{Bhardwaj.etal:07}. The data revealed a highly dynamic X-ray aurora, with multiple and variable intense arcs, and diffuse patches of X-rays at times visible and at times absent.  In at least one of the observations an isolated blob of emission is observed near the expected cusp location.  Lacking energy resolution, the HRC-I data could not probe directly the X-ray emission mechanism. However, one observation in 2004 January 24 during a bright arc seen by {\it Chandra} was accompanied, quite unplanned, by an overflight of the Sun-synchronous polar orbiting Defense Meteorological Satellite Program satellite F13 that was able to obtain simultaneous energetic particle measurements. \citet{Bhardwaj.etal:07} used those data to model the expected X-ray spectrum, finding that the observed soft X-ray signal was bremsstrahlung emission together with characteristic K-shell line emission of nitrogen and oxygen in the atmosphere produced by energetic electrons. 

A further source of X-rays from the Earth was only isolated in the mid-1990s: geocoronal SWCX emission resulting from interaction of the solar wind with neutrals in the geocorona.  SWCX also occurs throughout the heliosphere as neutral gas from
the interstellar medium flows by, contributing a significant fraction
of the soft X-ray background \citep{Cravens:00}. The contribution from the geocorona forms a variable background component in all X-ray observations from the Earth's vicinity \citep{Snowden.etal:95,Cravens.etal:01}.  \citet{Wargelin.etal:04} utilized {\it Chandra} observations of the night side of the Moon to show that the very faint signal detected by ROSAT was not due to solar wind impact on the moon, but to the geocoronal glow in the light of O~VII~K$\alpha$ and O~VIII~Ly$\alpha$ resulting from SWCX. The SWCX model for geocoronal X-ray emission was further confirmed in detail by \citet{Wargelin.etal:14} based on {\it Chandra} observations during times of strong solar wind gusts diagnosed by the {\it Advanced Composition Explorer (ACE)} spacecraft situated at the Sun-Earth L1 point about 0.01~AU toward the Sun.  

\subsubsection{The Moon}
\label{s:moon}

The Moon was first observed in detail in X-rays using proportional counters on Apollo 15 and 16 designed to detect fluorescent lines of abundant elements such as Mg, Al and Si from reprocessed solar X-rays \citep{Adler.etal:73}. The relative line strengths of  these elements is dependent on their bulk chemical content on the lunar surface---a direct measurement not provided by mineral-dependent albedo at other wavelengths.  While {\it Chandra} performed a pilot observation of the Moon using ACIS-I in 2001 and  fluorescent lines of O, Mg, Al and Si were detected by \citet{Wargelin.etal:04}, further lunar exploration that could have provided high resolution geochemical lunar maps was unfortunately curtailed by the realization that the ACIS contamination layer might be polymerized by geocoronal H~Ly$\alpha$ emission and potentially jeopardize any future attempts to remove the contaminant by thermal cycling (``bake out'').   

\begin{figure}[tbp]
\begin{center}
\includegraphics[angle=0,width=0.7\textwidth]{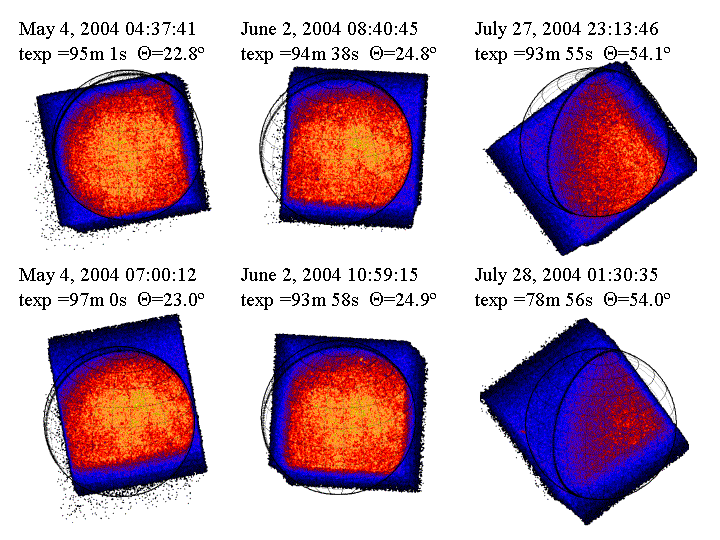}
\end{center}
\caption{Chandra HRC-I images of the Moon obtained in 2004 May, June, and July illustrating X-ray albedo reversal in which maria appear brighter than highlands (from \citealt{Drake.etal:04}).}
\label{f:moon}
\end{figure}

HRC-I observations performed in 2004 May, June, and July succeeded in detecting an albedo reversal with respect to images in visible light, in which optically dark maria are brighter in X-rays than optically bright highlands (\citealt{Drake.etal:04}; Figure~\ref{f:moon}). It is not yet known whether this effect results from differences in weathering or chemical composition.

\subsection{The Gas Giants}
\label{s:gasgiants}

\subsubsection{Jupiter}
\label{s:jupiter}

Like Earth, Jupiter has a comparatively strong magnetic field (7.8 G at the equator) and a well-developed magnetosphere. Coupled with the launching of gas from its moon Io that feeds the Io plasma torus it interacts with, in addition to interactions with the solar wind, Jupiter is arguably the most fascinating solar system object in X-rays.  It was first detected in X-rays by the {\it Einstein} satellite \citep{Metzger.etal:83} and, at the time of writing, Jupiter had amassed a total of 44 separate {\it Chandra} observations over several different campaigns spanning the full mission, from 1999 to 2019. 

X-rays from Jupiter arise from the full gamut of processes discussed in Sect.~\ref{s:mechanisms} above.  The first {\it Chandra} observations found soft X-rays to be concentrated in a hot spot pole-ward of latitudes expected to be magnetically connected with the inner magnetosphere. This surprisingly placed their origin beyond 30 Jupiter radii from the planet, with subsequent observations suggesting the precipitating plasma originates beyond 60 Jupiter radii ($> 4$ million km) from the planet \citep[e.g.,][]{Kimura.etal:16}.

\begin{figure}[tbp]
\begin{center}
\includegraphics[angle=0,width=0.66\textwidth]{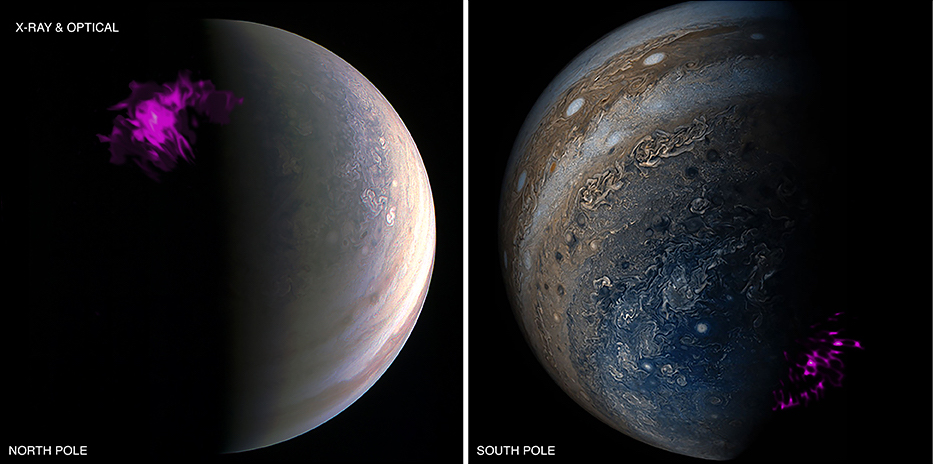}
\includegraphics[angle=0,width=0.33\textwidth]{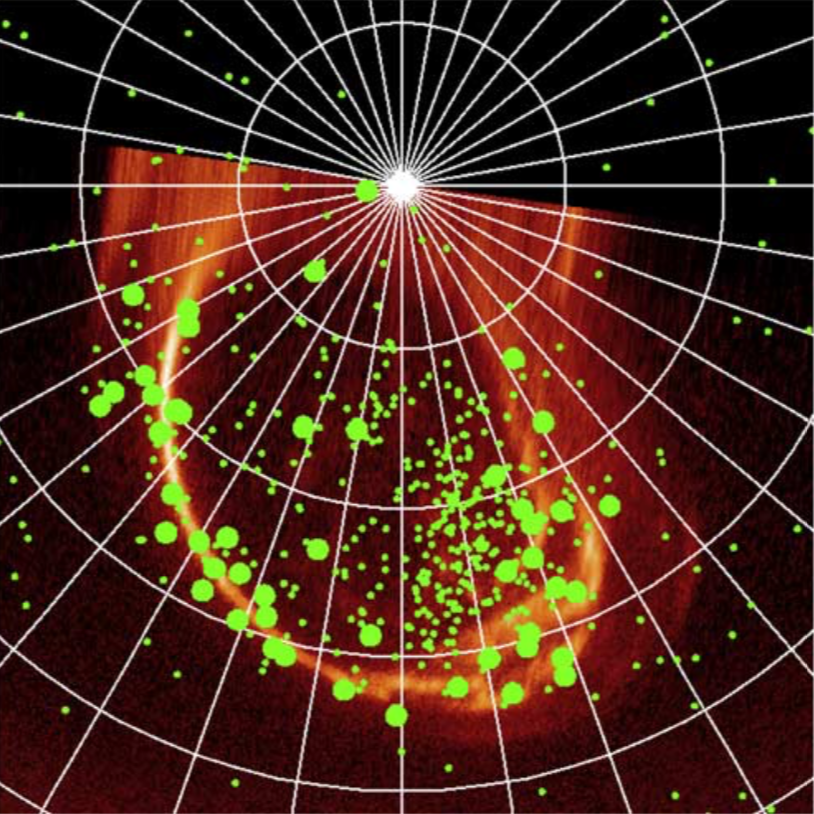}
\end{center}
\caption{Left: X-ray emission observed from the north and south aurorae of Jupiter by {\it Chandra}  (magenta) superimposed on visible light images from the {\it Juno} spacecraft. Based on \citet{Dunn.etal:17} from http://chandra.harvard.edu/photo/2017/jupiter/. Right: Polar projection of a Hubble STIS FUV image of Jupiter's northern aurora (orange) overplotted with the positions of X-ray photons detected by {\it Chandra} during simultaneous observations on 24 February 2003. Small and large green dots indicate photon energies  $< 2$~keV and $> 2$~keV, respectively (from \citealt{Branduardi-Raymont.etal:08}).} 
\label{f:jupiter}
\end{figure}

The hot spot was also observed to be spectacularly pulsating with a period of about 45 min, a phenomena that has also been seen in later observations \citep{Dunn.etal:16} and in energetic particle fluxes and Jovian radio emission \citep{Cravens.etal:03}.  Comparison of more recent observations from May--June 2016 with data obtained in March 2007 have also revealed a persistent southern X-ray hot spot that behaves independently of its northern counterpart and showed 11 min periodic pulsations and uncorrelated changes in brightness \citep{Dunn.etal:17}.   Subsequent studies have shown that Jupiter's aurora exhibits these intriguing regular pulsations of a few tens of minutes in at least 30\%\ of observations \citep{Jackman.etal:18}. Snapshots of the northern and southern Jovian aurorae are illustrated in Figure~\ref{f:jupiter}.  The origin of the pulsation behavior remains uncertain, though is thought to be due to either the ``bounce'' period for a magnetically trapped ion to repeat its north-south motion along a field line \citep[e.g.][]{Cravens.etal:03}, pulsed magnetic reconnection \citep{Bunce.etal:04,Dunn.etal:16}, or Kelvin-Helmholtz instabilities that cause Jupiter's magnetic field lines to resonate \citep{Dunn.etal:17}.  

Spectroscopic observations revealed the origin of the hot spot X-rays to be largely CX emission from energetic highly-charged ions of oxygen, sulfur and carbon \citep{Elsner.etal:05,Branduardi-Raymont.etal:07,Kharchenko.etal:08} that are co-spatial with the FUV emission seen by {\it Hubble} (see Figure~\ref{f:jupiter}, right panel).   The question then is what is the origin of the particles: the solar wind, or the Io plasma torus?  Ion abundances can in principle be used to distinguish between them, the Io plasma having a much larger sulphur abundance relative to oxygen than the solar wind owing to its volcanic origin.  Exploiting an elevated ram pressure from a CME  with a {\it Chandra} target of opportunity observation, \citet{Dunn.etal:16} found a factor of 8 enhancement in the Jovian X-ray aurora and periods of 26~min from S ions at lower latitudes and 12 min from C, O and S ions at the highest latitudes. \citet{Dunn.etal:16} trace the former to precipitation of trapped magnetospheric plasma of Io origin and the latter to the solar wind and more open field lines.

Jupiter also exhibits steady, fainter and more uniform non-auroral X-ray emission at low- and mid-latitudes, although with some surface magnetic field strength dependence \citep{Bhardwaj.etal:06}.  The temporal variation of this emission follows  variations in solar X-ray flux, tying the bulk of its origin to scattering of and fluorescence by solar X-rays, but with some additional component likely originating in ion precipitation from close-in radiation belts.

During {\it Chandra} studies aimed at Jupiter, detections of X-rays in the 0.25--2 keV band from the Galilean satellites Io and Europa, possibly Ganymede, and also the Io plasma torus, were made \citep{Elsner.etal:02}. The latter appeared to be due to bremsstrahlung emission from non-thermal electrons in the few hundred to few thousand eV range.  In contrast, \citet{Elsner.etal:02} found X-rays from the Galilean satellites most likely due to bombardment by energetic H, O, and S ions from the Io plasma torus. This fluorescent signature could hold tremendous promise for X-ray remote sensing of the composition of the Europa and perhaps Enceladus oceans with next generation X-ray missions, or from X-ray spectrometers on future giant planet missions.

\subsubsection{Saturn and Uranus}

\begin{figure}[tbp]
\begin{center}
\includegraphics[angle=0,width=0.48\textwidth]{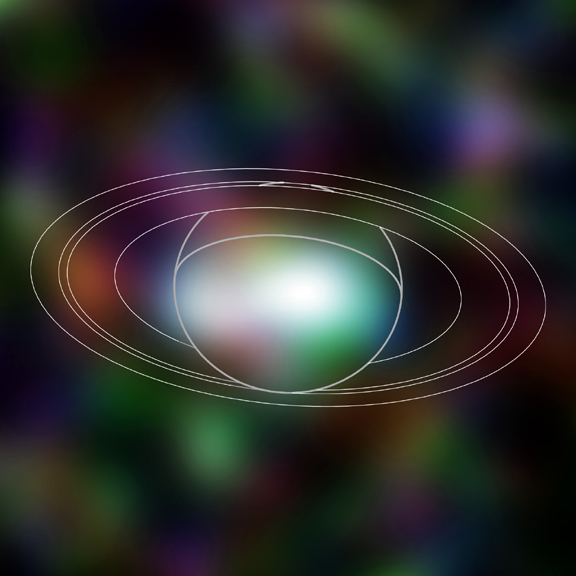}
\includegraphics[angle=0,width=0.48\textwidth]{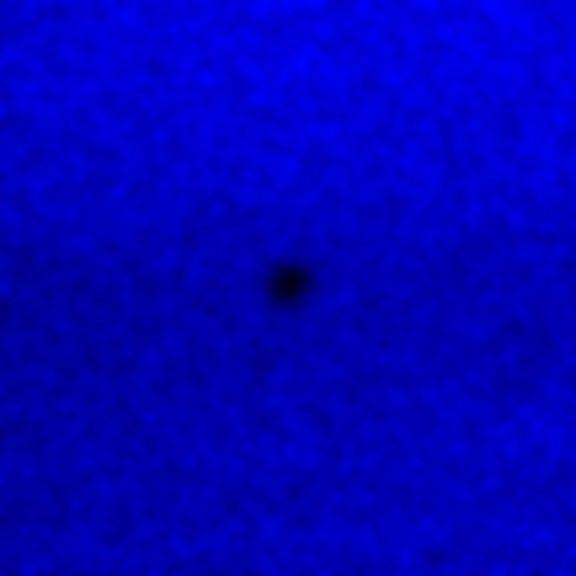}
\end{center}
\caption{Left:  A {\it Chandra}  X-ray image of the X-ray fluorescence and scattering from the surface of Saturn. Right: The X-ray shadow of Titan against the Crab Nebula observed during a transit on 2003 January 5 by \citet{Mori.etal:04}. Both images from http://chandra.harvard.edu/photo/category/solarsystem.html}
\label{f:saturn}
\end{figure}

Saturn was observed for 10ks using the {\it Einstein} Observatory on 1979 December 17 but no X-ray emission was detected \citep{Gilman.etal:86}.  A marginal detection by the {\it ROSAT} PSPC was reported by \citet{Ness.Schmitt:00}, but the first unequivocal detection of Saturn in X-rays had to await a 70~ks observation on 2003 April 14--15 by {\it Chandra} and the ACIS-S detector \citep[][Figure~\ref{f:saturn}]{Ness.etal:04}.  Unlike Jupiter, Saturn does not appear to drive strong X-ray emitting aurorae \citep{Branduardi-Raymont.etal:13}.  Instead, its X-ray signal is similar at both low and high latitudes, including the polar caps.  

In a {\it Chandra} observation obtained in 2004 January, \citet{Bhardwaj.etal:05} found the X-ray signal from Saturn to mirror the X-ray intensity from an M6-class solar flare originating in an active region associated with a sunspot that was clearly visible from both Saturn and Earth.  Analysing the full body of {\it Chandra}, {\it XMM-Newton} and {\it ROSAT} X-ray data on Saturn available until the end of 2004,  \citet{Bhardwaj.etal:05} noted that Saturn's X-ray signal is highly correlated with the solar 10.7~cm radio flux---a well known proxy for coronal emission.   This confirmed  Saturn' s X-ray emission as originating from fluorescence and backscattering of solar X-rays \citep{Bhardwaj.etal:05b,Branduardi-Raymont.etal:10}. 

The beguiling rings of Saturn might still hold some mysteries for future X-ray missions.  First confirmed to shine in the light of O~K$\alpha$ emission by \citet{Bhardwaj.etal:05b}, for which fluorescent scattering of solar X-rays from oxygen atoms in the H$_2$O icy ring material is the most obvious source mechanism, \citet{Branduardi-Raymont.etal:10} found the ring O~K$\alpha$ emission did not share the same dependence on the solar cycle as the disk emission, indicating a possible second excitation mechanism, such as lightning-induced electron beams.

\subsubsection{The X-raying of Titan's Atmosphere}

One of the most spectacular {\it Chandra} observations of the solar system was made in 2003 when Saturn made a very rare transit of the Crab Nebula, a conjunction that we will have to wait until 2267 to see again.  \citet{Mori.etal:04} noted that, although a similar conjunction
occurred in 1296 January, the Crab Nebula---the 
remnant of SN 1054---was probably too small to be occulted rendering the 2003 event the first such transit.

Unfortunately the transit of Saturn itself was during a passage of {\it Chandra} through the radiation belts and was missed. However,  \citet{Mori.etal:04} observed the occultation shadow of the largest of Saturn's moons, Titan, the only satellite in the solar system with a thick atmosphere.  The shadow, illustrated in Figure~\ref{f:saturn}, was clearly larger than the diameter of Titan's solid surface, indicating a thickness of the atmosphere of
$880 \pm 60$~km---essentially consistent with or perhaps slightly larger than estimated from earlier Voyager observations at radio, IR and UV 
wavelengths.  The difference could partly be explained by Saturn being slightly closer to the Sun during the {\it Chandra} observation.

\subsubsection{No X-ray Detection of Uranus Yet}

Uranus has so far proven to be elusive in X-rays.  {\it Chandra} has observed it twice without detecting it: in 2002 August for 30~ks using ACIS-S and in 2017 November for 50~ks with the HRC-I during a CME impact.  The observations aimed to see auroral emission, like that seen on Jupiter. Uranus' magnetic field is at 60 degrees to the rotation axis, and expected to produce a dynamic and constantly changing magnetosphere.  Detection of any X-ray emitting aurorae on Uranus must await longer exposures. While a fluorescent and scattering signal from processed solar X-rays is also expected, exposure times required for detection are also an order of magnitude longer than the existing observations. 

\subsection{Minor Planets and Comets}
\label{s:minor}

\begin{figure}[tbp]
\begin{center}
\includegraphics[angle=0,width=0.8\textwidth]{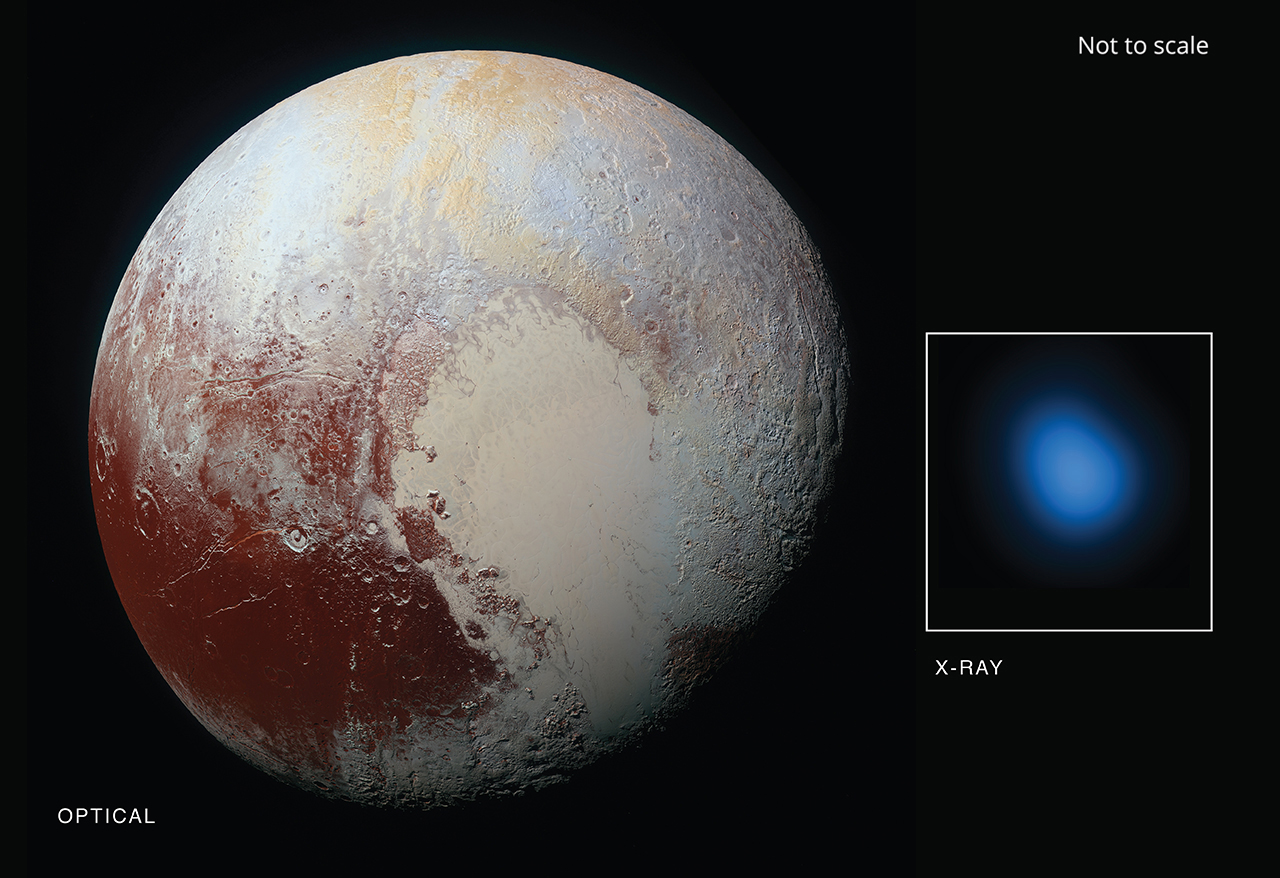}
\end{center}
\caption{An image of Pluto obtained by {\it New Horizons} in visible light, together with the X-ray image obtained by {\it Chandra} for observations obtained in 2014--2015. From http://chandra.harvard.edu/photo/2016/pluto/.}
\label{f:pluto}
\end{figure}

\subsubsection{The Remarkable X-rays from Pluto}
\label{s:pluto}

The ice giants, Uranus and Neptune, likely have to await the arrival of a mission with higher sensitivity than {\it Chandra} or {\it XMM-Newton} \citep[see, eg.,][]{Snios.etal:19}.  It is then quite remarkable that {\it Chandra} detected X-rays from Pluto.

Pluto was targeted in a campaign to support the New Horizons flyby in observations on 2014 Feb 24,
and in three further visits from 26 Jul to 03 Aug 2015, for total integration time of 174~ks.  A total of 8 X-ray photons were detected, all of which were in the 0.3--0.6~keV passband, amounting to a net signal of 6.8 counts after background subtraction \citep{Lisse.etal:17}.  The origin of this signal remains a mystery. \citet{Lisse.etal:17} noted that the X-ray power from Pluto---approximately 200 MW---was comparable to that of other Solar System X-ray emission sources such as auroral precipitation, solar X-ray scattering, and SWCX.  However, Pluto appears to lack a significant magnetic field that could drive aurorae, and no auroral airglow has been detected. Moreover,  the backscattered solar X-ray flux is expected to be 2--3 orders of magnitude below detection thresholds.  While SWCX can produce the appropriate X-ray photon energies, the lower than expected neutral escape rate from Pluto found by New Horizons \citep{Gladstone.etal:16} leaves the observed solar wind flux in Pluto's vicinity a factor of about 40 too weak to account for the observed signal.

\subsubsection{Comets: the Rosetta Snowballs of Charge-Exchange X-ray Emission}
\label{s:comets}

Comets were first detected in X-rays by ROSAT, beginning with C/1996 B2 (Hyakutake; \citealt{Lisse.etal:96}) rapidly followed by several others found by \citet{Dennerl.etal:97} in archival data.  As noted in Section~\ref{s:mechanisms}, these detections became of special general importance to X-ray astronomy because they established the charge exchange process as an important source of X-ray emission \citep{Cravens:97}. 
In turn, X-rays became an important diagnostic of cometary gas and dust production rates, as well as a means of probing the solar wind \citep{Dennerl.etal:97,Kharchenko.Dalgarno:00,Snios.etal:16}. The SWCX signal originates with the interaction of solar wind heavy ions with outgassing cometary neutrals and probes the gas in the coma independently of the dust, which cannot easily be done at other wavelengths \citep{Dennerl.etal:12}. 

\begin{figure}[tbp]
\begin{center}
\includegraphics[angle=0,width=0.45\textwidth]{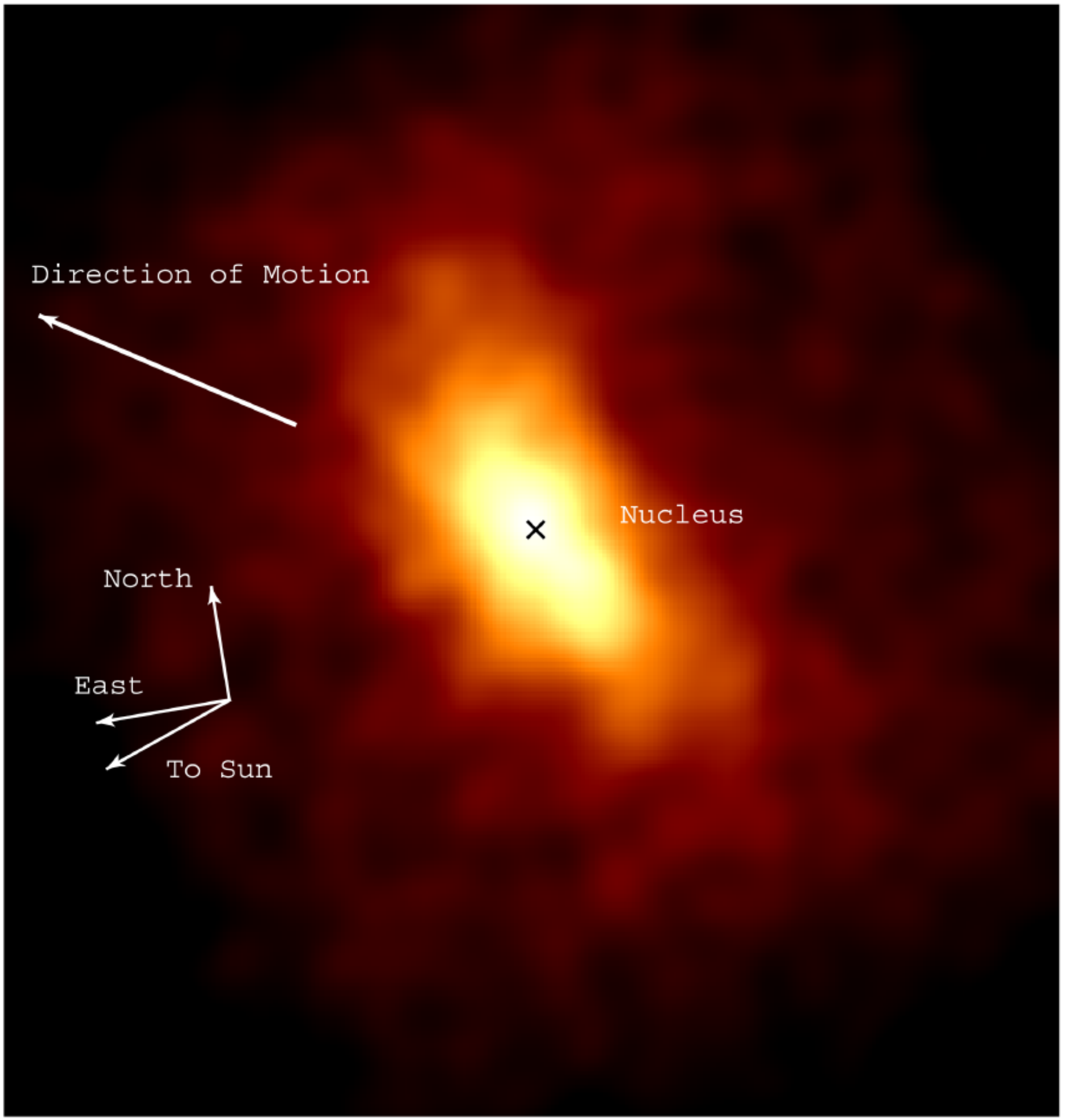}
\includegraphics[angle=0,width=0.53\textwidth]{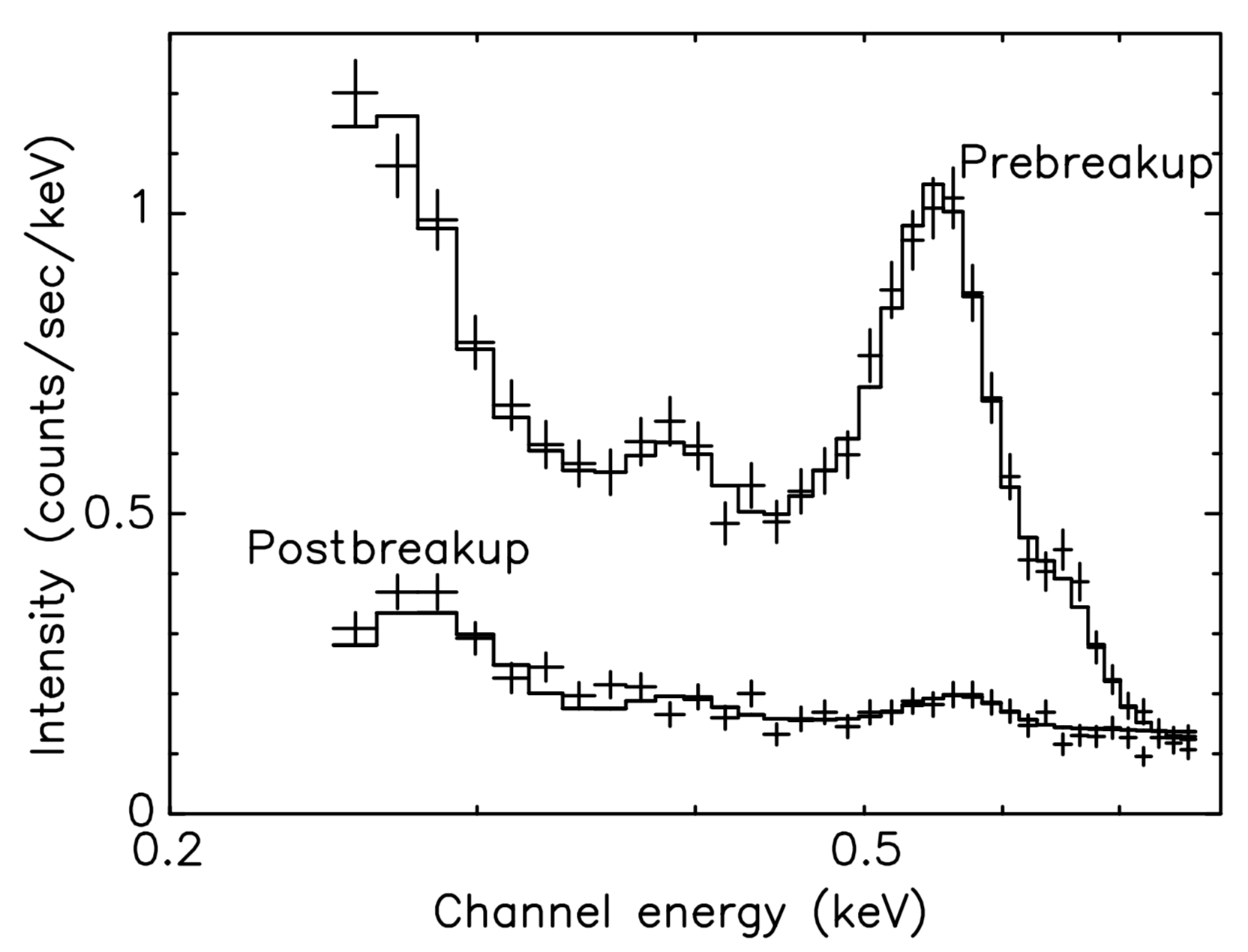}
\end{center}
\caption{Left: Image of the X-ray emission from comet C/1999 S4 (LINEAR) detected by {\it Chandra} in a  2.5 hour observation on 2000 July 14.  Right: The ACIS-S spectrum of C/1999 S4 (LINEAR) fitted with a spectral model comprising six lines arising from CX of solar wind C$^{5+}$, C$^{6+}$, N$^{7+}$, O$^{7+}$, and O$^{8+}$ ions with neutrals in the comet coma together with a weak bremsstrahlung component.  Both images from \citealt{Lisse.etal:01}.
} 
\label{f:clinear}
\end{figure}

{\it Chandra} has played a vital role in cometary X-ray studies. While the SWCX mechanism proposed by \citet{Cravens:97} remained the most likely explanation for cometary X-ray emission, the spectral resolution of the ROSAT PSPC was insufficient to resolve the expected line signatures from He-like and H-like C, N and O. It was not until {\it Chandra} observed  C/1999 S4 (LINEAR) with ACIS-S that definitive spectroscopic evidence of these lines was obtained \citep[][see Figure~\ref{f:clinear}]{Lisse.etal:01}.  The C/1999 S4 (LINEAR) spectrum also indicated the presence of a weak thermal bremsstrahlung component and subsequent {\it Chandra} observations of comets have confirmed that coherent scattering of solar X-rays by comet dust and ice particles contribute significantly to the signal at energies $> 1$~keV \citep{Snios.etal:14,Snios.etal:18}.

Cometary SWCX emission is dependent on the elemental composition, ionization state, number density and speed of the impacting solar wind, as well as the density of particles in the comet coma. Several studies have exploited this and used {\it Chandra} observations of comets to probe solar wind conditions.  \citet{Christian.etal:10} and \citet{Snios.etal:16} found very different X-ray signals from comets at high and low solar latitudes. 
The former interact with the low-density fast solar wind that is deficient in highly ionized species and characterized by lower ion freeze-in temperatures.  X-rays from high latitude comets were commensurately softer, and lacking in emission from species such as O$^{8+}$ and Ne$^{9+}$.

\section{X-rays from Low Mass Stars}
\label{s:coronae}

\begin{figure}[tbp]
\begin{center}
\includegraphics[angle=0,width=0.90\textwidth]{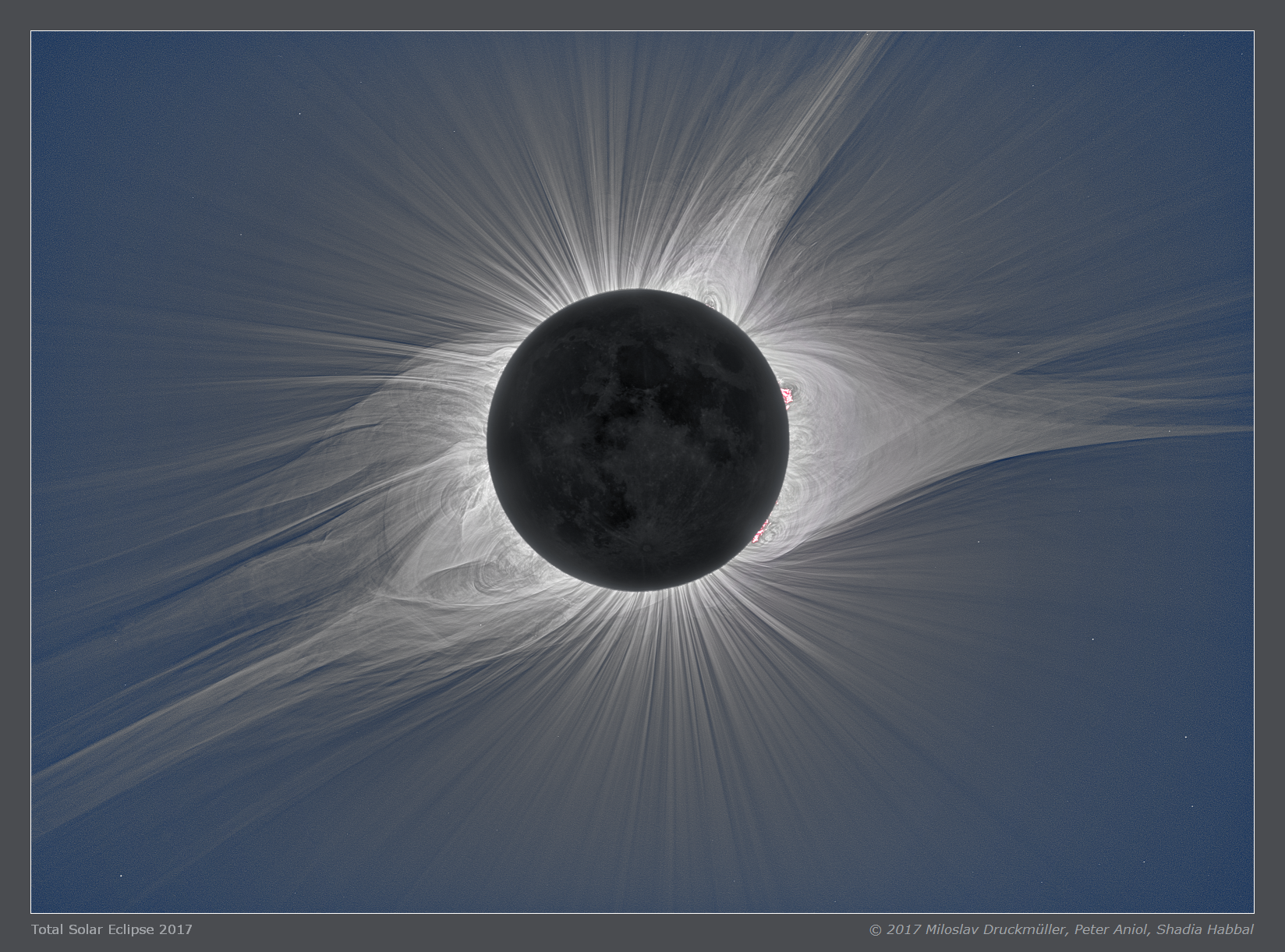}
\includegraphics[angle=0,width=0.45\textwidth]{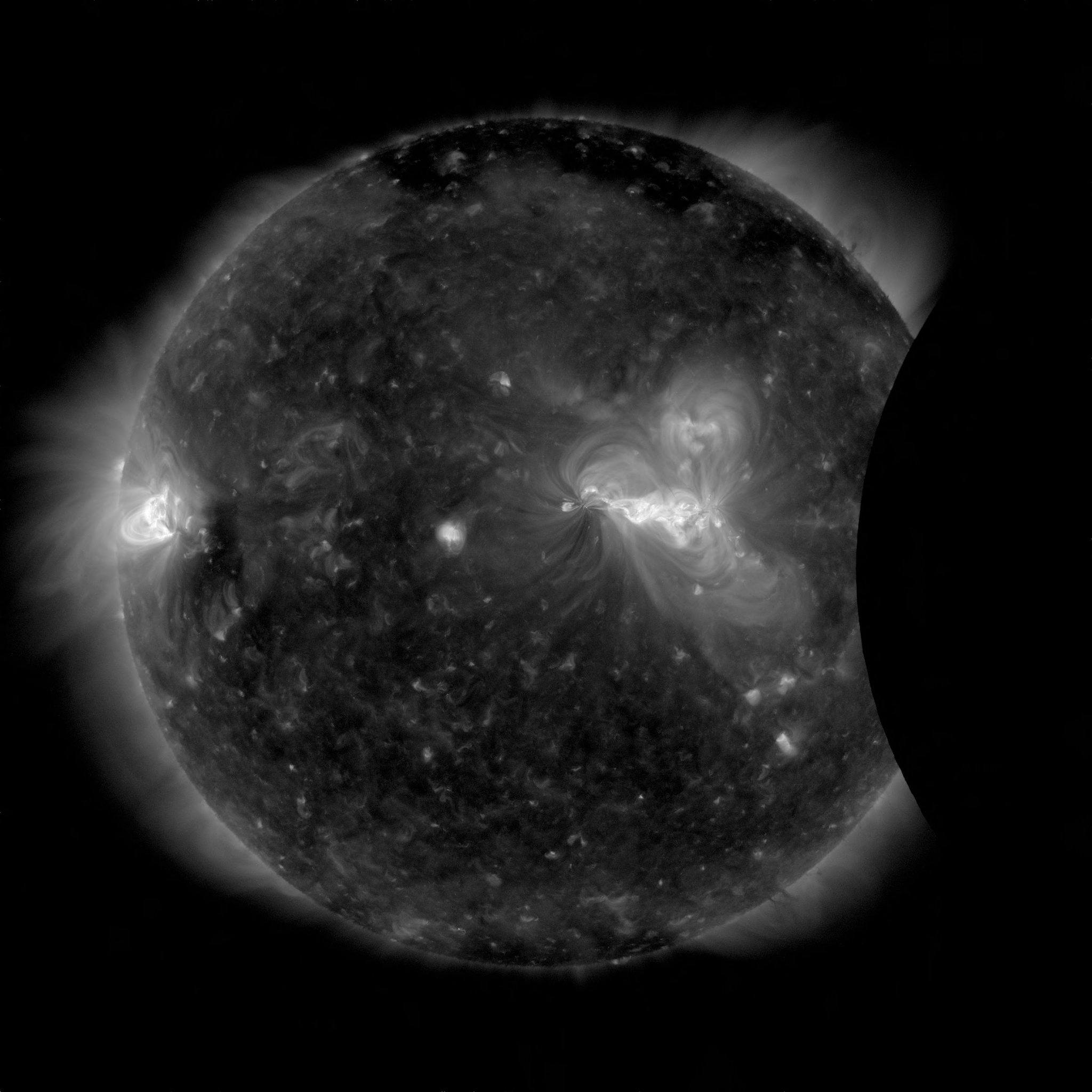}
\includegraphics[angle=0,width=0.45\textwidth]{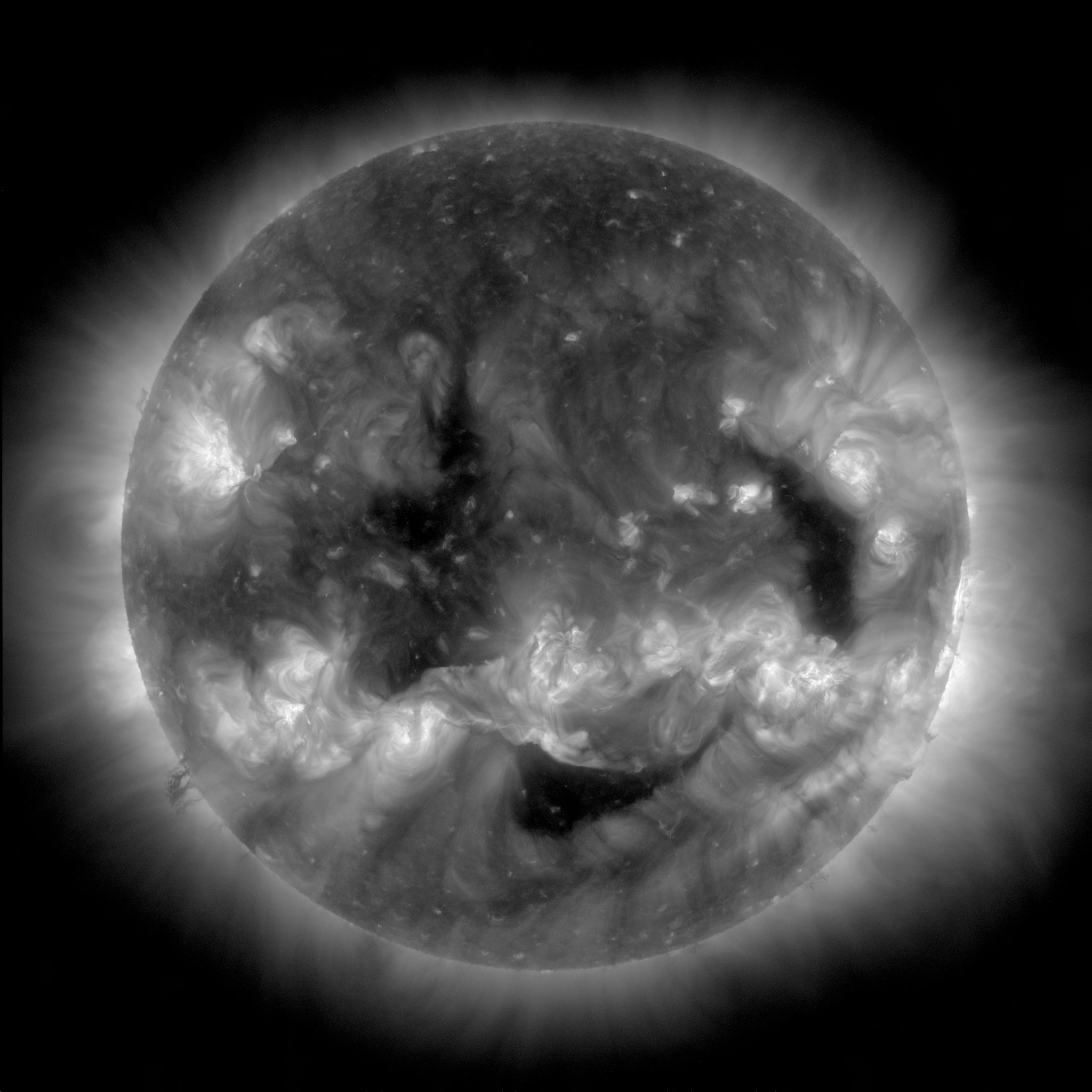}
\end{center}
\caption{Top: White light composite image of the solar corona showing scattering of solar visible light by hot electrons obtained during the north american eclipse of 2017 August 21 
(Image credit: Predictive Science Inc., Miloslav Druckm\"uller, Peter Aniol,
Shadia Habbal and NASA Goddard Space Flight Center, Joy Ng).   
Bottom left: The solar corona at the time of the eclipse, when the Sun was at a fairly low activity level, seen in the 211~\AA\ filter of the SDO AIA.  This filter is primarily sensitive to emission lines of Fe XIV formed at a temperature of approximately $2\times 10^6$~K. The shadow of the Moon can be seen encroaching to the right. Since SDO is in an inclined geosynchronous orbit, it did not experience the totality witnessed on the ground. Bottom right:  The AIA 211~\AA\ image near solar maximum exactly four years earlier on 2013 August 21, exhibiting a rich array of active regions.
} 
\label{f:whitelight}
\end{figure}

Humans have been studying the hot, million degree outer atmosphere, or ``corona", of the Sun for hundreds if not thousands of years, albeit without realising it until about 80 years ago.  A recent spectacular example of this ground-based activity from the 2017 north american solar eclipse is shown in Figure~\ref{f:whitelight}.  The white light corona we see with the naked eye, and rendered in spectacular detail in this composite exposure, is the light scatted by electrons in the million degree plasma that makes up the outer corona and the solar wind. The impressive filamentary structure brought out betrays the beautiful complexity of the magnetic field to which the fully-ionized plasma is tied and constrained.

While the white light corona has been appreciated for centuries, the beginning of the study of the hot outer atmospheres of stars can be traced to a breakthrough in the late 1930's and early 1940's when work by Edl\'en and Grotrian lead to identification of the ``red'' line at 6375~\AA\ seen during eclipses with a forbidden line of Fe X and betraying a temperature of a million Kelvin.  \citet{Hunter:42} commented ``The immediate reaction of most astronomers will no doubt be one of incredulity that such highly ionized matter as Edl\'en's proposals call for should exist in the outer envelope of a relatively cool star like the sun''.  While decades of familiarity have diffused the incredulity, a large part of the research on stellar coronae since then has been devoted to trying to understand how such temperatures arise---the ``coronal heating problem''.  While certain heating mechanisms are now known to operate in the solar corona, a definitive answer to the coronal heating problem has remained elusive. 

The million degree temperature of the solar corona was confirmed in suborbital experiments carried on V2 rockets captured from Germany following World War~II. The first solar X-ray image is traditionally attributed to a photograph from a flight on 1948 August 5 by \citet{Burnight:49}, though the X-ray nature of the photographic signal was disputed by \citet{Friedman:80} who argued that the rocket did not reach a high enough altitude.
The first actual detection of solar (and stellar!) X-rays was made a year later in 1949 September 29 with a V-2 flight carrying photon counting tubes \citep{Friedman.etal:51}.  Subsequent years have witnessed a myriad rocket and satellite instruments trained on the solar corona with increasing spectral and spatial resolution. Indeed, the impetus to improve spatial resolution imaging of the solar corona using grazing incidence X-ray telescopes \citet{Vaiana.etal:73} played a significant role in the train of development of X-ray optics that lead to the exquisite {\it Chandra} mirrors.

The history and development of the study of the solar corona is wonderfully summarised in the books by \citet{Golub.Pasachoff:09} and \citet{Mariska:92}, and in numerous reviews \citep[e.g.][]{Vaiana.Rosner:78} to which the reader is referred for more detailed accounts.

Also shown in Figure~\ref{f:whitelight} is an image of the Sun obtained at the time of the eclipse by the {\it Solar Dynamics Observatory} (SDO)  {\it Atmospheric Imaging Assembly} (AIA) in the 211~\AA\ band.  While this wavelength lies firmly in the extreme ultraviolet spectral region, the bandpass is designed to capture light from transitions of Fe~XIV formed at a temperature of 2 million K---a temperature at which we are also accustomed to observing in X-rays.  This can therefore be considered a good X-ray proxy image of the solar corona. 

The X-ray corona comprises bright active regions characterized by plasma trapped within magnetic loops, dark coronal holes, small-scale bright points and diffuse emission components.  As we shall discuss below, stars like the Sun also exhibit rapid variability and flares, resulting from the sudden release of stored magnetic energy in the form of accelerated particles, heat, and ejections of mass.

The 2017 eclipse occurred when the Sun had declined considerably in activity from the last solar maximum in the summer of 2013 and was more than half way toward minimum.  The corona captured by AIA in Figure~\ref{f:whitelight} is then rather weak, with very few active regions compared with conditions at solar maximum.  Also shown is an image in the same 211~\AA\ band near solar maximum exactly three years earlier, displaying a much richer array of bright active regions.  

The conspicuous aspect of the images in Figure~\ref{f:whitelight} is the magnetic nature of the corona.  It might then be surprising to learn that it was not actually until the first X-ray survey of stars was made by the {\it Einstein} observatory in the late 1970s that it was established that stellar coronae are magnetically-driven. Despite the coincidence of bright X-ray emitting regions with magnetically active regions on the solar disk, a common view up to the late 1970s was that the corona was heated acoustically \citep[e.g.][]{Biermann:46,Schwarzschild:48} and that the formation of coronae was distinct from magnetic field generation \citep[e.g.][]{Vaiana:80}.  \citet{Vaiana.etal:81} showed that the X-ray luminosities of stars of different spectral type were grossly inconsistent with acoustic heating models, while \citet{Pallavicini.etal:81} established that X-ray luminosity was strongly correlated with stellar rotation known to drive magnetic dynamos.  This picture of stellar coronae powered ultimately by a magnetic dynamo in the stellar interior had fully crystalized by 1980 \citep[e.g.][]{Rosner:80}. 

A number of observatories following {\it Einstein} have added details to the picture.  First and foremost is the 
{\it Roentgen} satellite that surveyed the sky at EUV and X-ray wavelengths and added many thousands of X-ray detections of stars \citep{Voges.etal:99}. The {\it Extreme Ultraviolet Explorer} (EUVE) obtained EUV spectra and plasma and chemical abundance diagnostics of stellar coronae \citep{Bowyer.etal:00,Drake.etal:96} that provided a precursor to the  X-ray grating spectroscopy  enabled by {\it Chandra} and {\it XMM-Newton}.  The {\it Advanced Satellite for Cosmology and Astrophysics} (ASCA) and {\it BeppoSAX} made extensive X-ray observations of stars at low spectral resolution (``CCD resolution"), providing valuable temperature and abundance measurements.  For reviews of these developments through to the {\it Chandra} and {\it XMM-Newton} era see, e.g., \cite{Drake:01} and \citet{Gudel:04}.

By the year 1999 and the launch of {\it Chandra}, the scene for stellar coronae was largely set and awaited the execution of observations to push to the next level of understanding.

\subsection{Properties of Stellar Coronal Emission}

Solar and stellar coronal spectra are characterized by emission lines from abundant ionized chemical elements superimposed on a continuum 
The lines are formed by collisional excitation and subsequent decay; the continua are produced in recombination free-bound transitions.  Solar and stellar outer atmospheres---the chromosphere, transition region and corona---span a temperature range of $10^4$-$10^8$~K. The EUV spectral range is where the dominant emission of the transition region and cooler plasma of the corona resides (although lines of very highly charged ions can be found, such as those of Fe up to Fe~XXIV; e.g.\ \citealt[][]{Drake:99}), while the X-ray range is the general regime of emission for plasma with temperatures $> 10^6$~K. 

The relatively low densities of the plasma in solar and stellar coronae (e.g.\ $n_e\sim 10^8$--$10^{10}$ in non-flaring plasma of
the solar corona) render the emission optically thin and
collision-dominated.  If the plasma is also in thermal equilibrium,
then the form of the emergent spectrum for a plasma at a given temperature 
is in principle determined by only three
parameters (albeit with the requirement of a substantial amount of
data describing the collisional excitation and ionization processes involved): temperature,
density and chemical composition.

\begin{figure}
\begin{center}
\includegraphics[width=0.7\textwidth]{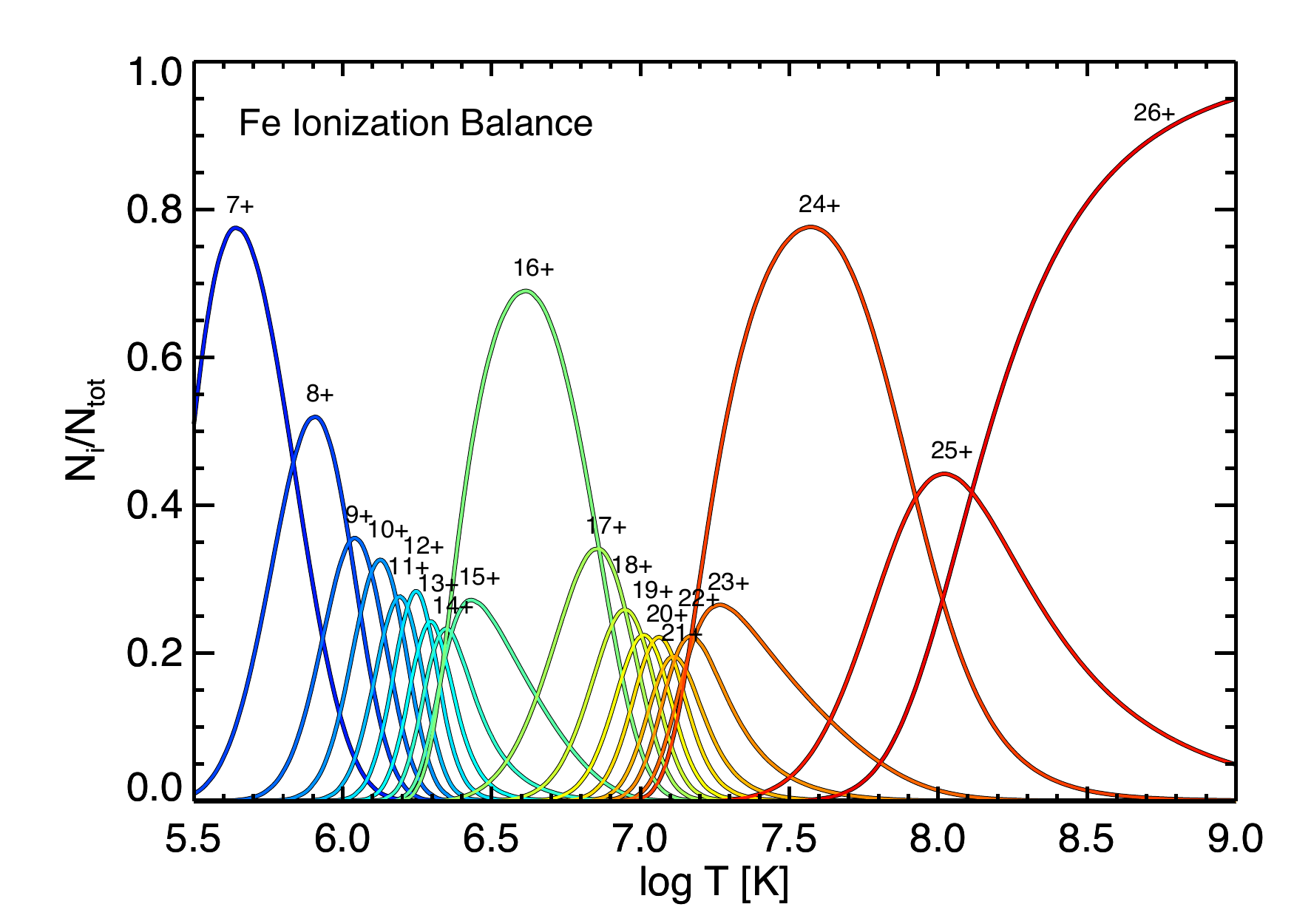} 
\end{center}
\caption{The ionization balance for Fe ions for the temperature range relevant for X-ray emission from the CHIANTI project \citep{Dere.etal:09,Landi.etal:13}. Note the broader temperature ranges over which the ions with filled outer electron shells dominate---in this case Fe$^{7+}$ (Ar-like isolectronic sequence), Fe$^{16+}$ (Ne-like) and Fe$^{24+}$ (He-like).
}
\label{f:feionbal}
\end{figure}

The lack of significant optical depth in a collision-dominated plasma
means that any one volume element of plasma is radiatively decoupled
from any other volume element; the plasma can then be thought of
simply as a collection of quasi-isothermal plasmas of different
temperatures, each occupying a different volume element.  
The emergent
intensity of a given spectral line from one of these isothermal plasma
elements simply depends on the volume integrated product of the number
density of the emitting ionic species, the number density of the
line-exciting species (mostly electrons), and the appropriate
excitation coefficient describing the efficiency of the line
excitation mechanism.  Through the ionization state of the plasma and
the relative abundance of the element in question, for a given
transition this product is proportional to the electron density
squared, $n_e^2$.  

More formally, for a transition $u\rightarrow l$, the intensity 
$I_{ul}$ is given by
\begin{equation}
I_{ul} = A K_{ul}\int_{\Delta T_{ul}} G_{ul}(T) n_e^2(T) \; dV(T)
\label{e:intensity} 
\end{equation}
where $A$ is the elemental abundance, $K_{ul}$ is a
known constant which includes the wavelength of the transition and the
stellar distance, and $G_{ul}(T)$ is the ``contribution'' function
of the line containing all the relevant atomic physics parameters  
(parent ion population and collisional excitation rates).

In the solar context, the integral in Equation~\ref{e:intensity} is often carried out over
the plasma depth rather than the volume.  The quantity
$n_e^2(T)V(T)$---the product of the volume and electron density squared for plasma at temperature $T$---is more applicable to the stellar context when the stellar disk is not resolved and is 
usually referred to as the {\em volume emission measure} (VEM).  Recasting
this integral into one over the temperature, $T$, of the emitting plasma
yields the expression for the total intensity of a spectral line:
\begin{equation}
I_{ul} = A K_{ul} \int_{\Delta T_{ul}} G_{ul}(T) \overline{n_e^2(T)}\,
\frac{dV(T)}{d\log T}\;d\log T
\,\,\, \mbox{erg cm$^{-2}$ s$^{-1}$}.
\label{e:flux}
\end{equation}
Here, the VEM has been transformed to its logarithmic differential form
the {\em Differential Emission Measure} (DEM; see also Sect.~\ref{s:dem}),
\begin{equation}
DEM(T)=n_e^2(T) \frac{dV(T)}{d\log T} \,.
\label{e:dem}
\end{equation}

The term $G_{ul}(T)$ is often referred to as the line contribution function. Stellar coronae are low density plasmas such that spontaneous decay rates usually greatly exceed collisional excitation rates. This ``coronal approximation" enables the greatly simplifying assumption that essentially all ions are in their ground states and that collisional de-excitation is negligible.
The contribution function can then be written 
\begin{equation}
G_{ul}(T)=\frac{n_i(T)}{n_{tot}} C_{lu}(T) B_{ul}, 
\label{e:gul}
\end{equation}
where $n_i$ is the number density of ionization state $i$ of the element in question with a total number density $n_{tot}$, $C_{lu}$ is the collisional excitation rate coefficient for the transition $l\rightarrow u$, and $B_{ul}$ is the branching ratio for the $u\rightarrow l$ transition in relation to all other possible radiative de-excitation routes from level $u$.  Under the coronal approximation, Equation~\ref{e:gul} differs from a two-level atom model only in the branching ratio term, $B_{ul}$. 

\begin{figure}
\begin{center}
\includegraphics[width=0.7\textwidth]{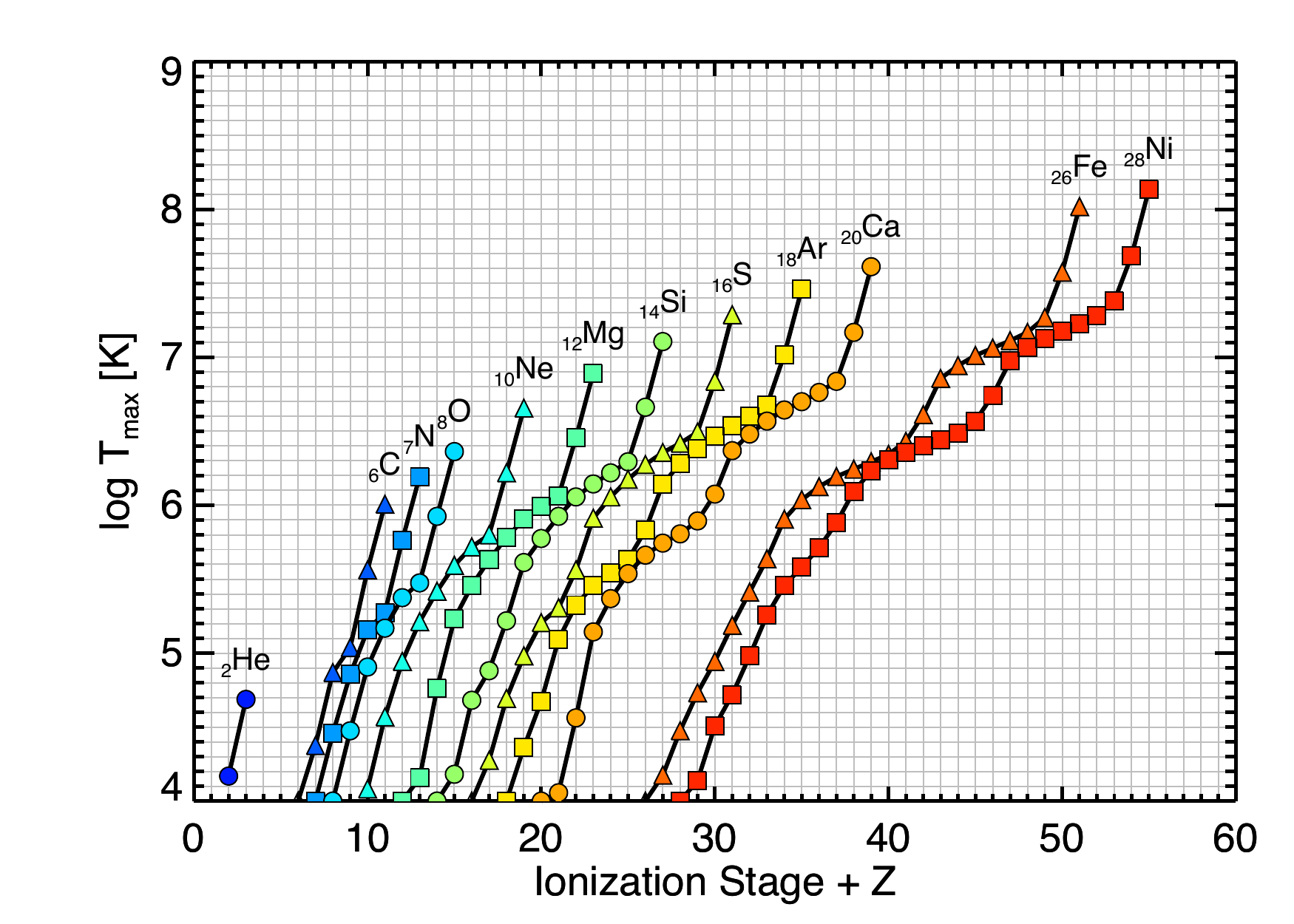} 
\end{center}
\caption{The temperature of maximum ion population for abundant elements. The x-axis is the sum of the ionization stage and atomic number, such that Fe$^{16+}$, (Fe~XVII), for example, lies at position $26+16=42$.  Species relevant for X-ray emission are those formed at temperatures of approximately $10^6$~K and higher.
}
\label{f:tmax}
\end{figure}

The rate coefficient $C_{lu}(T)$ represents the collisional excitation rate integrated over a Maxwellian velocity distribution for temperature $T$, and in modern assessments is usually the product of detailed quantum mechanical calculations and less commonly of laboratory experiments. Examples of the atomic data assessments and reviews have been presented by \citet{Dere.etal:97,Boyle.Pindzola:05,Kallman.Palmeri:07,Foster.etal:10,Foster.etal:12}. 

The dominant temperature dependence in Equation~\ref{e:gul} arises through the ionization balance term $n_i/n_{tot}$, which is computed from the equilibrium between collisional ionization and radiative and dielectronic recombination \citep[see, e.g.,][]{Jordan:70,Arnaud.Rothenflug:85,Bryans.etal:09,Dere.etal:09}.  For a given ion, $n_i/n_{tot}$ is strongly peaked at its characteristic temperature of formation.  This is illustrated for the case of Fe ions in the temperature range of relevance for X-ray emission in Figure~\ref{f:feionbal}.  Note in particular the somewhat broader population profiles of the ions with full outer shells, Fe$^16+$ and Fe$24+$.
The peak temperatures of formation of the ions of abundant elements are shown in Figure~\ref{f:tmax}, from which the ions relevant for X-ray emission ($T$ \gax $10^6$~K) can be seen. Thus, spectral lines of a given ion provide a strong indication of the plasma temperature that produces them.

\begin{figure}
\begin{center}
\includegraphics[width=0.7\textwidth]{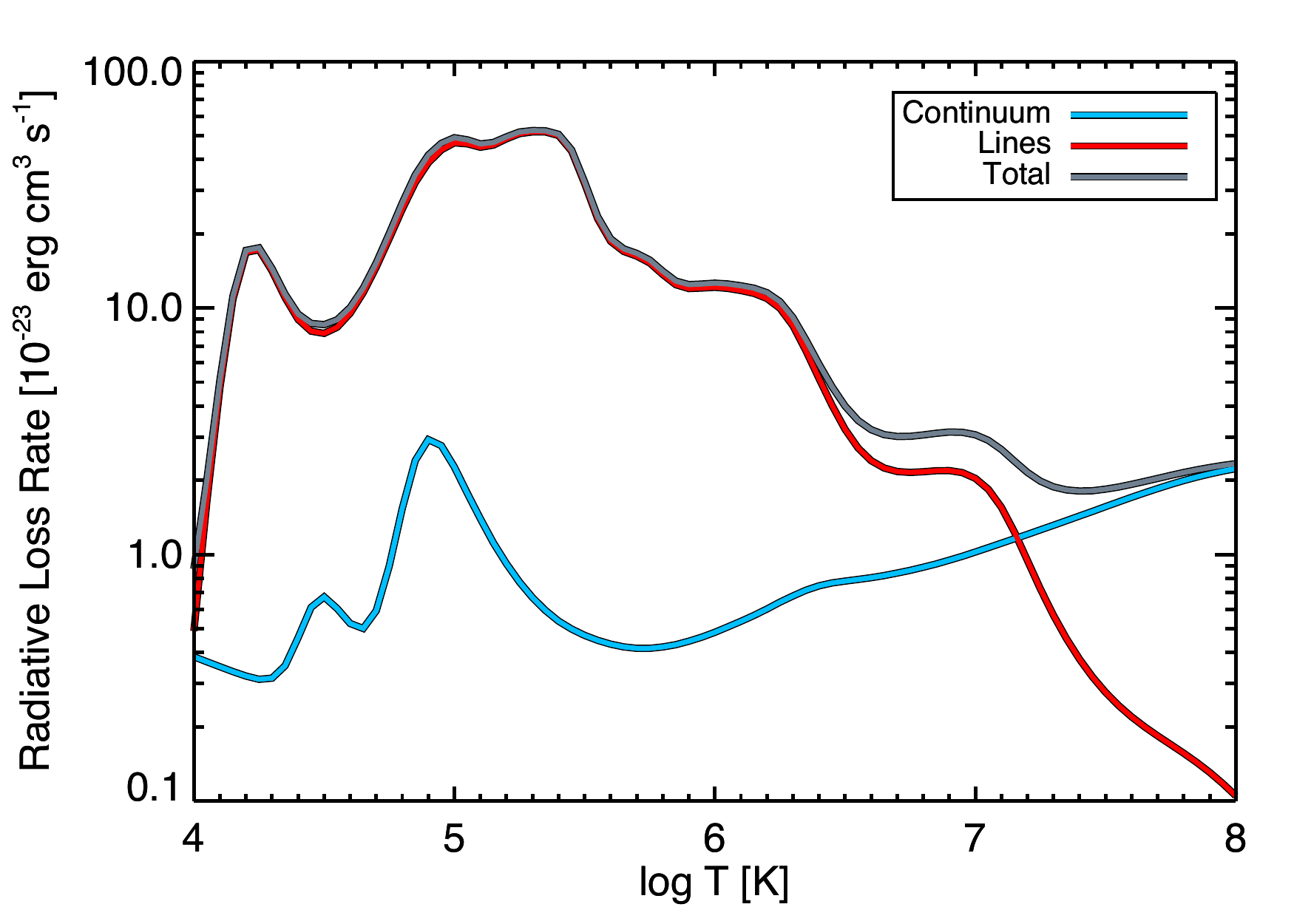} 
\end{center}
\caption{The radiative loss function vs.\ temperature, $\Lambda(T)$, computed using the CHIANTI database and the solar chemical abundance mixture of \citet{Grevesse.Sauval:98}.  Lines dominate for temperature below about 15 million K, at which point abundant metals are largely ionized and continuum emission begins to control cooling.
}
\label{f:radloss}
\end{figure}

The radiation from a hot, optically-thin plasma of a temperature $T$ then comprises line emission from the ions present at that temperature, combined with bound-free and free-free continuum radiation. A useful quantity for understanding the astrophysical behavior of hot plasmas is the radiative cooling time, $\tau_{rad}$,
\begin{equation}
\tau_{rad}\simeq \frac{3kT}{n_e\Lambda(T)},
\label{e:radloss}
\end{equation}
where $\Lambda(T)$ is the radiative loss function----essentially the total line + continuum emitted power as a function of temperature per unit emission measure. Figure~\ref{f:radloss} illustrates $\Lambda(T)$ for the solar chemical abundance mixture of \cite{Grevesse.Sauval:98}.  For a plasma density of the order of $n_e\sim 10^{10}$~cm$^{-3}$ the cooling time is of the order of 1000~s for a plasma with solar chemical composition.  This is an upper limit to the true cooling timescale, which for stellar coronae also suffers conductive losses down to the chromosphere with a cooling timescale that can be approximated for a coronal loop of length $L$ as
\begin{equation}
\tau_{cond}\simeq \frac{3n_ekT}{\kappa T^{7/2}/L^2} = \frac{3n_ekL^2}{\kappa T^{5/2}},
\label{e:condloss}
\end{equation}
where $\kappa$ is the plasma thermal conductivity. 

\begin{figure}
\begin{center}
\includegraphics[width=1.0\textwidth]{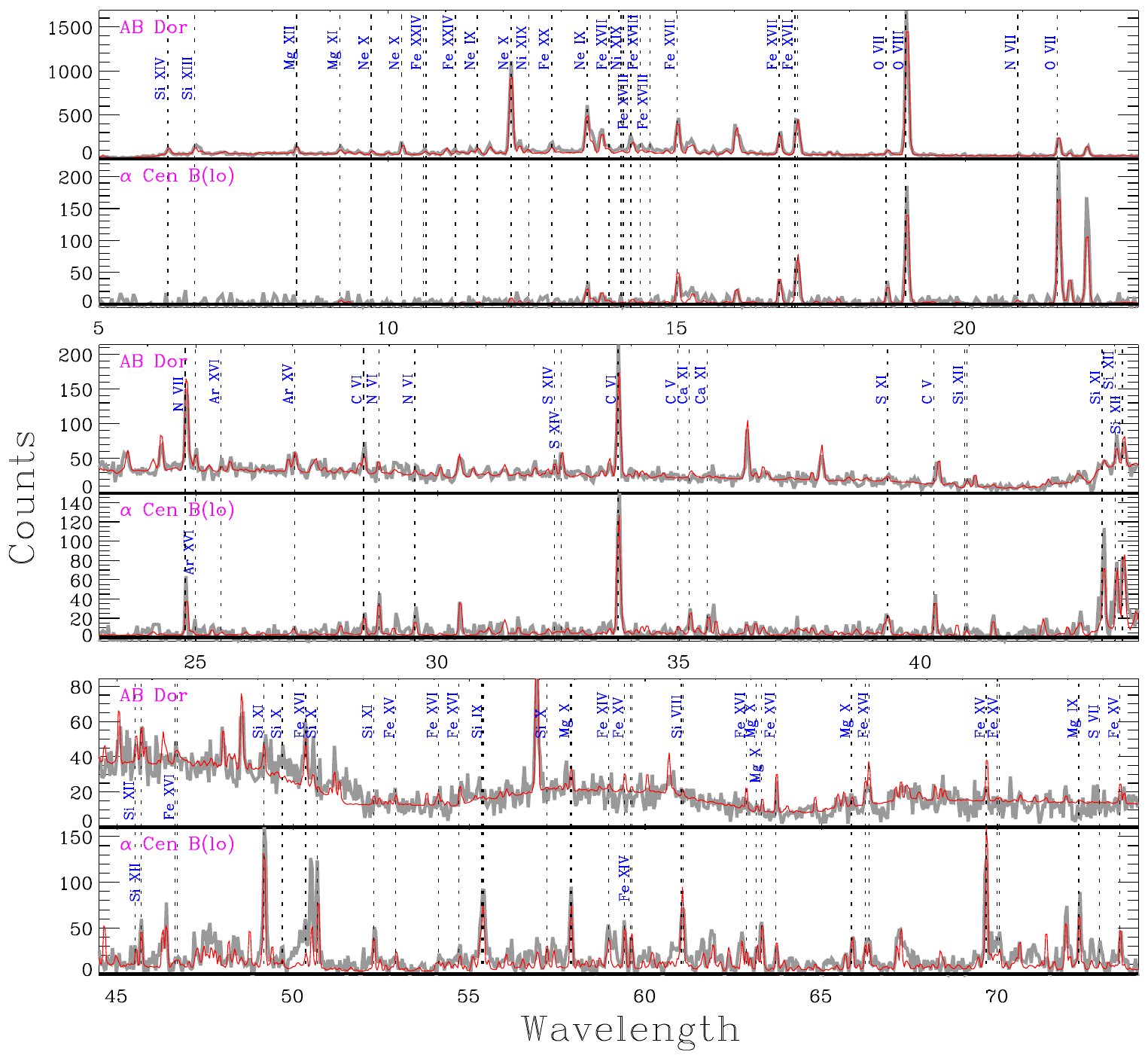} 
\end{center}
\caption{{\it Chandra} LETG+HRC-S spectra of the very active K0 V star AB~Dor and the inactive G2 V $\alpha$~Cen~A. Each shows a series of lines from abundant ions superimposed on a continuum (which is difficult to see here) formed in the optically-thin, collision-dominated plasma that makes up their coronae.  Ions responsible for prominent lines are indicated. Note that the spectrum of the active AB~Dor extends to shorter wavelengths and exhibits lines from plasma formed at much higher temperatures than the spectrum for the inactive $\alpha$~Cen~A.
The red traces are synthetic spectra computed from the DEMs derived from the spectra. From \citet{Wood.etal:18}.
}
\label{f:examplespec}
\end{figure}

Two example spectra obtained by the {\it Chandra} LETG+HRC-S for the very active K0 V star AB~Dor and the inactive G2 V $\alpha$~Cen~A from an analysis by \citet{Wood.etal:18} are illustrated in Figure~\ref{f:examplespec}.  These spectra demonstrate the presence of hotter plasma in the corona of the more active star, as betrayed by the presence of species such as Fe~XX, Mg~XII and Si~XIV that are very weak or absent in the spectrum of $\alpha$~Cen~A. Synthetic spectra generated using differential emission measure distributions derived from the observations show remarkably good agreement with the observed spectra, demonstrating the validity of the assumptions of emission from optically-thin, collision-dominated plasmas.

\subsubsection{The Differential Emission Measure Distribution}
\label{s:dem}

The sharp temperature range of formation of lines of a given ion is especially relevant for providing temperature-dependent  diagnostics for the DEM in Equation~\ref{e:dem}. The DEM was first formulated by \citet{Pottasch:63}, and in different forms by several
authors since. \citet{Craig.Brown:76} provided the first rigorous definition of the
DEM as a weighting function, or source term, in the integral equation
for the line intensity.  The DEM is a convenient parameterization of 
a potentially complex plasma, and in principle, once its form is defined, the spectrum from the chromosphere through to the corona can be synthesized.  

\begin{figure}
\begin{center}
\includegraphics[width=0.49\textwidth]{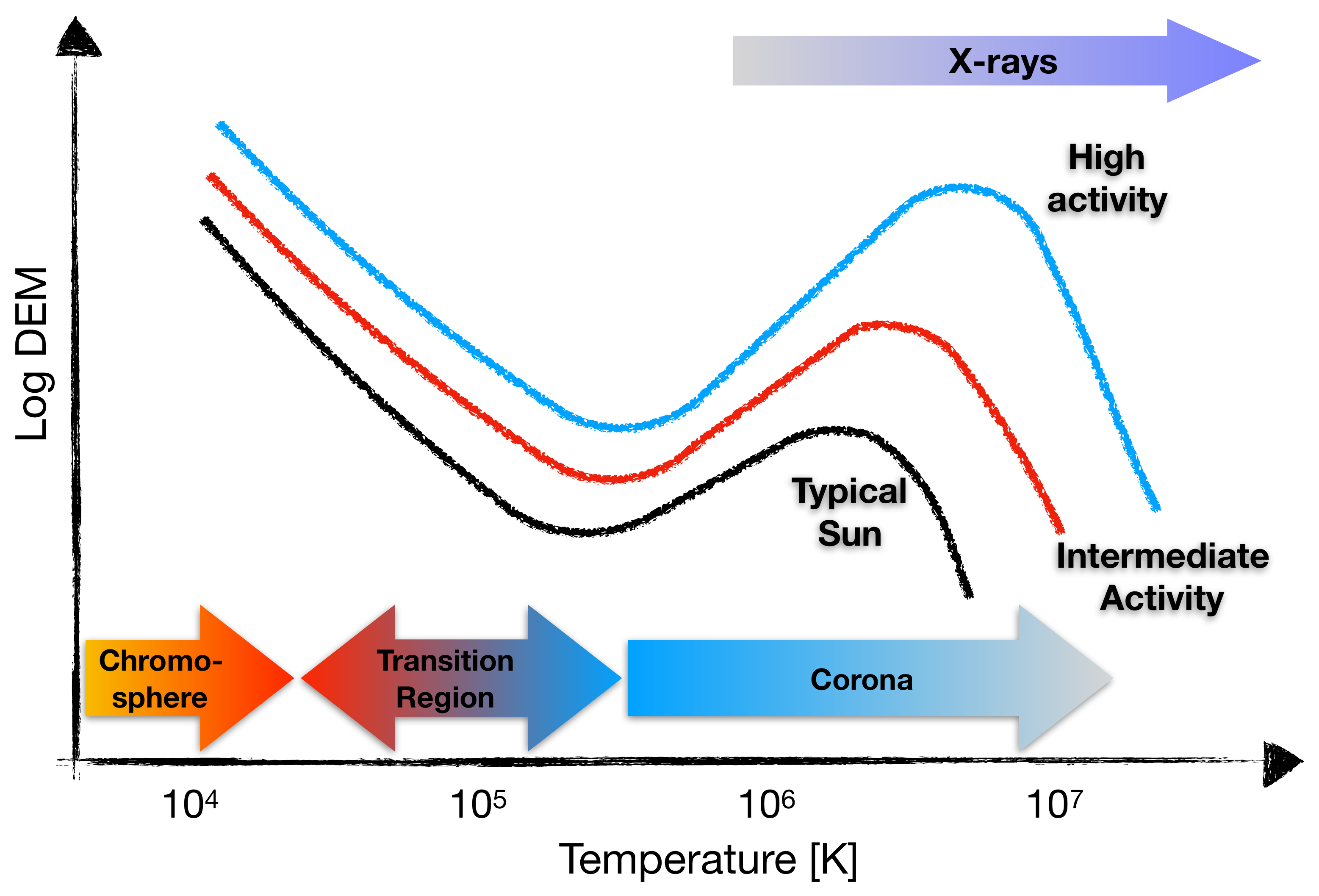} 
\includegraphics[width=0.49\textwidth]{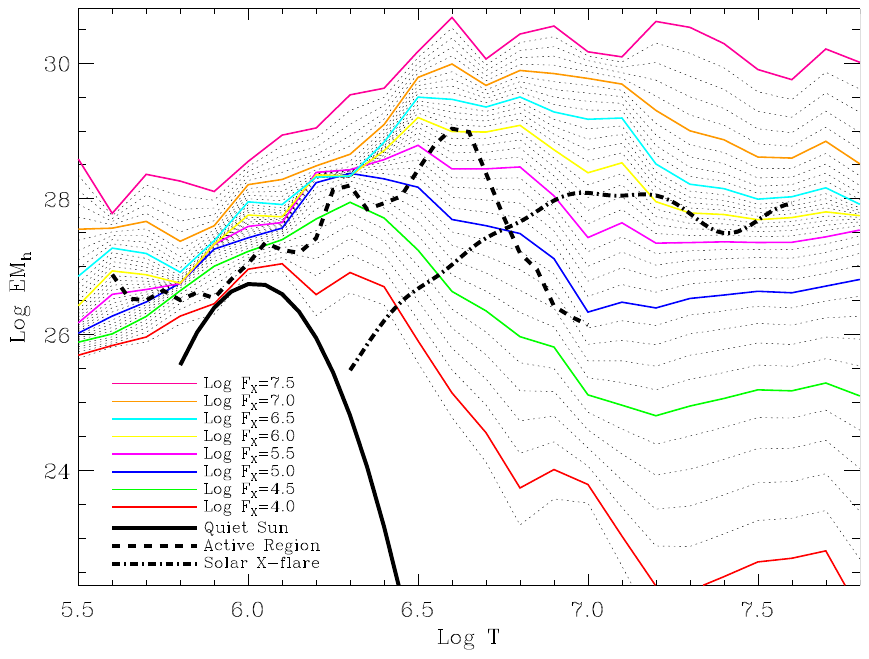} 
\end{center}
\caption{Left: Schematic illustration of the rough shape of the emission measure distribution from the chromosphere to the corona. More magnetically active stars have larger emission measures stretching to higher temperatures.  The form of the DEM is very uncertain in the transition region up until the corona and where temperatures reach in excess of $10^6$~K and are amenable to X-ray spectroscopy. Right: Main-sequence star DEM distributions as a function of surface flux, $F_X$, based on DEMs of 19 stars observed with the {\it Chandra} LETGS.  These are compared with distributions for a solar active region \citep{Warren.etal:12}, the quiet Sun \citep{Kamio.Mariska:12}, and an average solar flare \citep{Warren:14}. From \citet{Wood.etal:18}. 
}
\label{f:dem}
\end{figure}

While the DEM provides great simplification to describing coronal emission, determining the form of the DEM from observations of individual or groups of spectral lines is a notorious integral inversion problem (Eqn.~\ref{e:flux} being a Fredholm equation of the first kind), and generally requires some form of constraint to solve, such as enforcing artificial smoothness on the solution.  Numerous studies have employed DEM analysis in solar and stellar research.  For the former, see, e.g., \citet{Bruner.McWhirter:88}, \citet{Kashyap.Drake:98} and \citet{Jordan:00}.  Emission measure modeling of stars based on individual spectral lines was first applied to {\it EUVE} spectra \citep[see, e.g][]{Drake.etal:95,Sanz-Forcada.etal:02,Sanz-Forcada.etal:03,Bowyer.etal:00}.  
We shall return later in this section to DEM analysis based on {\it Chandra} observations.   Subsequent studies based on high resolution X-ray spectroscopy obtained by {\it Chandra} and {\it XMM-Newton} have helped to define the coronal emission measures for temperatures $> 10^6$~K.  A rough schematic of the approximate shape of the DEM, how it changes in stars of different activity level, and the temperature range in which EUV emission is dominant is shown in Figure~\ref{f:dem}.

Many emission measure distribution analyses have been carried out based on the strengths of emission lines seen in {\it Chandra} grating spectra.  One example is the work of \citet{Wood.etal:18} who use the X-ray spectra of 19 main-sequence stars from early F to mid-M spectral types observed using the {\it Chandra} LETG+HRC-S to investigate the emission measure distributions and chemical abundances in the corona.  They further used the data to determine the general behavior in the DEM as a function of activity level as described by surface X-ray flux using the Monte Carlo Markov Chain method  of \citet{Kashyap.Drake:98}. The resulting DEMs are illustrated in Figure~\ref{f:dem}.  It is important to note that the DEM at the ends of the temperature range covered by the available line diagnostics are generally very uncertain and depend on any applied smoothing and other assumptions made in the analysis.

\subsubsection{Density and Temperature Diagnostics of He-like Ions}
\label{s:helikes}

High-resolution X-ray spectra obtained by the {\it Chandra} diffraction gratings allow individual spectral lines to be resolved, opening up a wealth of plasma diagnostics afforded by the different behavior of some spectral lines to changes in plasma density and temperature.  Perhaps the most powerful of these are density diagnostics---lines whose excitation and resulting intensity depend to an observable extent on the plasma density in a different way to the simple $n_e^2$ dependence in Equation~\ref{e:flux}.  While numerous lines in the X-ray range offer some degree of density sensitivity, they are often faint, difficult to observe and blended with other lines, even at the resolution of {\it Chandra's} gratings.  We restrict our discussion here to the singlet and triplet lines of He-like ions that are by far the most commonly exploited density diagnostics in {\it Chandra} observations of collision-dominated, optically thin ``coronal" plasmas; see, e.g., \citet{Foster.etal:10,Foster.etal:12} for further details of X-ray plasma diagnostics. 

\begin{figure}
\begin{center}
\includegraphics[width=0.58\textwidth]{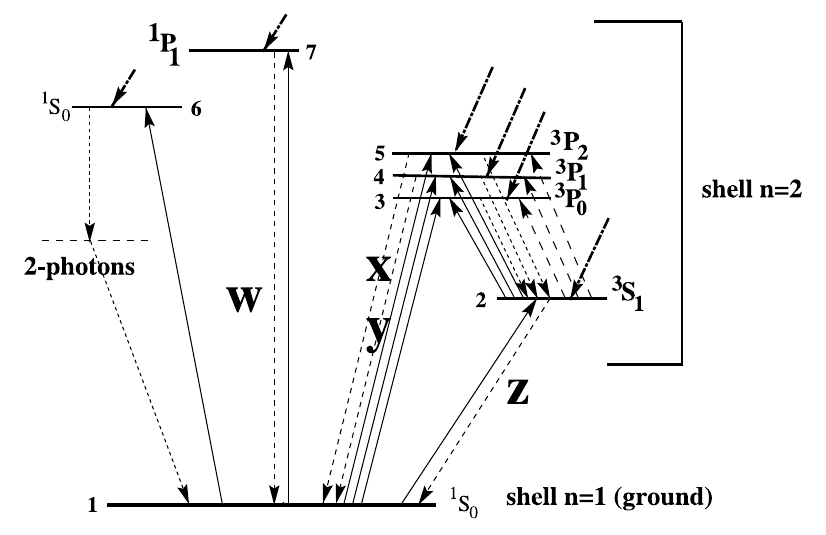} 
\includegraphics[width=0.40\textwidth]{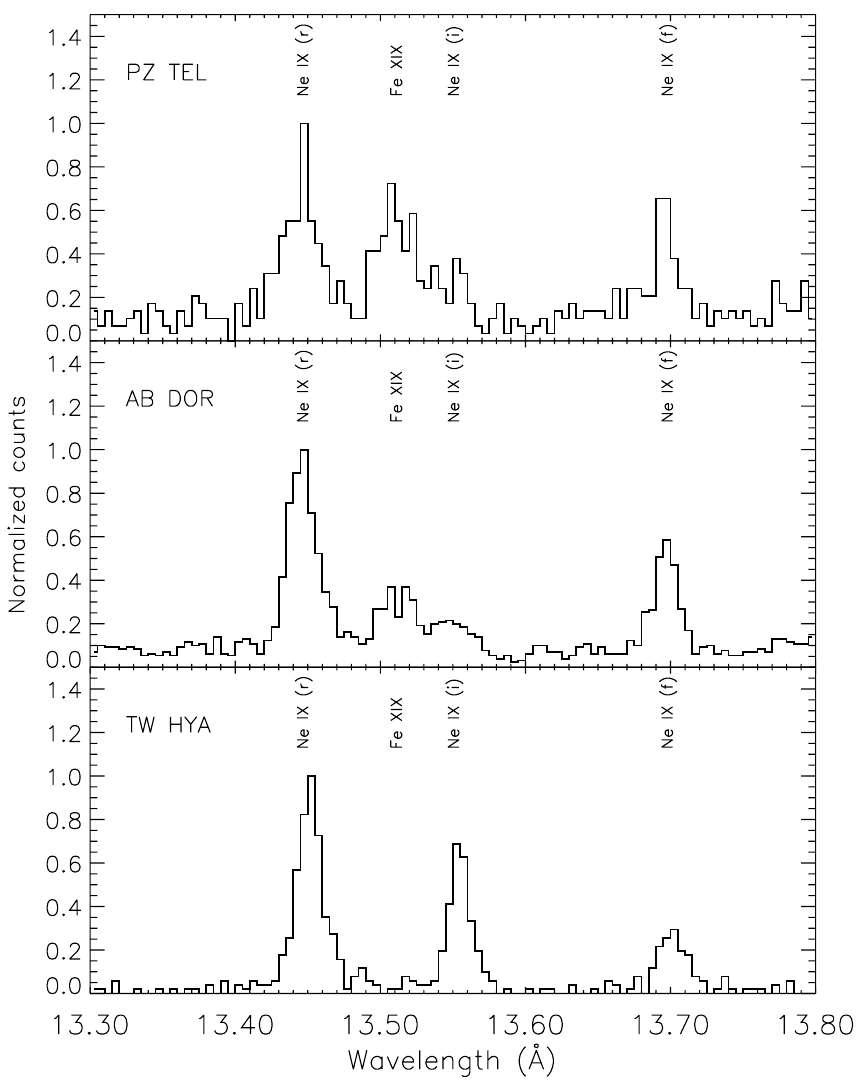} 
\includegraphics[width=0.48\textwidth]{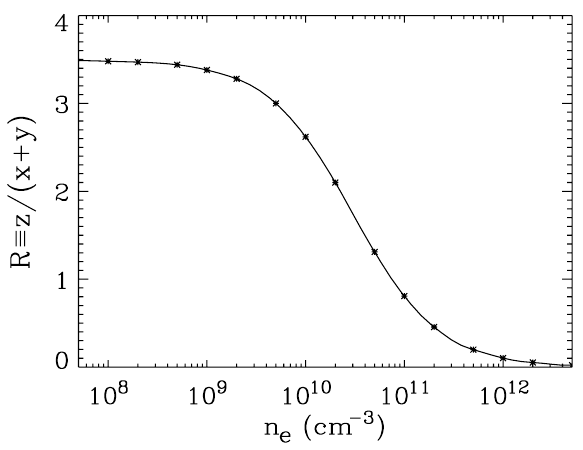} 
\includegraphics[width=0.48\textwidth]{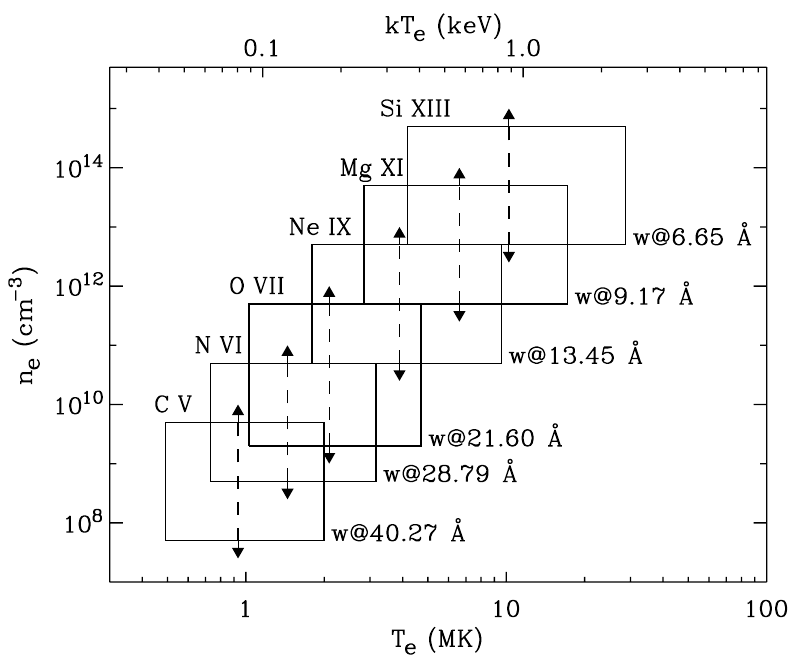} 
\end{center}
\caption{Top right: A simplified energy level diagram for He-like ions from \citet{Porquet.Dubau:00} illustrating the resonance $w$ (or $r$), intercombination $x, y$ (or $i$ ), and forbidden $z$ (or $f$ )  line transitions, respectively.  The various excitation and decay mechanisms are denoted by 
upward arrows for electron collisional excitation (solid arrows) and photoexcitation  (dashed arrows). Radiative transitions are denoted by downward arrows. The  thick dot-dashed downward arrows denote population of levels by radiative and dielectronic  recombination and  cascades.  Top right: the Ne IX He-like complex observed in the classical T~Tauri stars PZ~Tel and TW~Hya, together with the young active K1 dwarf AB~Dor.  TW~Hya exhibits high a density of $n_e\approx 10^{13}$~cm$^{-3}$ thought to arise in an accretion shock \citep[][see Section~\ref{s:accretion}]{Kastner.etal:02}, whereas PZ~Tel shows no signs of high densities and its spectrum resembles that of AB~Dor.  From \citet{Argiroffi.etal:04}.
Bottom: The density sensitivity of the ratio $R (n_e) = z / (x + y)$ for the He-like ion O~VII (left), and the ranges of density and temperature sensitivities of the He-like isoelectronic sequence for abundant ions in cosmic plasmas. Both figures from \citet{Porquet.etal:10}.
}
\label{f:helikes}
\end{figure}

Due to their closed shell structure, He-like ions are abundant over wider temperature ranges than other ions in collision-dominated plasmas (see Figures~\ref{f:tmax} and \ref{f:feionbal}). Owing to the large cosmic abundances of the elements C, N, O, their He-like lines are often the most prominent features of soft X-ray spectra.  He-like ions of Ne, Mg, Si, S and Fe are also commonly observed, although {\it Chandra's} gratings have insufficient resolving power to separate the different components for He-like Fe.  A simplified energy level diagram for He-like ions is illustrated in Figure~\ref{f:helikes}.

The strongest transitions in He-like ions are between the $n=2$ and $n=1$ ground state.  There are four lines of relevance: the resonance line, often referred to as  $w$ or $r$ ($1s\,2p\; ^1P_1 \rightarrow 1s^2\; ^1S_0$), the intercombination lines, often called $x$ ($1s\,2p\; ^3P_2 \rightarrow 1s^2\; ^1S_0$) and $y$ ($1s\, 2p ^3P_1 \rightarrow 1s^2\; ^1S_0$), or $i$ together, and the forbidden line $z$ or $f$ ($1s\, 2s\; ^3S_1 \rightarrow 1s^2\; ^1S_0$).  The  blended $x$ and $y$ lines are often counted as one line and the complex is commonly referred to as the He-like ``triplet".

The diagnostic utility of the He-like triplet was first pointed out by \citet{Gabriel.Jordan:69} in an analysis of this line complex for several elements in solar X-ray spectra.   There are two ratios of combinations of the He-like triplet that are particularly useful: the density sensitive ratio, $R (n_e) = z / (x + y)$ (sometimes written $R (n_e) =f/i$), and the temperature sensitive ratio $G (T_e) = (z + (x + y)) / w$ (sometimes $G (T_e) =(f+i)/r$).  

The ratio $R (n_e) =f/i$ is illustrated for He-like O (O$^6+$ or O~VII) in Figure~\ref{f:helikes}.  As density increases the ratio is fairly constant at $R\approx 3.5$ until $n_e\sim 10^9$~cm$^{-3}$, after which it declines quite steeply until it reaches close to zero at $n_e\sim 10^{12}$~cm$^{-3}$.  This behavior is a result of the small transition probability for the forbidden line $z$ and the collisional excitation route from its upper $^3S_1$ level  to the $^3P$ states which then decay to the ground state.  At densities of $n_e\sim 10^9$~cm$^{-3}$, this collisional excitation rate becomes comparable with the radiative decay rate from $^3S_1$ to ground, and at higher densities eventually dominates such that the forbidden line can be entirely quenched in favor of the $x$ and $y$ intercombination lines. 

The ratio $G (T_e)$ decreases with increasing temperature (exciting electron energy) owing to an increasing excitation rate for the singlet $^1P_1$ state relative to the triplet $^3S_1$ and $^3P_{0,1,2}$ states. 

The approximate ranges of density and temperature sensitivity of the He-like triplets of the abundant elements prominent in {\it Chandra} X-ray spectra are also illustrated in Figure~\ref{f:helikes}.  Higher $Z$ ions are only sensitive at higher temperatures and densities and this limits their utility to some extent.  For example, the Si~XIII triplet is not sensitive to density below several $10^{12}$~cm$^{-3}$, which is much higher than ambient densities in stellar coronae \citep[e.g.][]{Testa.etal:04,Ness.etal:04} and probably only realized during the strongest flares.

The $x$ and $y$ upper levels can also be excited from $^3S_1$ radiatively \citep{Gabriel.Jordan:69}, rendering the $R$ ratio also a potentially useful diagnostic of the intensity of the radiation field \citep[e.g.][]{Ness.etal:01}.  In the O~VII case, the transitions correspond to a wavelengths of 1624, 1638 and 1640~\AA; for lower Z He-like ions the analogous transitions are at approximate wavelengths of 1900 and 2275~\AA\ for N and C, respectively, and for higher Z ions at 1260, 1020 and 845~\AA\ for Ne, Mg and Si, respectively.  \citet{Ness.etal:02} has shown that the radiation fields of late-type stars are negligible for O and higher Z He-like triplets, but that for N and C radiative excitation can be significant for early G-type and hotter stars .

\bigskip

While above discussions of the DEM and of density diagnostics is by necessity brief, we are now in a position to understand the theoretical underpinnings of X-ray emission from low-mass stars. For a more complete treatment of coronal emission, see, e.g., \citet{Golub.Pasachoff:09} and \citet{Mariska:92}.

\subsection{The Rotation-Powered Magnetic Dynamo}
\label{s:rotation}

We noted in the introduction to this primer that the magnetic nature of stellar coronae was essentially established by the {\it Einstein} observatory and the realization that X-ray luminosity was highly correlated with stellar rotation \citep{Vaiana.etal:81,Pallavicini.etal:81,Walter.Bowyer:81}.  The groundwork for this was laid by chromospheric emission in Ca~II H \& K lines in the Sun showing line core emission fluxes varied linearly with surface magnetic field strength \citep{Frazier:70}, and the realization that H \& K fluxes declined linearly with rotation velocity \citep{Kraft:67} and with time $t$ approximately as $t^{1/2}$ \citep{Skumanich:72}.  The latter relation arises due to the gradual loss of angular momentum through the stellar analogue of the solar wind \citep{Kraft:67,Weber.Davis:67,Durney:72,Mestel.Spruit:87}. 

The remarkable short paper by \citet{Skumanich:72}, featuring only four Ca~II data points, is a seminal work on the stellar rotation-activity relation and spawned a large volume of research on this topic.

\subsubsection{A Magnetic Activity Rossby Number}
\label{s:rossby}

The next major development was the inclusion of a ``Rossby number" into flux-rotation relations.  Ca~II flux vs.~rotation period for late-type stars is subject to a systematic spectral-type dependent scatter that  \citet[][see also \citealt{Noyes:83}]{Noyes.etal:84} noticed was removed if the rotation period is replaced by the ratio of the convective turnover time to rotation period, $Ro=P_{rot}/\tau_c$, where this definition of the Rossby number is analogous to its more common fluid dynamics usage.  

\begin{figure}
\begin{center}
\includegraphics[width=0.49\textwidth]{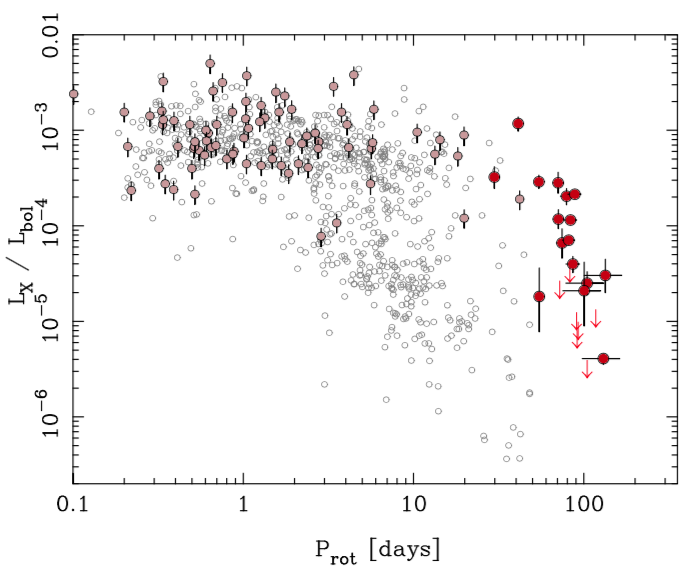} 
\includegraphics[width=0.49\textwidth]{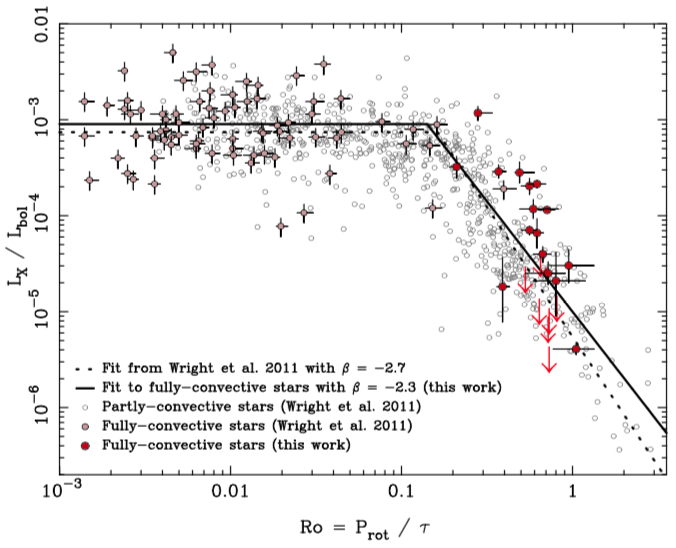} 
\end{center}
\caption{X-ray to bolometric luminosity ratio, $L_X/L_{bol}$, plotted against stellar rotation period, $P_{rot}$ (left) and Rossby number, $Ro=P_{rot}/\tau_c$ (right). Fully convective stars observed by \citet{Wright.etal:16} and \citet{Wright.etal:18} are denoted with large red points, with other fully convective stars in the saturated regime shown as light red points.  Partly convective stars are represented by grey empty circles. Error bars are shown for all fully convective stars. Upper $3\sigma$ limits are shown for the undetected fully convective stars observed as part of this work as red arrows. From \citet{Wright.etal:18}.
}
\label{f:wright18}
\end{figure}

When the early {\it Einstein} X-ray--rotation results were bolstered with the advent of the {\it ROSAT} all-sky survey and the addition of many more X-ray luminosity data points for stars with a large range of ages and rotation periods, the scatter in the X-ray--rotation activity relation was also found to be greatly reduced when the ratio of stellar X-ray to total bolometric luminosity, $L_X/L_{bol}$ was plotted as a function of $Ro$ \citep[see, e.g.][]{Pizzolato.etal:03,Wright.etal:11}. This was borne out more dramatically when \citet{Wright.etal:16} and \citet{Wright.etal:18} added {\it Chandra} observations of slowly rotating late M-dwarfs that are too faint to have been detected by {\it ROSAT}.  The data are illustrated in Figure~\ref{f:wright18}, plotted against both rotation period and Rossby number.

The association of chromospheric emission and coronal X-rays with an internal magnetic dynamo posits that some fraction of the magnetic energy created within the star by dynamo action and subject to buoyant rise is dissipated at the stellar surface and converted into particle acceleration and plasma heating.
It must be emphasized that none of these processes are fully understood!  In this context, 
Figure~\ref{f:wright18} is remarkable, showing that the extremely complex physical system of plasma and magnetic fields that make up a stellar corona, fed by a complex internal dynamo, behaves in bulk in a very simple way: at slower rotation rates, $L_X/L_{bol} \propto Ro^\beta$ up until a threshold at which point X-ray emission saturates, $L_X/L_{bol} \sim 10^{-3}$, close to $Ro=0.13$. This saturation behavior was already apparent based on {\it Einstein} data \citep{Vilhu:84,Micela.etal:85}, though its origin has still not been firmly demonstrated.  It is likely that it represents saturation of the dynamo itself, but other mechanisms such as centrifugal stripping of coronal plasma by rapid rotation, or saturation of the coronal surface filling factor, have also been suggested to play a role \citep[see, e.g., the discussion in][and \citealt{Blackman.Thomas:15}]{Wright.etal:11}.  

The implication from the correlation of stellar rotation with magnetic activity is that stellar dynamos are driven by rotation, or more properly by {\em differential} rotation that results from convective transport in a rotating reference frame, providing an elegant qualitative confirmation of elementary ``$\alpha\Omega$" dynamo theory proposed by \cite{Parker:55} and demonstrated in the solar context by, e.g., \citet{Babcock:61}.  An $\alpha\Omega$ dynamo comprises differential rotation that stretches an initially poloidal field to produce a toroidal field (the $\Omega$ effect), and cyclonic convection which stretches the field as it rises due to magnetic buoyancy (the $\alpha$ effect), regenerating poloidal field.  The topic of stellar dynamos now comprises and immense literature and its full discussion requires an entire book of its own; the reader is instead referred to reviews by \citet{Ossendrijver:03} and \citet{Charbonneau:14}.

The Rossby number approach of \citet{Noyes.etal:84} leads to a dynamo efficiency (or ``dynamo number") proportional to $Ro^{-2}$, such that on the unsaturated part of the $L_X/L_{bol}$--$Ro$ relation $L_X$ should depend on $Ro$ as $L_X\ \propto Ro^{-2}$. \citet{Montesinos.etal:01} further refined the expression for the dynamo number and \citet{Wright.etal:11} showed that this can be approximated by 
\begin{equation}
N_D\propto \frac{1}{Ro^2} \frac{\Delta \Omega}{\Omega},
\label{e:dynnum}
\end{equation}
where $\Omega$ is the angular rotation rate $2\pi/P_{rot}$, and $\Delta \Omega$ is an effective mean differential rotation.  Based on the stellar data, \citet{Wright.etal:18} found $L_X \propto Ro^\beta$ with $\beta=2.3^{+0.4}_{-0.6}$. This implies that, to within the accuracy of measurement, the differential rotation rate relevant to stellar dynamos is proportional to the rotation rate, 
$\Delta \Omega/\Omega \propto \Omega$, thus recovering $L_X\ \propto Ro^{-2}$.  

\begin{figure}
\begin{center}
\includegraphics[width=0.4615\textwidth]{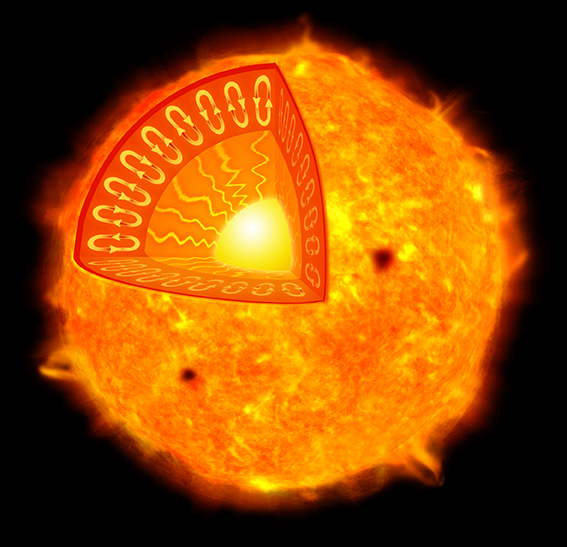} 
\includegraphics[width=0.46\textwidth]{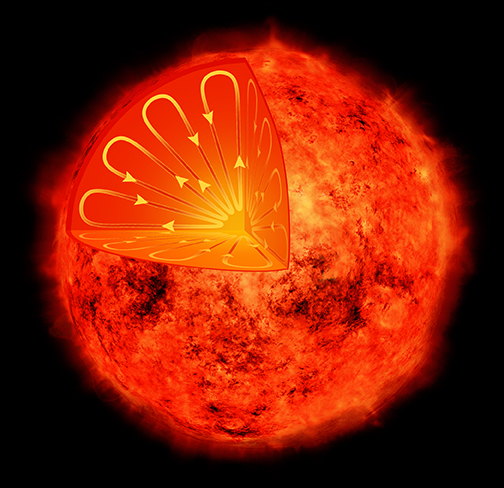}
\end{center}
\caption{Illustration of the internal structure of a Sun-like star, with an outer convection zone and a radiative, convectively stable, core (left), and a low-mass, late M dwarf whose interior is unstable to convection through to the center. Conventional  dynamo theory seats the dominant differential rotation amplification $\Omega$ of magnetic field at the base of the convection zone. From http://chandra.harvard.edu/photo/category/stars.html.
}
\label{f:structure}
\end{figure}

It is presently thought that the site of the differential rotation responsible for the $\Omega$ effect in $\alpha\Omega$ dynamos is the ``tachocline"---a thin shear layer at the base of the convection zone at the radiative-convective boundary found through helioseismology \citep[][see, however, the critique of \citealt{Spruit:11}]{Brown.etal:89,Goode.etal:91}. This poses an interesting issue for main sequence stars with masses $M$\lax $0.35 M_\odot$, corresponding to spectral types later than M3.5 V, whose internal structure is fully convective: since they possess no tachocline their dynamo behavior might naively be expected to be different.  The internal structure of stars with and without a tachocline are illustrated schematically in Figure~\ref{f:structure}. The slowly rotating late M dwarfs observed by {\it Chandra} show the same dependence of $L_X$ on Rossby number as higher mass stars, implying that a tachocline is not a necessary ingredient in solar-like dynamos.

\subsubsection{Supersaturation}
\label{s:supersaturation}

The X-ray luminsosity data for stars in Figure~\ref{f:wright18} show one further interesting feature: at very fast rotation rates and small values of $Ro$ there is a hint of a {\em decline} in $L_X/L_{bol}$ with increasing rotation in G and K dwarfs.  Dubbed ``supersaturation", this phenomenon first came to light in ROSAT stellar surveys \citep[e.g.][]{Randich.etal:96}.  \citet{Wright.etal:11} examined two possible machnisms for supersaturation, centrifugal stripping of coronal loops, and poleward migration of magnetic flux due to rotation induced polar updrafts, but found the data insufficient to distinguish between them.

The answer might have been provided by \citet{Argiroffi.etal:16}, who used {\it Chandra} to investigate the 
rotation-activity relation in the young $\sim 13$~Myr old cluster h~Persei.  Stars in h~Persei 
have ended their T~Tauri accretion phase (see Sections~\ref{s:young} and \ref{s:accretion}) and those of approximately solar mass have developed a radiative core.   The expectation is that their dynamos should operate with the same $\alpha\Omega$ mechanism as the solar dynamo, as discussed in Section~\ref{s:rotation}.  Some of them have already experienced rotational breaking such that the cluster samples from the most rapid rotators through to the slower rotators expected to be in the unsaturated regime.

\citet{Argiroffi.etal:16} found that h~Per members in the mass range $1.0 M_\odot < M < 1.4 M_\odot$ indeed sample all three regimes of the rotation-activity space: unsaturated, saturated and supersaturated.  Supersaturation in $L_X/L_{bol}$ was better described by rotation period, $P_{rot}$,  than by $Ro$, and that the distribution in $L_X/L_{bol}$ at the fastest rotation periods were compatible with the centrifugal stripping of coronal loops.  Compact loops with sizes significant less than the stellar radius should be stable to centrifugal stripping; \citet{Argiroffi.etal:16} concluded that a significant fraction of the X-ray luminosity in active stars originates in structures as large as two stellar radii above the stellar surface.

\subsubsection{X-ray Magnetic Cycles}

A salient aspect of solar magnetic activity is the 22 year cycle in which the amplitude of activity indices is strongly modulated with an 11 year period and the magnetic field polarity undergoes a full reversal and return cycle.  The X-ray luminosity of the Sun varies over the cycle from sunspot minimum to maximum by a factor of 5--10, depending on the bandpass of measurement \citep[e.g.][]{Ayres:14}.

Stellar magnetic cycles have been search for since 1966 and the beginning of the ``HK Project" at Mount Wilson Observatory to monitor the chromospheric activity diagnostic Ca~II H and K line cores in stars \citet{Wilson:78}.  Cycles can in principle provide key information on the dynamo process at work in the stellar interior.  In most cool stars (F--M) they are thought to arise from the interplay of large-scale shear arising from differential rotation, small-scale convective helicity, and meridional circulation in an $\alpha\Omega$ dynamo \citep{Charbonneau:14}.

\begin{figure}
\begin{center}
\includegraphics[width=0.4\textwidth]{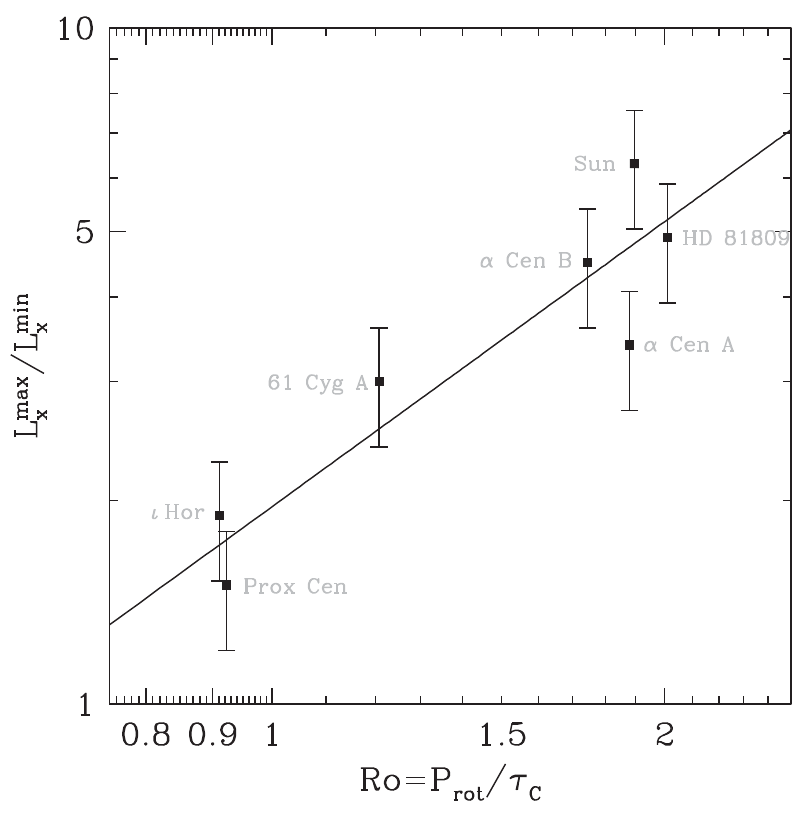}
\includegraphics[width=0.58\textwidth]{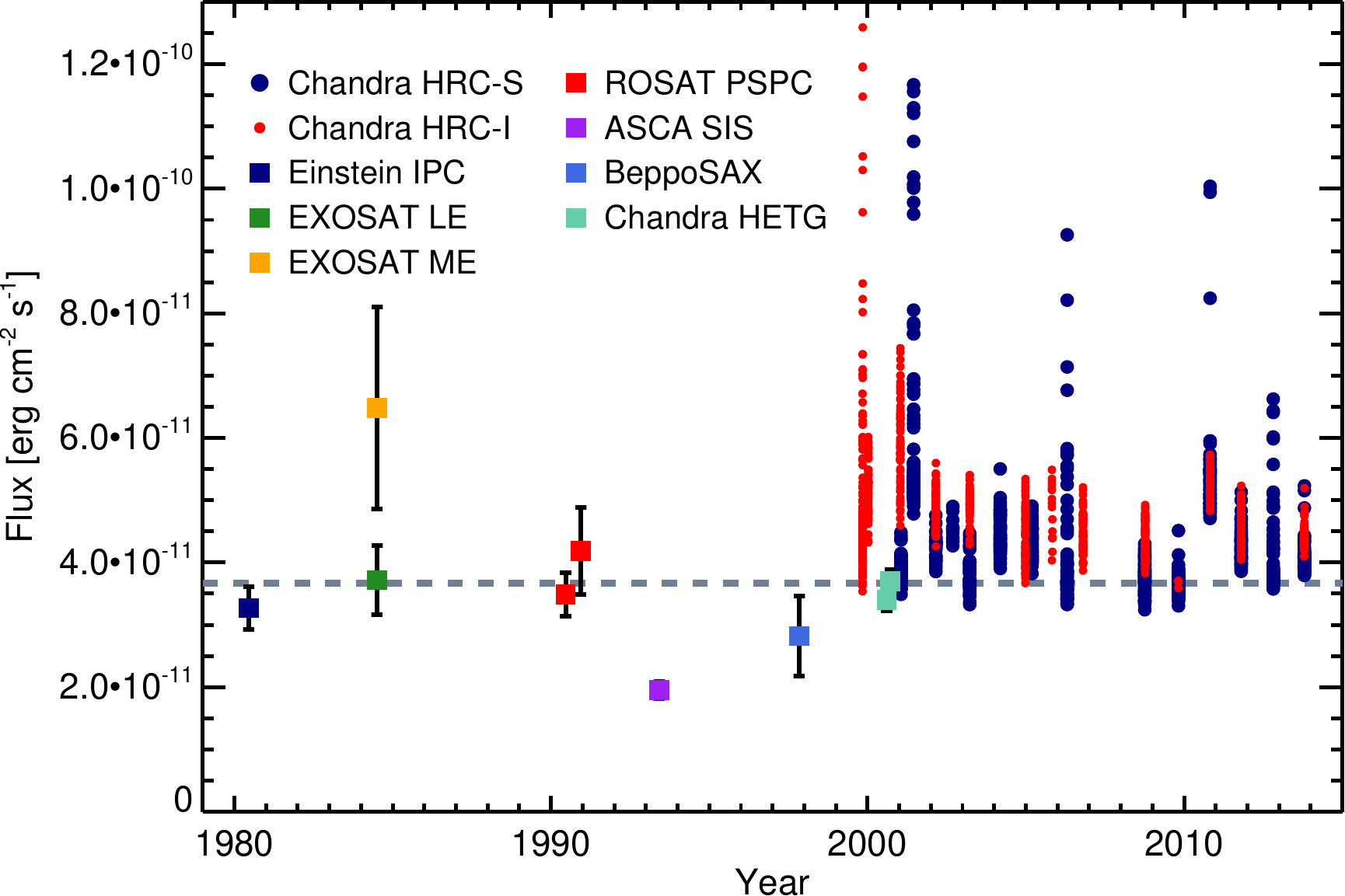}
\end{center}
\caption{Left: X-ray cycle amplitude versus Rossby number for stars with X-ray cycle detections, using data from Table 3. The fitted power law corresponds to $L_X^{max}/L_X^{min} =2Ro^{1.4}$.  Cycle amplitude uncertainties have not been properly determined for these data and $\pm 20$ per cent error bars are shown for illustrative purposes. From \citet{Wargelin.etal:17}. Right: The X-ray flux measured for the RS~CVn-type active binary AR~Lac, showing a remarkably constant level of coronal activity over a period of 33 years. The sharp peaks in the {\it Chandra} data correspond to flares on top of this steady quiescent base level emission. From \citet{Drake.etal:14c}.}
\label{f:xcycles}
\end{figure}

Clear magnetic cycles from H and K lines turn out to be less common than flat or chaotic activity trends, with the occurrence rate tending to increase with stellar age and rotation period.   The first detection of an X-ray cycle was made based on {\it XMM-Newton} monitoring of the G2V star HD 81809, which also has a clear Ca~II cycle \citep{Wilson:78}.  The stars with detected X-ray cycles at the time of writing are illustrated in Figure~\ref{f:xcycles} from \citet{Wargelin.etal:17}.  {\it Chandra} has contributed the key data of the $\alpha$~Cen system to this collection (\citealt{Ayres:14}, \citealt{Wargelin.etal:17}).  

The X-ray amplitude of cycles tends to decrease with increasing stellar activity such that they become challenging to detect; the cycle for Proxima Cen, for example, required careful filtering out of flaring in order to see the underlying cyclic trend.  \citet{Wargelin.etal:17}  compared X-ray amplitude to various stellar parameters, such as mass and rotation period, and found that the best correlation was with Rossby number, $Ro$. The best fit power law to the data 
yields $L_X^{max}/L_X^{min} =2Ro^{1.4}$.

That X-ray cycle amplitude decreases with increasing magnetic activity level meshes with the observation that the most active stars seem to have essentially constant levels of quiescent (non-flaring) X-ray emission. The longest sequence of X-ray observations of a star other than the Sun is of the very active RS~CVn-like binary system AR~Lac, that is used for short {\it Chandra} calibration and monitoring observations.  AR~Lac comprises  has a and is well inside the saturated activity regime. 

\citet{Drake.etal:14c} combined {\it Chandra} data with older ASCA, Einstein, EXOSAT, ROSAT, and BeppoSAX observations (see Figure~\ref{f:xcycles}) and found the level of quiescent, non-flaring coronal emission at X-ray wavelengths to have remained remarkably constant over 33 yr, with no sign of variation due to magnetic cycles. Variations in base level X-ray emission seen by Chandra over 13 yr were only $\sim 10\%$, while variations back to pioneering Einstein observations in 1980 amounted to a maximum of 45\% and more typically about 15\%.

\subsubsection{X-rays in Time}
\label{s:xrays_time}

Magnetic activity turns out to be the cause of its own demise. Figure~\ref{f:wright18} can be thought of as a roadmap of the X-ray activity of a star through time, beginning somewhere toward the left at the zero-age main-sequence and slowly evolving in Rossby number to the right as angular momentum is leached from the star by its wind that is itself powered by magnetic dissipation. Stars of different mass evolve in Rossby number at different rates, with F stars evolving much more quickly than M stars.  This can be seen from the Rossby number definition itself: for a given rotation period, $Ro$ is proportional to the inverse of the convective turnover time, which is a few days for stars of spectral type F but up to 1000 days for late M dwarfs \citep[e.g.][]{Wright.etal:11}.   Higher mass stars of a given rotation period then lie further to the right in Figure~\ref{f:wright18} than lower mass stars rotating at the same rate.  A solar mass star, with $\tau_c\sim 10$~days,  desaturates rapidly in a time of about 200~Myr.
The lowest mass stars instead do not leave the saturated regime until their rotation periods reach of the order of 100 days, which can correspond to timescales of several Gyr \citep[e.g.][]{Newton.etal:16}.

\subsection{Inference of Coronal Structure from Density Diagnostics}

Plasma density diagnostics such as those described in Section~\ref{s:helikes} have now been applied to a wide range of stars based on high-resolution {\it Chandra} grating spectra \citep[e.g.][]{Ness.etal:04,Testa.etal:04}.  Their diagnostic power lies partly in helping understand the geometry and morphology of stellar coronae.  As we have seen in Section~\ref{s:rotation}, the Sun is a comparatively inactive star and the most active stars can have X-ray luminosities 10,000 times brighter than that the average solar X-ray output.  Since their first detection, an important question has been how are these coronae a structured? Are their coronal volumes 10,000 times larger, such that coronae are significant extended relative to the stellar radius? or are they more dense and therefore compact?  Through understanding the density, $n_e$, the emitting volume drops readily out of the volume emission measure, $n_e^2(T)V(T)$.

\citet{Ness.etal:04} derived densities for a sample of 42 stars observed in 22 {\it XMM-Newton} RGS spectra, and in 16 {\it Chandra} LETGS  and 26 HETGS spectra that scattered between $\log n_e \approx 9.5$--11 based on the  O~VII triplet and between $\log n_e \approx 10.5$--12 from Ne~IX.  
\citet{Testa.etal:04} analyzed  O~VII, Mg~XI and Si~XIII X-ray spectra of a sample of 22 active stars observed with the {\it Chandra} HETG.  Mg~XI lines indicated the presence of high plasma densities up to a few times $10^{12}$~cm$^3$ for most of the more active sources with X-ray luminosity $L_X > 10^{30}$~ergs~s$^{-1}$, where stars with higher $L_X$ and $L_X/L_{bol}$ have higher densities at high temperatures. These densities indicate remarkably compact coronal structures. O~VII lines instead yielded much lower densities of a few $10^{10}$~cm$^{-3}$, showing that cooler and hotter plasmas occupy physically different structures.  

\begin{figure}
\begin{center}
\includegraphics[width=0.49\textwidth]{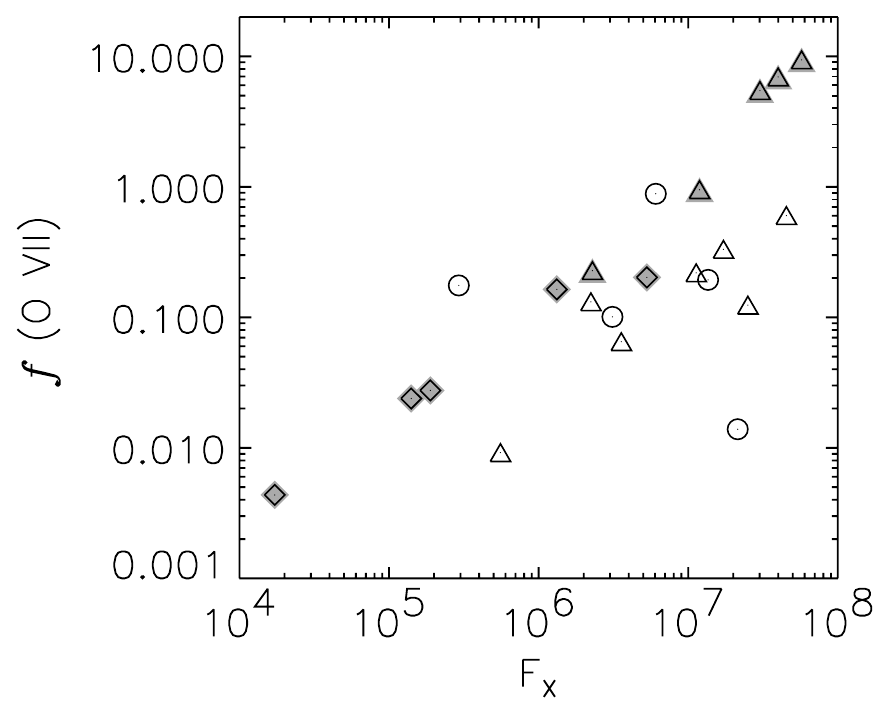}
\includegraphics[width=0.49\textwidth]{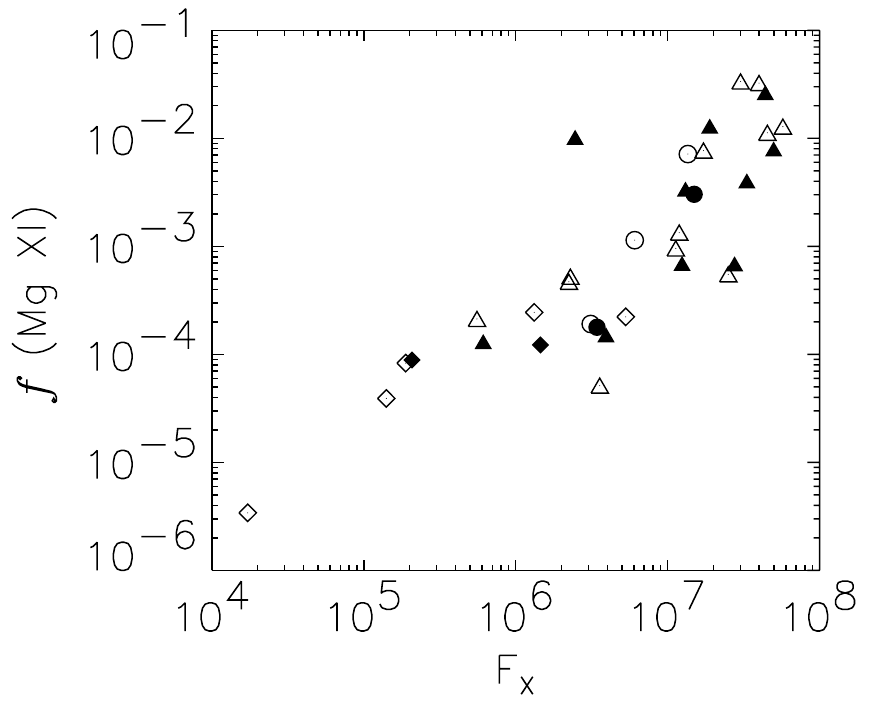}
\end{center}
\caption{Surface filling factors derived from the He-like density diagostics O~VII (left) and Mg~XI (right) vs.\ stellar X-ray surface flux in a sample of active stars observed by the {\it Chandra} HETG and analysed by \citet{Testa.etal:04}.  Circles denote dwarfs, diamonds denote giants and triangles refer to binaries. For the O~VII data, grey symbols refer to stars whose O~VII lines were not measurable and a density $n_e=2\times 10^{10}$~cm$^{-3}$ was assumed instead. The filling factors greater than 1 are unphysical and are likely a result of the true density being higher than this for those stars. In the case of the Mg~XI data, filled symbols are results from the HEG while empty symbols denote MEG results.  Remarkably, the filling factor for hot $10^7$~K plasma indicated by Mg~XI lines increases sharply once the lower temperature plasma reaches a filling factor close to unity. From \citet{Testa.etal:04}.}
\label{f:densities}
\end{figure}

Based on understanding the coronal emitting volume, coronal ``filling factors"---the fraction of the stellar surface covered by X-ray emission---can be derived by assuming a scale height for the coronal plasma. \citet{Testa.etal:04} adopted a scale height based on the lengths of quasi-static uniformly heated coronal loops \citep{Rosner.etal:78} and found filling factors ranging from $f_{MgXI}\approx 10^{-4}$ to $10^{-1}$ for Mg~IX and the hot ($\sim 10^7$~K) plasma,  and $f_{OVII}$ from a few $10^{-3}$ up to 1 for O~VII and the cooler ($2\times 10^6$~K) plasma.  These results are shown as a function of X-ray surface flux in Figure~{f:densities}.  Remarkably, $f_{OVII}$ approaches unity at the same stellar surface X-ray flux level as characterizes solar active regions ($F_X\sim 10^7$~erg~cm$^2$~s$^{-1}$), suggesting that at this activity level these stars become completely covered by active regions.  At the same surface flux level, $f_{MgXI}$ increases more sharply with increasing surface flux. 
\cite{Testa.etal:04} concluded that hot, dense $10^7$~K plasma in active coronae arises from flaring activity and that this flaring activity increases markedly once the stellar surface becomes covered with active regions which then increases surface magnetic interactions.

\subsection{Magnetic Reconnection Flares}
\label{s:flares}

Stellar coronae are almost all observed to undergo flaring in X-rays, much like the Sun.  The flares are caused by the impulsive release of magnetic energy that is gradually built up by convective and other surface motions on the stellar surface within which magnetic field is anchored.  Magnetic reconnection flares then represent a resetting of the corona to a lower magnetic potential energy state.  The theoretical challenge of flare studies is to understand the processes involved and how the energy is partitioned between the different loss mechanisms---optical  through to X-ray radiation, energetic particles, MHD waves, mass motions and associated CMEs.

Flares on the Sun are observed over a very wide range of X-ray energies, ranging from less than $10^{27}$~ergs in the Geostationary Operational Environmental Satellite (GOES) 1--8~\AA\ band ($\sim 1.5-12$~keV) to $10^{31}$ergs, and likely more for historical solar events such as the 1859 Carrington event \citep[e.g.][]{Moschou.etal:19}.  Shorter events appear to decay on radiative cooling timescales (Eqn.~\ref{e:radloss}) of tens of minutes to an hour or so, whereas the largest flares on the Sun---so-called ``two-ribbon flares" can last significantly longer than this.  These events occur in more complex loop or arcade systems and imply that continued heating after the flare onset is applied.  In the canonical flare picture this is by continuous reconnection of nested,  initially open, magnetic fields at successively greater heights.  The literature on solar flares is vast; the reader is referred to \citet{Benz:08} and \citet{Shibata.Magara:11} for thorough reviews.  

There are many examples of flares in {\it Chandra} observations of stars and we touch on this topic again in Section~\ref{s:ttauflares} below. Here we cite two examples. 

Figure~\ref{f:flares} illustrates the {\it Chandra} X-ray light curve showing flares on the active M dwarf ``flare star'' EV Lac based on a 96~ks observation with the HETGS obtained by \citet{Huenemoerder.etal:10} in 2009 March.  EV Lac is a nearby (5 pc) dM3.5e single star and among most X-ray of its type with a mean $L_X \sim 4\times 10^{28}$~erg~s$^{-1}$.   The term ``flare star" originates from the early to mid-19th century when nearby red dwarfs, such as AT Mic and UV Ceti, were first noticed to undergo unpredictable and dramatic increases in brightness.  These variations in the optical are the stellar analogues of the white light component of flares---essentially the photospheric and chromospheric response to fluxes of energetic electrons and protons accelerated in the magnetic reconnection process \citep[e.g.][]{Benz:08}.

The X-ray brightness variations in Figure~\ref{f:flares} show structure on different scales.  The beginning of the observation is characterized by the decay of a large flare whose X-ray peak likely occurred prior to the observation start. The decay timescale of this event is of the order of 10~ks.  In contrast to this, several smaller events are seen, all with much shorter decay timescales of 1~ks or less.  Surprisingly, \citet{Huenemoerder.etal:10} found the shorter flares to be hotter than large, long events.  To place these EV Lac flares in the solar context, their peak flare power is about 
$3\times 10^{29}$~erg~s$^{-1}$, and their total energies are $\sim 10^{32}-10^{34}$~erg---10 to 1000 times more energetic than the largest solar flares ever recorded. 

\begin{figure}
\begin{center}
\includegraphics[width=0.59\textwidth]{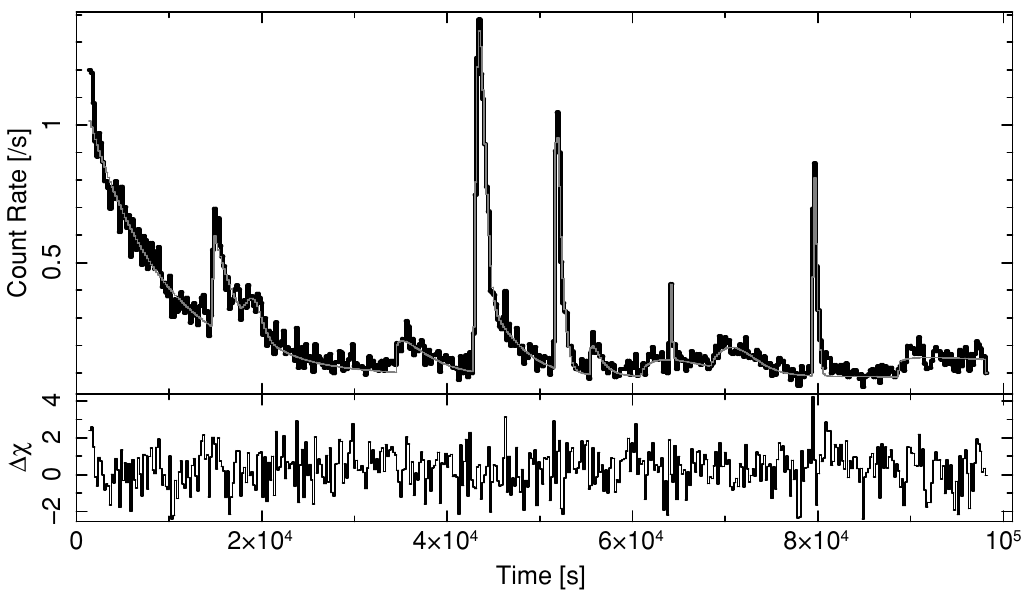}
\includegraphics[width=0.39\textwidth]{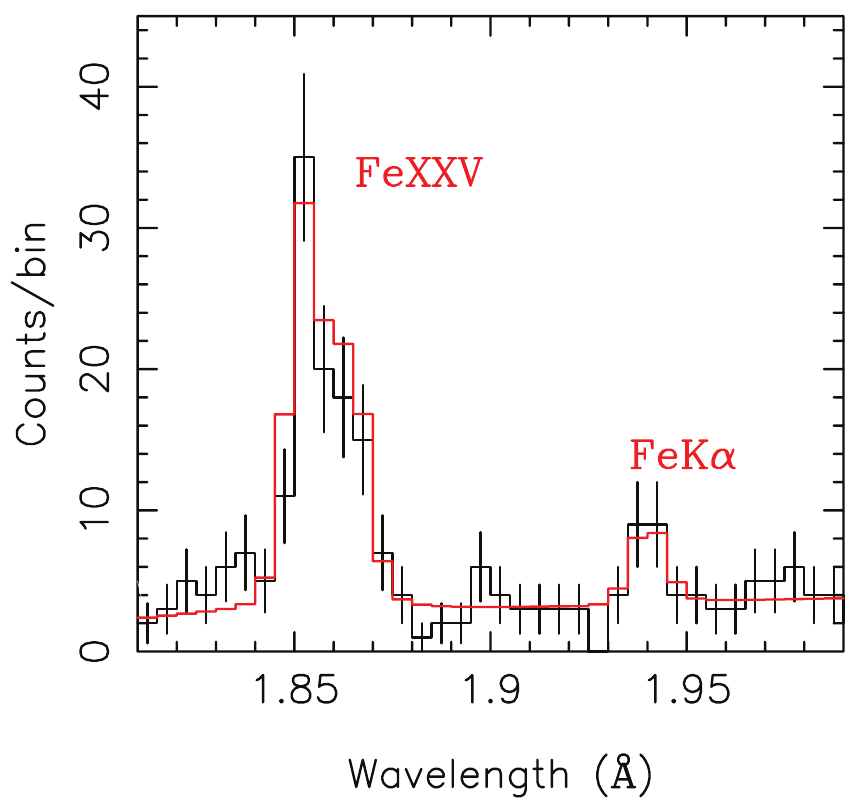}
\end{center}
\caption{Left: HETG 1--25~\AA\ events light curve of EV Lac from a 2009 March observation by \citet{Huenemoerder.etal:10}.  Flares are frequent with an approximate rate of 0.4~hr$^{-1}$ for easily discernible events.  An empirical function modelling the flares is superimposed together with residuals (lower panel). Right: A portion of the HEG spectrum of HR~9024 on which a large flare was observed during the {\it Chandra} observation. In addition to the prominent He-like resonance line of  Fe~XXV,  a cool fluorescence line resulting from reprocessing of the X-rays by the stellar photosphere was seen than provided geometrical constraints on the flare size, limiting the height of the source above the photosphere to $0.3R_\star$ or less. From \citet{Testa.etal:08b}.
}
\label{f:flares}
\end{figure}

An even more energetic flare was observed by the {\it Chandra} HETG on the single intermediate mass active K1 giant HR~9024 by \citet{Testa.etal:07}, peaking at $3.5\times 10^{32}$~erg~s$^{-1}$.  The flare lasted about 50~ks and the integrated X-ray energy was approximately $6\times 10^{36}$~erg---five orders of magnitude more energetic than the largest solar events.  \citet{Testa.etal:08b} detected the fluorescent line of Fe produced by inner shell ionization of the iron in the chromosphere.  The strength of this line depends on the height of the ionizing source, and \citet{Testa.etal:08b} were able to use the line to constrain the height of the flaring loop, or loops, to within 0.3 stellar radii. \citet{Testa.etal:07} note that the radius of HR~9024 is about $14R_\odot$, giving a flaring loops heigh of up to 4 solar radii. This was broadly consistent with a hydrodynamic flare model investigated by \citet{Testa.etal:07}.

\citet{Argiroffi.etal:19} subsequently detected Doppler shifts in S~XVI, Si~XIV and Mg~XII lines
that betrayed upward and downward plasma motions with velocities of
100--400 km~s$^{-1}$ within the flaring loop, again in broad agreement with a hydrodynamic model. Perhaps even more fascinating was a later blueshift seen in the O~VIII reveals a line of sight  upward motion with velocity 90~km~s$^{-1}$ that \citet{Argiroffi.etal:19} ascribed to a CME, representing the first direct X-ray detection of a CME event on a star other than the Sun.  The estimated CME mass was $10^{21}$~g and the inferred kinetic energy was of the order of $5\times 10^{34}$~erg. Again, comparison with the solar case is instructive: the largest observed solar CMEs have masses of $10^{17}$~g and kinetic energies of the order of $10^{32}$--$10^{33}$~erg.  The HR~9024 CME candidate was then more massive than the largest solar CMEs by four orders of magnitude, but only fifty times more energetic.  

At face value the \citet{Argiroffi.etal:19} result suggests that the partitioning of energy is different for CMEs on the Sun and on very active stars, and in particular that CME kinetic energy is much lower than might be expected by extrapolating solar flare-CME relations. While it must be cautioned that solar CME parameters show a large scatter, the relatively low HR~9024 CME candidate kinetic energy compared with its larger inferred mass is consistent with the general picture from a compilation of stellar CME candidates on active stars analysed by \citet{Moschou.etal:19}.

The HR~9024 flare is at present the only example of a potential CME inferred from Doppler shifts in {\it Chandra} observations; the challenge for future missions will be to garner a larger sample of such events and build a more secure foundation for theoretical models.

The vigor of flare activity is strongly related to the underlying magnetic activity level of the star itself: more active stars experience more energetic and more frequent flaring owing to the larger reservoir of stored magnetic energy in their coronae.  Red dwarf ``flare stars" are conspicuous for two reasons: their timescales for activity decline are much longer than for higher mass stars (Section~\ref{s:xrays_time}), and their photospheres are comparatively red and faint, such that the contrast between white light produced during flares and photospheric emission is much stronger and more readily visible.

\subsection{Stellar Coronal Chemical Compositions}

Following the launch of the {\it ASCA} and {\it EUVE} satellites in the early 1990's a picture began to emerge of abundances of elements in stellar coronae differing from those in the underlying photosphere \citep[e.g.][]{Drake:96,Drake:02}.  There is a considerable history of  abundance anomalies occurring in the solar corona \citep[see, e.g., the review by][]{Laming:15}, so in some respects the results were not a surprise.  In the Sun, the abundance anomaly is referred to as the ``First Ionization Potential (FIP) Effect'': elements with low FIP (FIP$ \leq 10$~eV, e.g., Si, Mg, Fe) are enhanced to by factors of 2--4 relative to elements with high first ionization potentials (FIP$ \geq 10$~eV, e.g., N, Ne, Ar).  

The FIP effect was indeed seen in a small handful of low activity stars, such as $\alpha$~Cen~AB \citep{Drake.etal:97}.  But more commonly, since they are brighter in the EUV and X-rays, active stars were selectively observed more and their abundances appeared to show the {\em reverse} of this!  Fe in particular appeared depleted in the coronae of active stars.  

High resolution spectra obtained by the {\it Chandra} and {\it XMM-Newton} gratings that enabled individual lines for different elements to be easily measured provided more definitive observations of coronal abundance anomalies, and one of the first results was of a what is now termed an {\em inverse FIP} (iFIP) effect in the coronae of the active RS~CVn-type binary HR~1099 \citep{Brinkman.etal:01,Drake.etal:01}.  Elements with low FIP (FIP$ \leq 10$~eV, e.g., Si, Mg, Fe) were observed to be depleted relative to elements with high first ionization potentials (FIP$ \geq 10$~eV, e.g., N, Ne, Ar).   This pattern has also now been extensively observed in the very active T~Tauri stars \citep{Maggio.etal:06,Flaccomio.etal:19}.

\begin{figure}
\begin{center}
\includegraphics[width=0.49\textwidth]{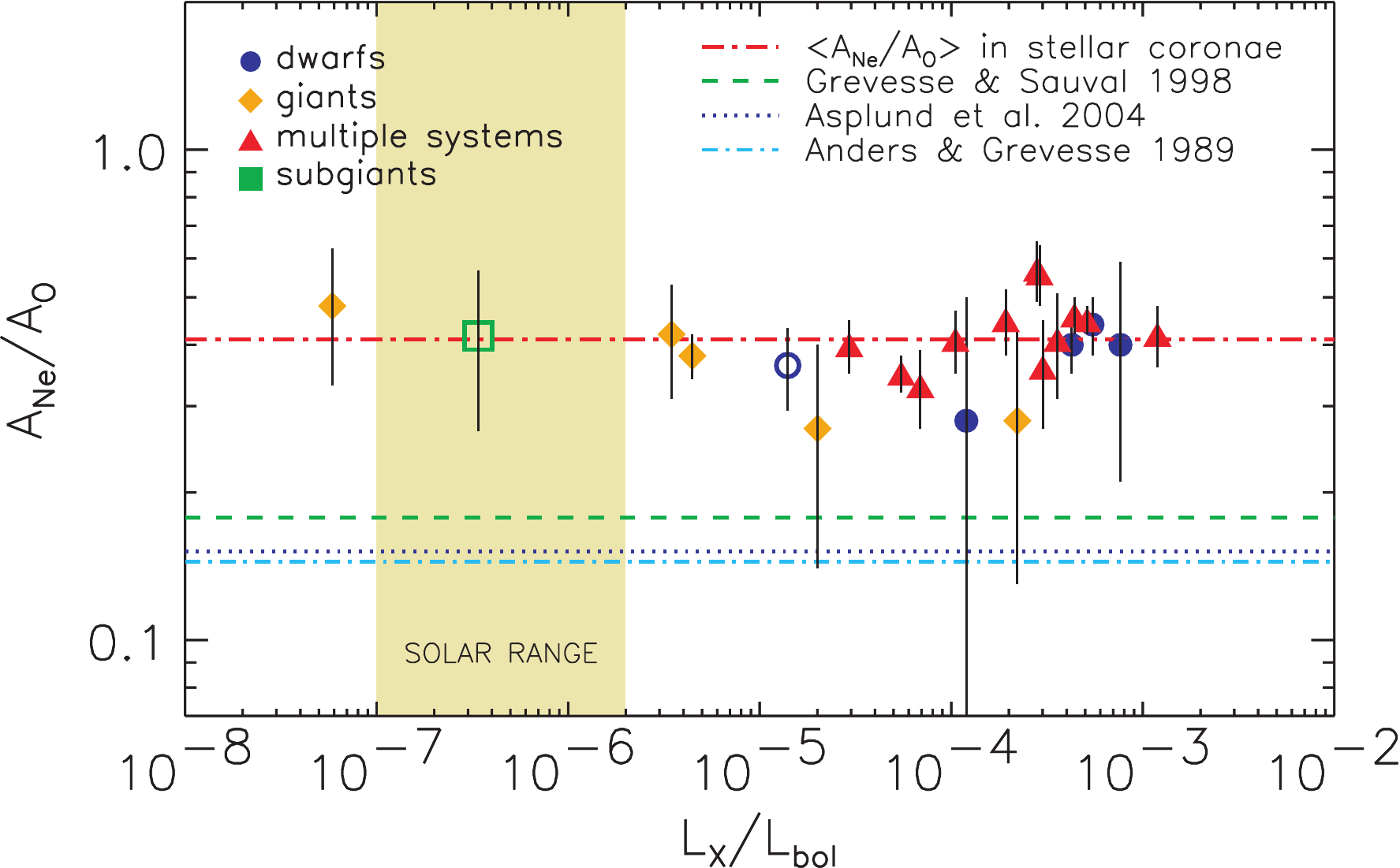}
\includegraphics[width=0.49\textwidth]{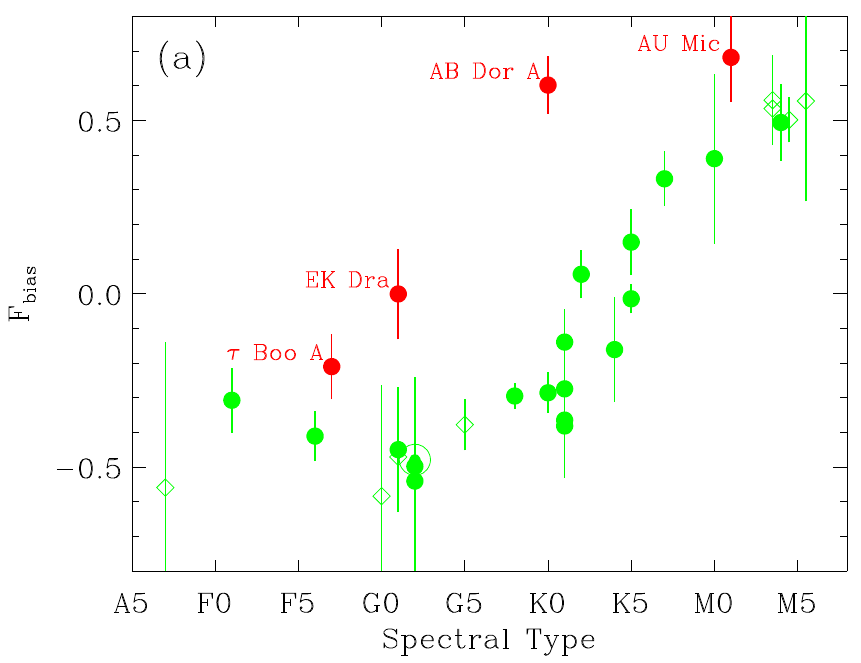}
\end{center}
\caption{Left: Ne/O abundance ratios in coronae with different activity levels from the analysis of \citet{Drake.Testa:05} demonstrating a ``saturation" in the Ne/O ratio.  Some more recent measurements, including of the Sun, indicate that Ne/O declines toward the lowest activity levels.
Right:  the ``FIP bias'', the ratio of an average of high FIP element abundances relative to Fe, plotted as a function of spectral type, illustrating a spectral type and activity dependence of the FIP effect. Active stars are shown as red points and also define a dependence of FIP bias on activity level. From \citet{Wood.etal:18}.}
\label{f:fipeffect}
\end{figure}

The Ly$\alpha$ resonance line of Ne~X was particular conspicuous in the {\it Chandra} grating spectra, and \citet{Drake.Testa:05} showed that stars over a large range of activity levels appeared to have Ne/O abundance ratios about twice that typically seen in the Solar corona (Figure~\ref{f:fipeffect}), although the solar Ne/O ratio is also observed to vary from region to region. This picture has evolved and it now appears that other low activity stars also exhibit more solar-like Ne/O ratios.

As more stars were observed, an additional dependence of the FIP and iFIP effects on stellar spectral type emerged \citep[see][and references therein]{Wood.etal:18}. The ``FIP bias''---an average of high FIP element abundances relative to the low FIP element Fe compared with a solar photospheric mixture---based on an analysis of a sample {\it Chandra} LETG+HRC-S spectra by \citet{Wood.etal:18} is illustrated in Figure~\ref{f:fipeffect}.  The FIP bias is negative for the solar FIP effect in which low FIP elements are relatively enhanced, and positive for the inverse FIP effect. A strong trend of increasing FIP bias with later spectral type is observed, together with a strong increase in FIP bias with activity level which is reminiscent of the early stellar results.

The mechanism underlying the chemical fractionation responsible for FIP-based effects has not yet been identified with certainty. The characteristic FIP at which the abundances change 
corresponds to the temperatures in the chromosphere ($\sim 10,000$~K).  The most promising explanation is based on the ponderomotive forces experienced by ions in the chromosphere as a result of Alfv\'en waves initiated both from the photoshere through convection and turbulence and above in the corona through magnetic reconnection \citep{Laming:15}. The variation in FIP effect with spectral type might then be due to the change in Alfv\'en wave spectrum and intensity as the characteristics of the convection zone change with effective temperature. Since Alfv\'en waves are thought to be important sources of coronal heating and for driving the solar wind, the possibility of using coronal abundance anomalies as diagnostics of Alfv\'en wave is potentially very important.

\subsection{The End of the Main Sequence and Beyond}
\label{s:bd}

We noted in Section~\ref{s:rossby} that stars with masses below that at which they become fully convective appear to behave the same as a function of Rossby number to higher mass, partially convective stars.  An important questions is whether or not this behavior persist all the way into the substellar brown dwarf (BD) regime?  Magnetic activity of BDs is not only of fundamental astrophysical interest, but is also important for understanding the possible influence of magnetic star spots in the interpretation of surface features on BD generally interpreted as clouds. 

BDs never achieve sufficient core temperatures and densities to fuse hydrogen and subsequently cool down and become fainter and fainter as they age.  From a H$\alpha$ survey of mid-M to L field dwarfs, \citet{Mohanty.Basri:03} found a drastic drop in activity and a sharp break in the rotation-activity relation. The H$\alpha$ emission levels were found to be much lower than in earlier types, and often undetectable, even in very rapidly rotating objects. They argued that chromospheric emission may shut down below a critical temperature because the atmosphere becomes too neutral to provide sufficient coupling between the gas and magnetic field.

Very young BDs have been quite regularly detected in sensitive X-ray surveys of young star-forming regions \citep[e.g.][]{Preibisch.etal:05b}.  In these cases, the objects are still in the contracting phase and are of higher luminosity and earlier spectral type---the stellar/substellar boundary being at a spectral type of about M6 at an age of 1~Myr---than mature, fully-collapsed objects.  Their X-ray emission might also be in part due to accretion.  \citet{Preibisch.etal:05b} found the BDs in the {\it Chandra} survey of the Orion Nebula Cluster (see Section~\ref{s:ttau} below) to have similar X-ray properties to field M dwarfs of the same spectral type, suggesting that the effective temperature, rather than mass, is the most important parameter for dynamo action. 

\begin{figure}
\begin{center}
\includegraphics[width=0.53\textwidth]{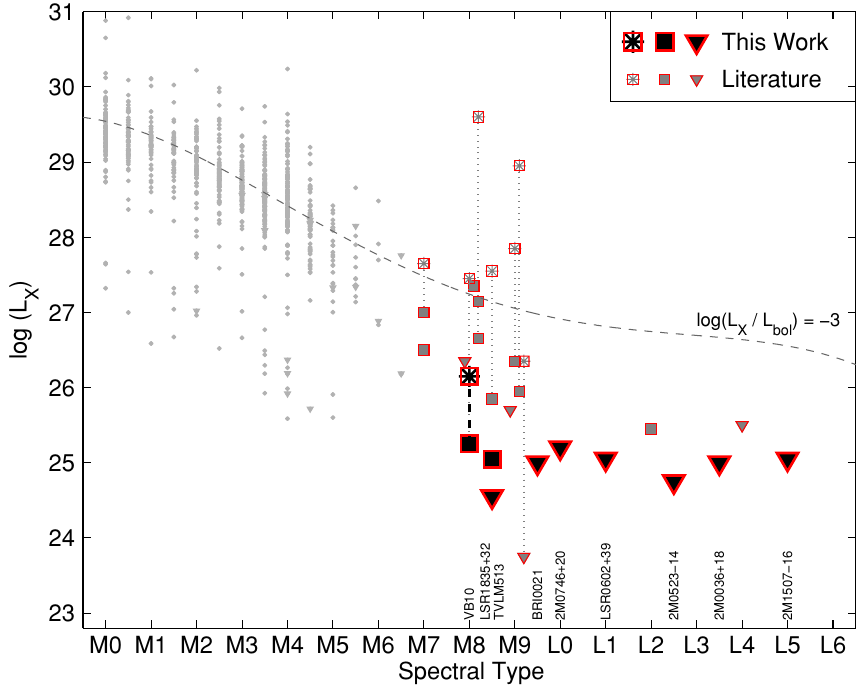}
\includegraphics[width=0.45\textwidth]{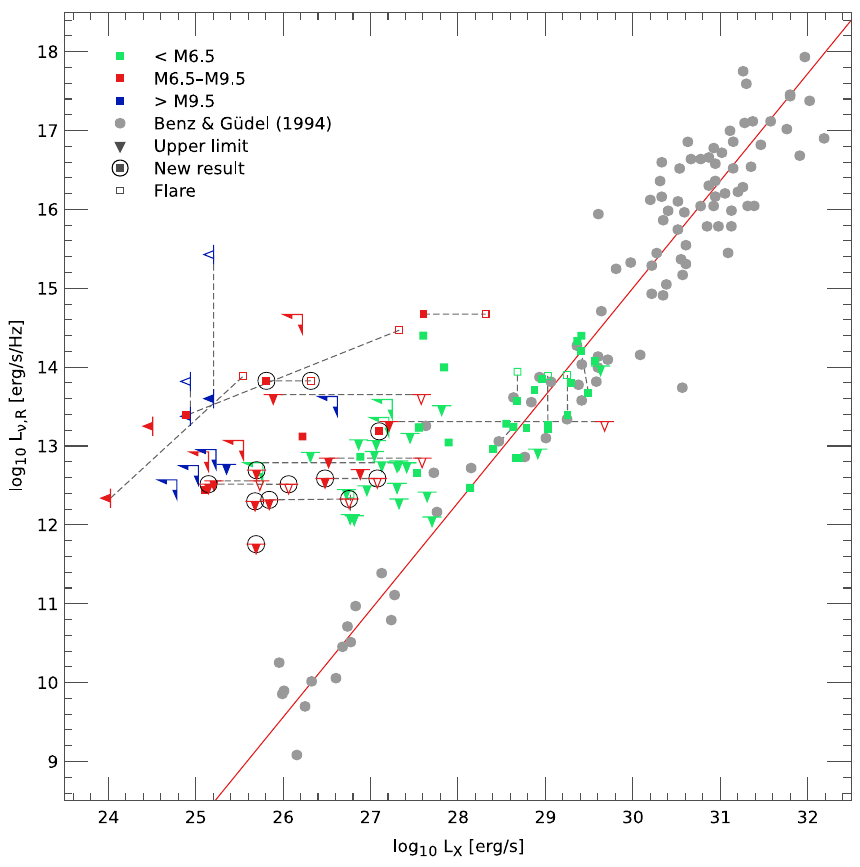}
\end{center}
\caption{Left: X-ray luminosity of M and L dwarfs based largely on {\it Chandra} surveys as a function of spectral type. Literature values for earlier M dwarfs are shown as grey dots.  Quiescent emission of ultracool dwarfs (spectral type $> $M7) is denoted by squares, upper limits by inverted triangles, and flares by asterisks. Dotted lines connect objects for which both flares and quiescent emission were observed. 
The dashed line marks the canonical saturation ratio for F-M stars of $\log L_X/L_{bol} = -3$.  From \citet{Berger.etal:10}.
Right: Relationship between X-ray, $L_X$, and radio, $L_{\nu,R}$, luminosities from \citet{Benz.Guedel:94} (grey) overlaid with data for ultracool dwarfs from the survey of \citet{Williams.etal:14}. Upper limits are denoted by triangles pointing down, to the left, or both down and left.  Grey points with $L_{\nu,R} <12$ are from solar flares. From \citet{Williams.etal:14}.}
\label{f:bd}
\end{figure}

{\it Chandra} made the first X-ray detection of an isolated, fully-contracted mature BD when a flare was detected on LP~944$-$20 in a 44~ks observation on 1999 December 15 \citep{Rutledge.etal:00}.  LP~944$-$20 is a $\sim 500$~Myr old, rapidly rotating ($v \sin i =28$~km~s$^{-1}$, $p\sim 4.4$~hr~$\sin i$) BD at a distance of 5~pc with an effective temperature of approximately 2500~K and a bolometric luminosity of $\sim 6\times 10^{29}$~erg~s$^{-1}$.  A total of 15 counts were detected from the source during a period of 1--2~hr, with an expectation of 0.14 background counts.  The estimated temperature of the signal was $0.26$~keV ($3\times 10^6$~K), which is rather cool compared with the flares typically observed on stars by {\it Chandra} (Section~\ref{s:flares}).  The source was not detected prior to the flare, setting a $3\sigma$ upper limit to the relative X-ray and bolometric luminosities of $L_X/L_{bol} < 2\times 10^{-6}$~erg~s$^{-1}$, to be compared with $L_X/L_{bol} \approx 2\times 10^{-4}$~erg~s$^{-1}$ at the flare peak.  

The detected flare is important as it requires that a relatively strong persistent magnetic field be present on LP~944-20 and that at least occasionally this field must be perturbed into a configuration in which reconnection and flare dissipation of stored magnetic energy occurs.  However, despite the fast rotation of LP~944$-$20, its extremely low quiescent $L_X/L_{bol}$ ratio indicates that it is severely out of line with the saturation value of $L_X/L_{bol}\sim 10^{-3}$ seen in fully-convective M dwarfs (Figure~\ref{f:wright18}).  While LP~944-20 is obviously very faint in X-rays, the upper limit to its fractional X-ray luminosity is still consistent with that of the Sun, which is about $L_X/L_{bol} \sim 10^{-6}$~erg~s$^{-1}$ at solar maximum. 

Further surveys of field BDs with {\it Chandra} have fleshed out the details of this $L_X/L_{bol}$ difference and highlighted a precipitate decline in X-ray emission at a spectral type of about M9 which is illustrated in Figure~\ref{f:bd} \citep[see, e.g.,][]{Stelzer.etal:06,Berger.etal:10}.  \citet{Cook.etal:14} found an anti-correlation between rotation and X-ray activity reminiscent of supersaturation (Section~\ref{s:supersaturation}), and that the scatter $L_X/L_{bol}$ X-ray activity at a given rotation rate is three times larger than for earlier-type stars.  

While X-ray emission appears to be comparatively weak in BDs, radio emission persists and the luminosity remains relatively unchanged from spectral types M0 to L4, reflecting a substantial increase in $L_{rad}/L_{bol}$ with later spectral type, corresponding to radio over-luminosities of up to a factor of 100 (Figure~\ref{f:bd}; see, e.g., \citealt{Williams.etal:14}). The persistence of radio emission is a testament to the continued presence of magnetic fields and particle acceleration. 

\citet{Williams.etal:14} and \citet{Cook.etal:14} suggested that magnetic field topology could be key, and that the large scatter in X-ray fluxes reflects the presence of two dynamo modes that produce distinct magnetic topologies.  While more detections of BDs in X-rays would help understand the relative partition of radio and X-ray magnetic dissipation, substantial progress likely must await a next-generation mission with greater sensitivity.

\subsection{Young Stars, Protostars, Disks and Jets}
\label{s:young}

Young stars in the pre-main-sequence phase---the so-called T~Tauri stars named after the prototypical example in Taurus---are observed to rotate rapidly, with typical periods of from one to several days \citep[e.g.][]{Herbst.Mundt:05}, and they are also vigorously convecting with typical convective turnover times in the 100-200~day range \citep{Preibisch.etal:05}.  On this basis, with Rossby numbers $Ro$\lax$0.1$, they would be expected to be vigorous X-ray emitters in the saturated regime. This is just what was found when the {\it Einstein} observatory was pointed toward nearby star-forming regions and many T~Tauri stars were detected with relative X-ray luminosities of $L_X/L_{bol}\sim 10^{-3}$  \citep[e.g.][]{Ku.Chanan:79,Feigelson.DeCampli:81,Walter.Kuhi:81}.  Some years in advance of the detailed picture of coronal X-ray emission we are privileged to have today, it was  \citet{Walter.Kuhi:81} who first made the connection between solar-like coronal emission and T~Tauri X-rays.

The early work using {\it Einstein} was a watershed moment: since the X-ray luminosity of a solar-mass star declines rapidly over time by orders of magnitude---within a few hundred Myr---X-rays provide an extremely efficient method for distinguishing between young stars and the myriad field stars along the lines of sight to star-forming regions.  Moreover, X-rays from young stars can penetrate through all but the most opaque lines of sight in the Galaxy and deep into molecular clouds whose obscuration stymies observations at other wavelengths. 

The power of X-ray observations was cogently demonstrated when \citet{Walter:86} used X-ray-selected objects in the Taurus, Ophiuchus, and Corona Australis star forming regions to first classify the ``Naked" T~Tauri stars.
The main defining features of T~Tauri stars had been the presence of infrared excesses from a circumstellar accretion disk (also known as a protoplanetary disk, or ``proplyd" for short) and emission lines from the accretion process. Many of the sources being detected in star forming regions did not exhibit these characteristics and where initially thought to be ``post-T~Tauri'' stars, on their way to the main sequence.  \citet{Walter:86} showed that these stars are still in the T~Tauri evolutionary phase and are simply bereft of circumstellar material.  The ``naked" monicker eventually passed out of common usage and the terms ``classical" and ``weak-lined" are now used to refer to T~Tauri stars with or without substantial circumstellar material and accretion.

{\it ROSAT} added substantially to the study of T~Tauri stars and X-rays from star-forming regions \citep[see, e.g.][]{Krautter.etal:94,Neuhauser:97} and by the end of the 1990s the time was ripe for {\it Chandra} to bring its full arsenal of high angular resolution and X-ray imaging spectroscopy to bear on crowded regions of star formation.   {\it Chandra} results on star forming regions are now so extensive that here we must limit ourselves to some ``greatest hits".

\subsubsection{The Detection of X-rays from Protostars}
\label{s:protostars}

One key question in the magnetic activity of young stellar objects is the evolutionary phase at which X-ray emission sets in.  Ionization of infalling gas by X-rays can tie it to magnetic field lines and potentially slow the speed of gravitational collapse.  More locally to the protostar itself, entrainment of the gas to field lines could be important for the launching of outflows and jets.  The lesson from main sequence stars in Section~\ref{s:rotation} is that rotation and convection are the key ingredients for dynamo activity and associated X-ray emission, both of which should be present in protostars.

In the evolutionary framework defined by the shape of the infrared spectrum \citep[e.g.][]{Lada:87} running from Class 0 that are highly obscured and only appear in the far-infrared, to Class III---the weak-lined T Tauri stars with blackbody-like infrared spectra---Class 0 protostars have never been definitively detected, although it is often difficult to distinguish between classes 1 and 0 without far infrared data. 

Class~I protostars, despite often being still deeply embedded in natal clouds and circumstellar material, are instead commonly picked up as heavily absorbed, and therefore spectrally quite hard, X-ray objects \citep[e.g.][]{Imanishi.etal:01,Winston.etal:07,Romine.etal:16}. These objects, with typical ages of 1--$5\times 10^5$~yr, are still deriving a significant fraction of their luminosity from accretion.  Upper imits to X-ray fluxes for Class 0 objects, whose lifetimes are an extremely short $10^4$~yr or so, are similar to those for Class I objects and it is still possible that X-ray do turn on during the Class 0 phase.

\begin{figure}
\begin{center}
\includegraphics[width=0.6\textwidth]{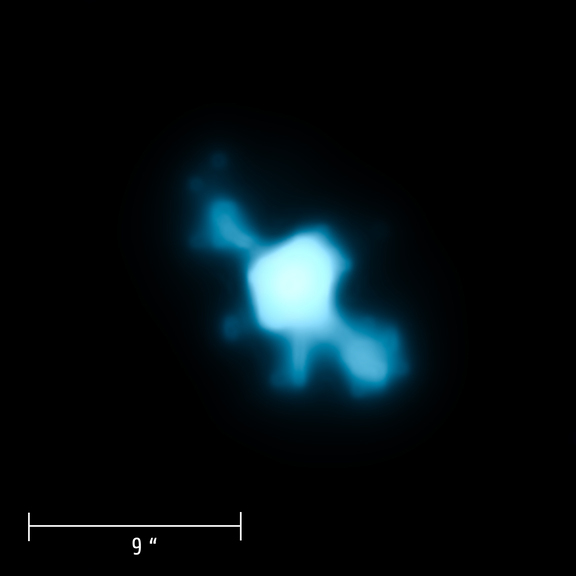} 
\end{center}
\caption{The {\it Chandra} X-ray image of the spectacular X-ray emitting jet of the classical T Tauri star DG Tau from the study by \citet{Gudel.etal:08}.   At the 140~pc distance of DG~Tau, the 9" scale bar shown corresponds to a size of 1260~AU. Image from http://chandra.harvard.edu/photo/2008/dgtau/.
}
\label{f:dgtau}
\end{figure}

Class 0 and I protostars are also observe to drive bipolar outflows with velocities up to several hundred km~s$^{-1}$ that interact and shock-heat gas in the ambient interstellar cloud, producing what are known as Herbig-Haro objects \citep[e.g.][]{Reipurth.Bally:01}.  The pre-shock velocities are often sufficiently high to heat gas to million degree X-ray-emitting temperatures \citep[e.g.][]{Raga.etal:02} and {\it Chandra} has detected these shocks in several Herbig-Haro flows \citep[e.g.][]{Pravdo.etal:01,Favata.etal:02,Pravdo.etal:04}, as well as from the base of the flow in other cases \citep{Gudel.etal:07}.  The X-ray detections represent only a few percent of the Herbig-Haro objects observed, likely because the emission is too soft to penetrate the surrounding gas.

X-ray-emitting outflows are also observed from more evolved objects. 
The classical T Tauri star DG Tau shows X-ray emission from a jet complex extending several hundred AU in either direction, in addition to an inner X-ray emission region located only a few 10s of AU from the star itself  \citep{Gudel.etal:08,Schneider.Schmitt:08} (Figure~\ref{f:dgtau}).

\subsubsection{X-ray Properties of T Tauri Stars}
\label{s:ttau}

The field of view of {\it Chandra} is well-matched to typical angular sizes nearby regions of star formation and many have been observed during the course of the mission. Arguable the most important for understanding the X-ray properties of T~Tauri stars has been the Orion Nebula Cluster (ONC; sometimes also known as the Trapezium Cluster or Ori Id OB Association).
At a distance of approximately 450~pc, the ONC is the most well-studied star forming region in the sky and has been extensively catalogued at all wavelengths. It contains a dense and rich population of approximately 2000 pre-main sequence stars  within a spherical volume 2~pc across, 80\%\ of which are younger than 1 Myr \citep{Hillenbrand:97}. Its combined unresolved X-ray emission was originally detected by the {\it Uhuru} satellite \citep{Giacconi.etal:72}. 

{\it Chandra} has observed the ONC on several different occasions, using both HRC-I \citep{Flaccomio.etal:03} and ACIS-I.
\citep{Flaccomio.etal:03}  detected 742 X-ray sources in a 63~ks HRC-I pointing and established that X-ray luminosities of low-mass stars and brown dwarfs ($M\leq 3 M_\odot$) increase with increasing mass and decreasing stellar age.  Intermediate mass stars of mass 2--$4M_\odot$ were conspicuously fainter, which \citep{Flaccomio.etal:03} attributed to their being essentially fully radiative and unable to sustain a magnetic dynamo (see Section~\ref{s:intermediate}).  Similar results were obtained in 83~ks of data ACIS-I data obtained by \citet{Feigelson.etal:02} in 1999 October and 2000 April, with the addition of photon energy information afforded by the ACIS-I detector.  T Tauri plasma temperatures were found to be often very high even outside of obvious flares, sometimes reaching $10^8$~K and beyond.

The rich data reaped by the initial {\it Chandra} observations provided impetus for a much deeper 1~Ms exposure with ACIS-I in what was dubbed the {\it Chandra Orion Ultradeep Project} (COUP; \citealt{Getman.etal:05}, \citealt{Feigelson.etal:05}).  COUP garnered 1616 X-ray sources in a region of the sky the size of the ACIS-I detector ($16^\prime \times 16^\prime$), of which 1408 were identified with cluster members, providing a large co-eval sample of pre-main-sequence stars down to below the stellar limiting mass.

\begin{figure}
\begin{center}
\includegraphics[width=0.46\textwidth]{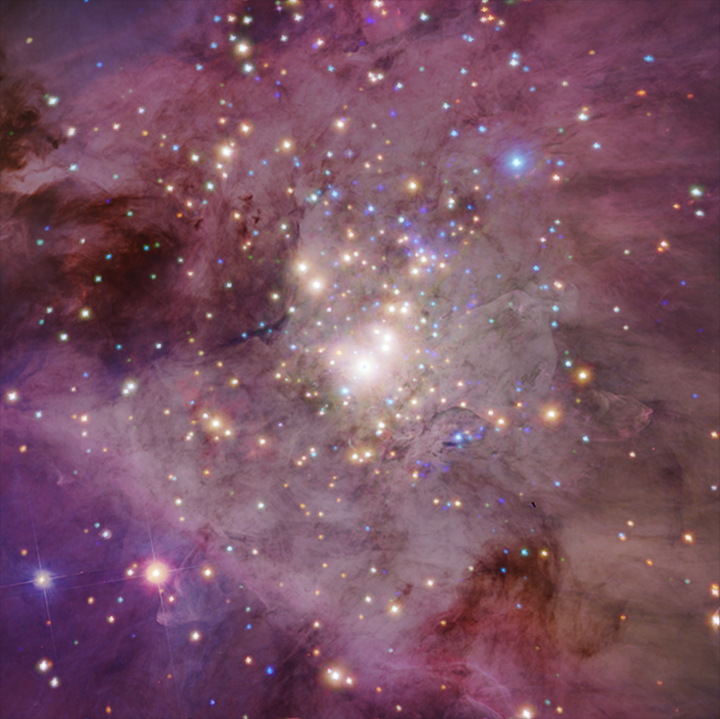} 
\includegraphics[width=0.46\textwidth]{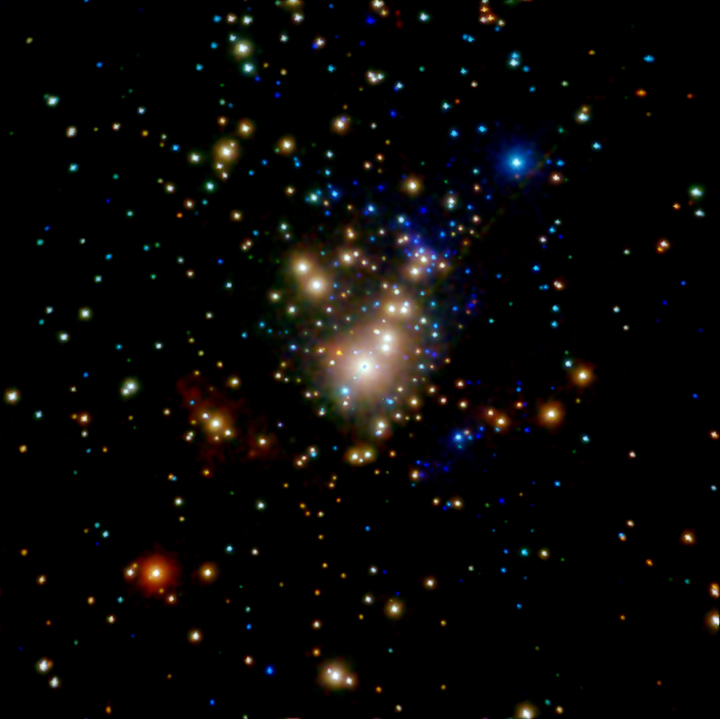}
\end{center}
\caption{The Orion Nebula Cluster seen in a composite {\it Hubble} optical (red-purple) and {\it Chandra} X-ray image (left), and just in X-rays observed by {\it Chandra} (right).  X-ray false colors represent the energy bands 0.3--1.0 keV (red), 1.0--3.0 keV (green), and  3.0--8.0 keV (blue).  A total of 1616 X-ray sources were detected, of which 1408 were identified with cluster members  \citep{Feigelson.etal:02}. Images from http://chandra.harvard.edu/photo/2007/orion/.
}
\label{f:}
\end{figure}

While the completeness of the COUP survey was still mass-dependent---higher mass, X-ray brighter stars were preferentially detected compared with lower mass fainter stars---it provided the most unbiased census of T~Tauri stars yet obtained.  \citet{Preibisch.etal:05} studied nearly 600 of the COUP X-ray sources that were reliably identified with well-characterized T Tauri stars.  The detection limit was $L_X \geq 10^{27.3}$~erg~s$^{-1}$ for the least absorbed sources, leading to a detection completeness of 97\% of cluster stars with spectral types in the range F--M. The immediate deduction here is that {\it all T Tauri stars are X-ray bright and very active} and there is no hidden faint population with suppressed magnetic activity. 

\begin{figure}
\begin{center}
\includegraphics[width=0.32\textwidth]{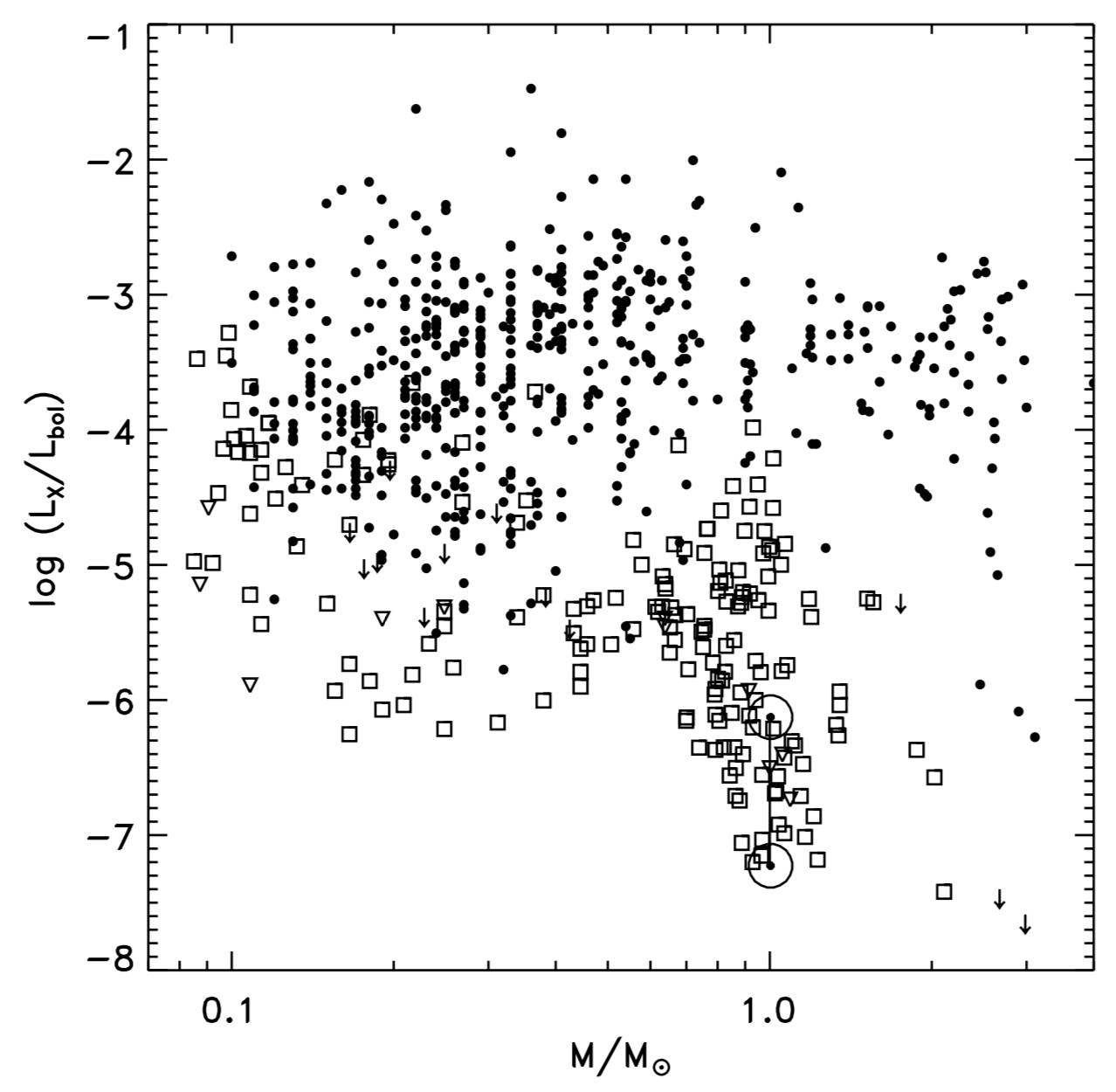} 
\includegraphics[width=0.32\textwidth]{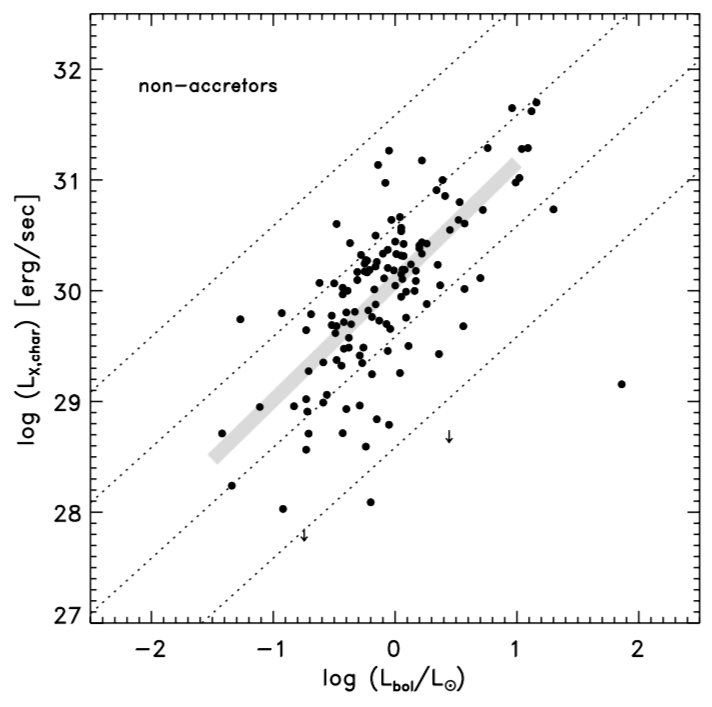}
\includegraphics[width=0.32\textwidth]{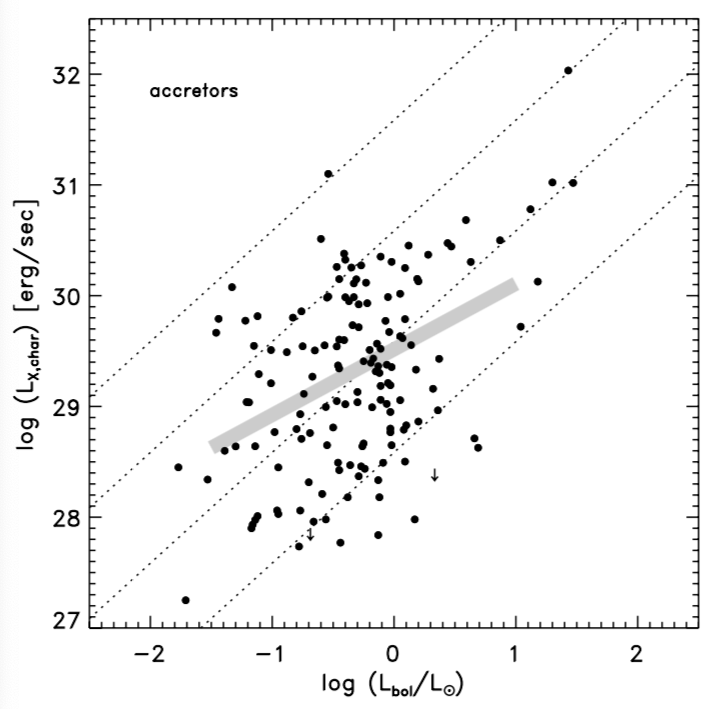}
\end{center}
\caption{The X-ray luminosities for T Tauri stars in Orion from the COUP survey from \citet{Preibisch.etal:05}. The left panel illustrates the fractional luminosity, $L_X/L_{bol}$, as a function of stellar mass for COUP stars (solid dots, arrows for upper limits) and  for a sample of nearby field stars observed by {\it ROSAT} (open squares, triangles for upper limits). 
The right two panels show what \citet{Preibisch.etal:05} term the ``characteristic'' X-ray luminosity, computed to be that remaining after obvious flares are removed, as a function of stellar luminosity for weak-lined T~Tauri stars (``non-accretors'') and classical T~Tauri stars (``accretors").  The range of $L_X/L_{bol}$ exhibited by the Sun through its cycle \citep{Judge.etal:03} is denoted by the joined hollow circles.
}
\label{f:ttaulxlbolvm}
\end{figure}

The fractional X-ray luminosity, $L_X/L_{bol}$, for the \citet{Preibisch.etal:05} sample is compared as a function of stellar mass with field stars detected by {\it ROSAT} \citep{Schmitt.Liefke:04} in Figure~\ref{f:ttaulxlbolvm}.   There are several important results from these data.  From a solar perspective, the most conspicuous feature is the absolute value of the luminosities in comparison to the Sun: T~Tauri stars are 3-4 orders of magnitude brighter than the Sun in X-rays.
 There is no trend in T~Tauri $L_X/L_{bol}$ with stellar rotation velocity, as expected because all T~Tauri stars are in either the saturated or supersaturated regime.  The fractional X-ray luminosities are relatively flat as a function of stellar mass, with a slight rising trend of $\log L_X/L_{bol}=  -3.65(\pm 0.05) + 0.40(\pm 0.10) \log(M/M_\odot)$ but with a very large scatter over the range $10^{-5} \leq L_X/L_{bol} \leq 10^{-2}$.  While in the saturated regime, these X-ray luminosities are slightly below those for saturated main sequence stars in Figure~\ref{f:wright18}.
The best estimate for how the X-ray luminosity itself varies with stellar mass is to $\log L_X=27.58(\pm 0.07) + 1.25(\pm 0.15)\log(M/M_\odot)$.

The origin of the large X-ray luminosity scatter, and whether T~Tauri stars  change their X-ray luminosities by factors of 100 or more over timescales longer than the observation length (approximately 13 days total elapsed time in the case of COUP), remains a subject of debate. Some fraction can be attributed to stochastic variability and \citet{Flaccomio.etal:12} established that in all subsamples the variability amplitudes increase with increasing timescale at least up to the elapsed time over which the observations were taken.  However, comparison of the COUP X-ray luminosities to those in the earlier shorter pilot study by \citet{Feigelson.etal:02} revealed only a factor of 2 or so difference over the 3 years between the respective observations. 

Part of the scatter can be attributed to a difference between classical (accreting) and weak-lined (non-accreting) T~Tauri stars.  Figure~\ref{f:ttaulxlbolvm} also shows the X-ray luminosities of the \citet{Preibisch.etal:05} sample as a function of stellar luminosity, divided into accreting and non-accreting stars.  The former are typically a factor of 2--3 brighter than the latter, confirming results based on earlier studies of smaller samples of T~Tauri stars in Taurus observed with {\it ROSAT} and {\it Einstein} (\citealt{Neuhaeuser.etal:95}, \citealt{Damiani.etal:95}).  

Also clear from Figure~\ref{f:ttaulxlbolvm} is the much larger scatter in the X-ray luminosities of the classical T~Tauri stars. In fact, \citet{Preibisch.etal:05} and \citet{Flaccomio.etal:12} argue that the scatter in the weak-lined T~Tauri star X-ray luminosities are consistent with the variability observed over short and long timescales combined with the uncertainties in derived parameters, such as mass and luminosity.  However, this does not explain the classical T~Tauri stars.  \citet{Flaccomio.etal:12} found that classical T~Tauri stars also showed more variability and the most promising explanation to date for the luminosity scatter is that circumstellar material is responsible for obscuration of some fraction of their coronae, leading to both a lower $L_X$ on average, and to larger secular change in $L_X$.

\subsubsection{Giant Flares and the Ionization of Protoplanetary Disks}
\label{s:ttauflares}

\begin{figure}
\begin{center}
\includegraphics[width=0.53\textwidth]{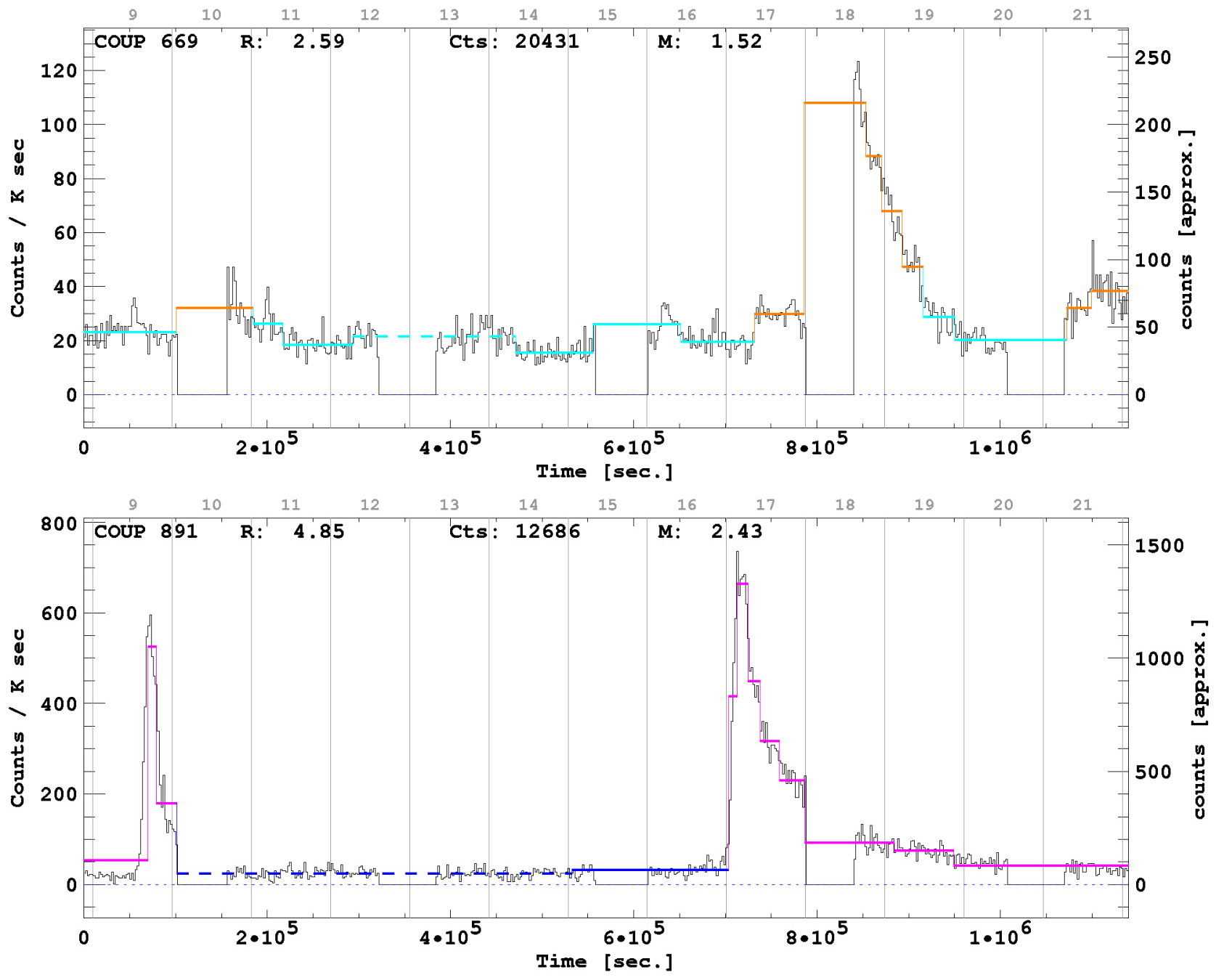}
\includegraphics[width=0.45\textwidth]{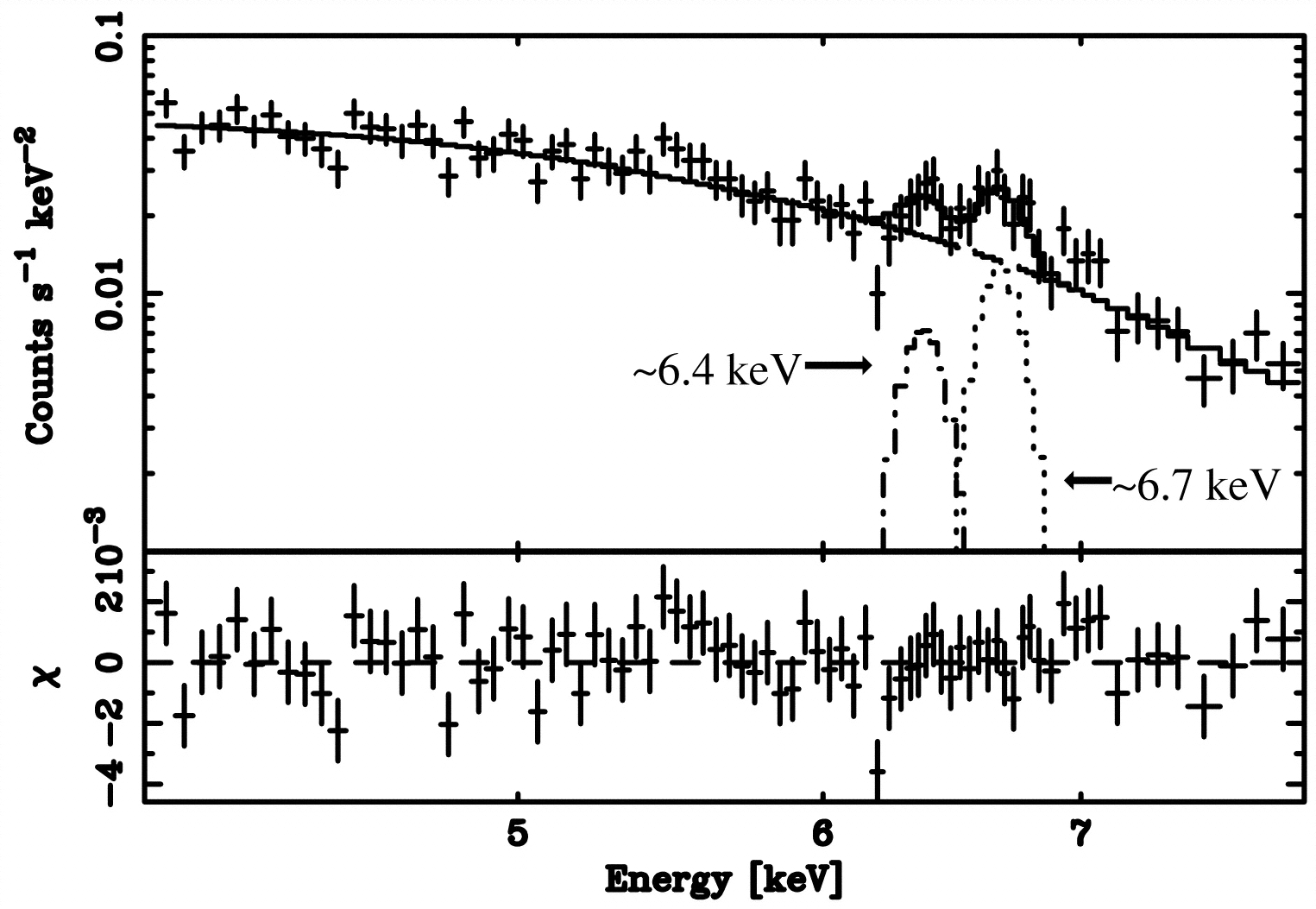}
\end{center}
\caption{Left: The X-ray light curves of two bright flares from ONC stars analysed by \citet{Favata.etal:05}.  In the upper light curve, structure reveals a low-level variability that can be attributed to a superposition of continual flaring. The corresponds to an X-ray flux of about $2\times 10^{32}$~ergs, or 5\%\ of the solar X-ray luminosity. Right: The {\it Chandra} ACIS X-ray spectrum of the Class~I protostar  YLW 16A in $\rho$~Ophiuchi Cloud exhibiting a strong Fe~K$\alpha$ fluorescence line originating from flare ionization of the protoplanetary disk. From \citep{Imanishi.etal:01}.}
\label{f:ttauflares}
\end{figure}

The high, saturated activity level of T~Tauri stars is accompanied by vigorous flaring as expected from analogous behavior of their active main sequence counterparts. Figure~\ref{f:ttauflares} illustrates the {\it Chandra} X-ray light curves of two T~Tauri stars observed to flare during the COUP survey from the study of \citet{Favata.etal:05}.  
The peak count rate in the largest flare (left lower panel; source COUP 891) of 0.75 count~s$^{-1}$ corresponds to an X-ray flux of about $2\times 10^{32}$~ergs.  This is 5\% of the solar bolometric luminosity!  In the context of large solar flares, denoted as ``X-class" and having a peak flux $\geq 10^{-4}$ Watts~m$^{-2}$ in the 1--8~\AA\ band, the COUP 891 flare amounts to $10^2$~Watts~m$^{-2}$, or a million times more energetic.

\citet{Favata.etal:05} found a broad range of decay times for the brightest flares, ranging from 10 to 400~ks. Peak flare temperatures were often extremely high, with half having $T >10^8$~K.  Analysing the flares in terms of simple quasi-static loop models, \citet{Favata.etal:05} found significant sustained heating was required for the majority of the flares, and that flaring loops were extremely large, with semi-lengths up to $10^{12}$~cm---several times the stellar radius.  Such loops would not be stable due to centrifugal forces, and it was speculated that loops could be anchored between the star and its disk.

The  flare frequency as a function of energy for the Orion T~Tauri flares was investigated by 
\citet{Caramazza.etal:07}. The frequencies of flares on the Sun and stars are typically distributed as a function of energy in a power law:
\begin{equation}
\frac{dN}{dE} = k\,E^{-\alpha}   \;\;\;\;\; {\rm where} \;\;\;\;\; \alpha > 0
\label{e:flarefreq}
\end{equation}
where $N$ is the number of flares with energies between $E$ and $E+dE$, emitted in a given time interval. If the index of the power law, $\alpha$, is larger than 2, then a minimum flare energy must be introduced in order to keep the total emitted energy finite and, depending on this cut off, even very high levels of apparently quiescent coronal X-ray emission can be obtained from the integrated effects of many small flares. 

\citet{Caramazza.etal:07}  found a  power law index of $\alpha=2\pm$ for the Orion T~Tauri flares, consistent with the conclusion that T~Tauri X-ray the light curves are entirely composed by overlapping flares. While the more intense flares can be individually detected, weak flares merge together and form a pseudo-quiescent emission level.  Low mass and solar mass stars of similar X-ray luminosity have very similar flare frequencies. 

The energetic radiation of T~Tauri stars is important for understanding their protoplanetary disks within which planets form. It had been thought that the dominant source of ionization of protoplanetary disks would be by cosmic rays, but an analysis by \citet{Glassgold.etal:97} found that stellar X-rays are dominate ionization source through much of the disk volume.   The disks are predominantly heated by optical and IR light from the stellar photosphere, which gives them a flared geometrical structure with a scale high that increases with increasing radial distance \citep[e.g.][]{Armitage:11}.  This geometry favors penetration of X-rays, ionization and coupling of the gas to magnetic fields, and ultimately likely contributes to disk dispersal through photoevaporation \citep[e.g.][]{Ercolano.etal:08}.  X-ray ionization influences chemistry and drives ion-molecular reactions in the outer disk layers \citep[e.g.][]{Semenov.etal:04,Walsh.etal:12}. 
Gas--magnetic field coupling is important for induction of magnetohydrodynamic turbulence via the magnetorotational instability \citep{Balbus:03} and allows angular momentum transport that can drive accretion.  More recent studies have favored the action of magnetized disk winds in driving angular momentum loss, accretion, and contributing to disk dispersal. \citep{Bai:16}.  Regardless of the exact mechanisms and their relative importance, X-ray ionization appears to be crucial. 

That flare X-rays are indeed absorbed by the surrounding disk is illustrated graphically in the {\it Chandra} ACIS X-ray spectrum of the Class~I protostar  YLW 16A in $\rho$~Ophiuchi Cloud \citep{Imanishi.etal:01}. A prominent fluorescence line is present at 6.4~keV originating from inner-shell ionization of cold, neutral, or near-neutral, Fe atoms in the disk by flare X-rays (see also Section~\ref{s:mechanisms}).  

\subsubsection{X-ray Evidence for Accretion from Protoplanetary Disks}
\label{s:accretion}

The protoplanetary gas accretion disk of T~Tauri stars expected to be truncated by the stellar magnetic field at a distance of several radii from the stellar surface near the corotation radius 
\citep{Uchida.Shibata:84}.
From there, accretion is thought to proceed down open magnetic field lines, and form an accretion shock at the stellar surface. This accretion is ultimately the source of the elevated UV emission of T~Tauri stars for their spectral type and the ``veiling" and filling in of absorption lines \citep[e.g.][]{Calvet.Gullbring:98}.

The free-fall velocity, $v_{ff}$, from several stellar radii can amount to 150--400~km~s$^{-1}$ or so, and since the Mach number is high the gas shock temperature, $T_s$, can be estimated under strong shock conditions and is given for a star of mass $M$ and radius $R$ by \citep[e.g.[][]{Calvet.Gullbring:98}
\begin{equation}
T_s=\frac{3\mu m_H}{16k} v_{ff}^2=8.6\times 10^5\; K\left(\frac{M}{0.5\,M_\odot}\frac{2R_\odot}{R}\right).
\label{e:tshock}
\end{equation}
For a typical T~Tauri star, the term in the brackets on the right of Equation~\ref{e:tshock} is approximately unity, and the immediate post-shock temperature is about a million degrees K.  

For a strong shock, the post-shock density is four times the pre-shock density, and for a mass accretion rate $\dot{M}$ in solar masses per year through an accretion column covering a fraction $f$ of the stellar surface, is given by 
\begin{equation}
n_H \approx 2\times 10^{13} \left( \frac{0.01}{f}\right) \left(\frac{\dot{M}}{10^{-8}}\right)\left(\frac{0.5\,M_\odot}{M}\right)^{1/2} \left(\frac{2R_\odot}{R}\right)^{3/2}\;\; \rm cm^{-3}.
\label{e:shockdens}
\end{equation}
Typical accretion spot filling factors, $f$, estimated from UV observations are a few percent, while accretion rates tend to fall in the range $10^{-8}$--$10^{-10} M_\odot$~yr$^{-1}$. With the rest of the terms in Equation~\ref{e:shockdens} being approximately unity, the post-shock density is expected to be of the order of $10^{11}$--$10^{13}$~cm$^{-3}$---considerably higher than typical ambient plasma densities in stellar coronae (Section~\ref{s:helikes}). 

The first, and as it turns out, best example of accretion shock X-ray emission was found in the closest known classical T~Tauri star TW~Hya ($d= 60$~pc) by \citet{Kastner.etal:02}.  The He-like Ne~IX lines were illustrated earlier in Figure~\ref{f:helikes}; the intercombination lines $x$ and $y$ are clearly enhanced relative to the coronal case (AB Dor in the case of Figure~\ref{f:helikes}), and the $w/(x+y$ ratio indicates a density of $n_e\approx 10^{13}$~cm$^{-3}$.  Pushing {\it Chandra} HETG spectral resolution to its limits by means of comparative spectroscopy with non-accreting coronal X-ray sources \cite{Argiroffi.etal:17} measured the line-of-sight redshift in the post-shock accreting gas of $38.3 \pm 5.1$~km~s$^{-1}$, indicating the accretion occurs at low latitudes. TW~Hya possesses a well-resolved face-on dusty disk, and the Doppler shift indicates accretion streams are close to the disk plane.

High density accretion signatures have been seen in several other classical T~Tauri stars, but they are not ubiqitous among the class. The most likely reason for this is the shocks being formed too far into the photosphere and getting obscured from view.  The shock height depends on the accretion rate through both the ram pressure of the stream, and the cooling time of the shocked plasma: higher accretion rates have larger ram pressure that penetrates further into the photosphere, and smaller scale heights due to higher densities and more rapid cooling \citep{Drake:05}. Since not all the X-rays from accretion necessarily escape, determining accretion rates from X-rays is hazardous. \citet{Bonito.etal:14} have examined the observability of accretion shocks based on the results of detailed hydrodynamic simulations and found lower fluxes than expected because of the complex absorption by the optically-thick chromosphere within which the shocks can be embedded,  and of the unperturbed accretion stream. 

The TW~Hya accretion shock Ne/O abundance ratio is anomalously high \citep{Drake.etal:05} and perhaps due to the accretion of grain-depleted gas in a mature old protoplanetary disk.  However, other accretion shocks do not necessarily show the same pattern \citep{Argiroffi.etal:07}, and the origin of the anomaly remains unclear.  More extensive {\it Chandra} observations of TW~Hya using the HETG have also revealed gas emitting in O~VII that appears to be ambient corona plasma heated indirectly by the accretion process \citep{Brickhouse.etal:10}, although this might also be explained by complex accretion shocks \citep{Bonito.etal:14}. An excess in O~VII emission had indeed been noticed based on earlier {\it XMM-Newton} observations of T~Tauri stars in the Taurus Molecular Cloud that likely arises from the same processes. 

\subsubsection{Protoplanetary Disk Destruction in High-Mass Star-forming Regions}

The ability of {\it Chandra} to see all the young stars in a star-forming region in a reasonably unbiased fashion enables an assay of the {\it fraction} of stars that have retained protoplanetary disks.  The disks are conspicuous by their infrared excess emission on top of the stellar photospheric radiation field and can be readily detected in infrared surveys by {\it Spitzer} and to some extent by ground-based near-infrared K-band excesses \citep[e.g.][]{Guarcello.etal:13}.

\begin{figure*}[ht!]
\begin{center}
\includegraphics[width=0.98\textwidth]{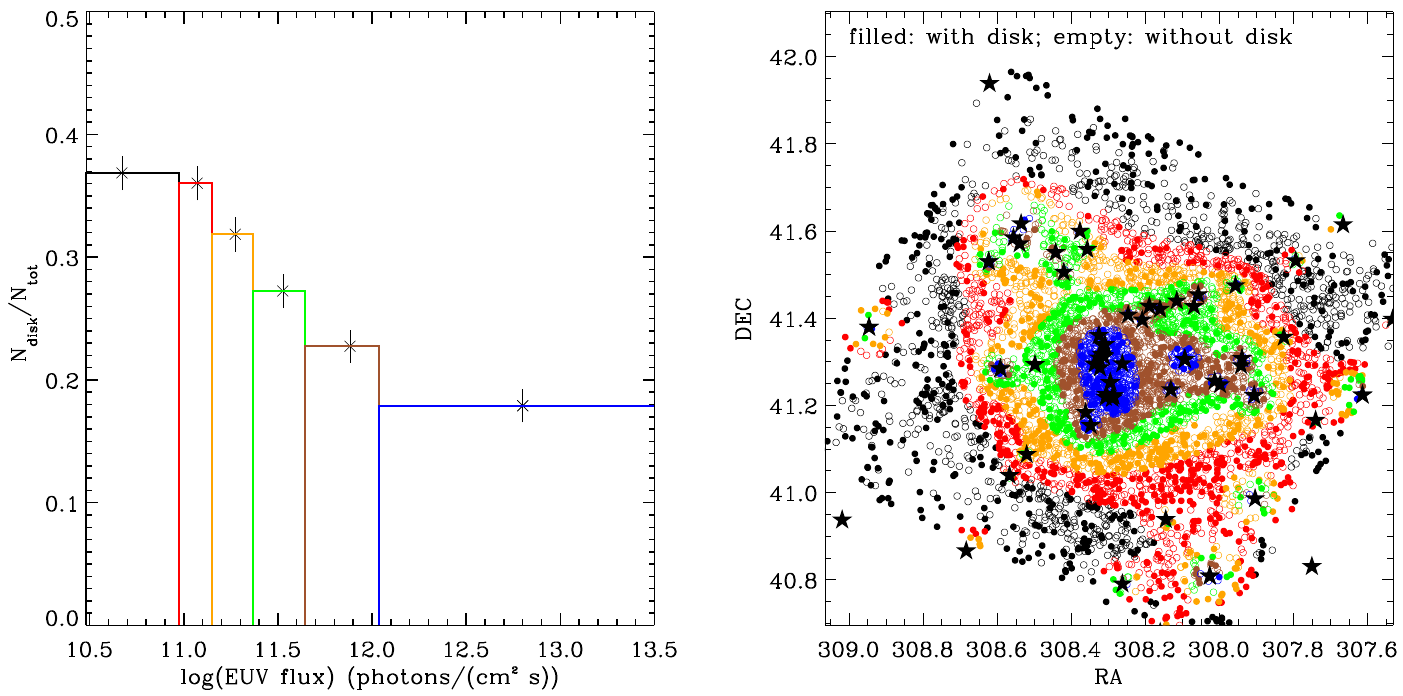}
\end{center}
\caption{The protoplanetary disk fraction as a function of the local EUV radiation field in the massive Cygnus OB2 Association from a {\it Chandra}-based on a multiwavelength survey. The spatial distribution of the stars in the {\it Chandra} survey field falling in each bin is shown in the right panel, with the different colors marking those stars falling in the corresponding bin in the left panels. The star symbols mark the positions of the O stars.  A sharp decline in disk fraction is seen with increasing EUV flux from high mass stars. From \citet{Guarcello.etal:16}.
}
\label{f:photevap}
\end{figure*}

Protoplanetary disks are gradually dissipated by accretion, disk winds and photoevaporation driven by host star UV-X-ray irradiation, by dynamical interactions in very dense star-forming regions, and if intense enough, by photoevaporation by the external radiation field.  Typical disk lifetimes are a few Myr, with essentially all disks having dissipated by 10~Myr \citep[e.g.][]{Ribas.etal:14}.  Modeling of dynamical interaction effects indicate they only become important in the most dense stellar environments  exceeding $\sim 10^3$~pc$^{-3}$ \citep[e.g.][]{Rosotti.etal:14}. 

The relative importance of radiative processes leading to disk dispersal has been more difficult to determine.  This poses problems for understanding the types and configurations of planetary systems that remain after the disk is gone because they likely have a large influence on both the planet formation process itself, which depends on the reservoir of building material, and on planet migration due to planet/planetesimal-gas interaction \citep[e.g.][]{Ribas.etal:15}.  A key part of the problem is whether or not the strong UV and EUV radiation fields of high mass stars are significant sources of disk photoevaporation and loss. 

Evidence of photoevaporation in regions of clusters close to high mass stars based on {\it Spitzer} and combined  {\it Spitzer}  and {\it Chandra} observations was uncovered by, e.g.,  \citet[][NGC~2244]{Balog.etal:07} and \citet[][NGC~6611]{Guarcello.etal:10}.  However, the most definitive evidence to date for disk photoevaporation was uncovered by \citet{Guarcello.etal:16} based on a multiwavelength analysis as part of the {\it Chandra} Cygnus OB2 Legacy Survey \citep{Drake.etal:19}.  At a distance of 1.4~kpc and a stellar content of about 17,000~$M_\odot$, including at least 52 O-type stars and 3 Wolf-Rayet stars \citep{Wright.etal:15}, Cygnus OB2 is the nearest truly massive star forming region to Earth.  Despite being so massive, it is of quite low stellar density and consequently is dynamically relatively unmixed and dynamical interactions are negligible.
\citet{Guarcello.etal:16} found the protoplanetary disk fraction to decline from  40\% to $\leq 20$\% as a function of local FUV and EUV fluxes. The FUV radiation dominates disk dissipation timescales within a parsec of the O stars. In the rest of the association, EUV photons likely induce a significant disk mass loss across the entire association, modulated only at increasing distances from the massive stars due to absorption by the intervening intracluster material.  

At face value, the \citet{Guarcello.etal:16} study indicates that massive star forming regions could be hostile to habitable planet formation.  However, if terrestrial planet formation timescales are shorter than photoevaporation timescales, the disk gas loss might only affect giant planet formation \citep[e.g.][]{Ribas.etal:15}.

\subsubsection{Large-Scale and Small-Scale Diffuse X-ray Emission in Star Forming Regions}

\label{s:diffuse}

The sharp {\it Chandra} PSF not only provides high sensitivity for detecting point sources, but also enables accurate {\em removal} of point sources in survey data. This capability is crucial for understand the origin of residual diffuse or unresolved X-ray emission.  

Star forming regions hosting high-mass stars present tantalizing targets for probing diffuse X-ray emission.  From a theoretical perspective, high-mass stars with supersonic winds and high mass loss rates are expected to blow stellar wind bubbles in the interstellar medium.  A  reverse shock from interaction with the ambient ISM should heat the gas to temperatures of $10^6$~K and higher \citep[e.g.][]{Freyer.etal:06}. Such emission is in principle detectable with {\it Chandra} provided the surface brightness is sufficiently high. 

Diffuse and mostly thermal X-ray emission has indeed now been detected from several star-forming regions \citep[see, e.g., the summary by][]{Townsley.etal:11b},  although it is not always clear that such emission originates from shocks induced by massive stellar winds.  \citet{Wolk.etal:02}, for example, find evidence for non-thermal synchroton diffuse emission in the embedded region RCW~38 possibly associated with a supernova remnant, and similar non-thermal emission is present in NGC~3576 \citep{Townsley.etal:11b}.  Both the thermal and non-thermal emission is seen to fill voids between bright dust emission revealed by Spitzer, as might be expected by a flow seeking the path of least resistance.

\begin{figure*}[ht!]
\begin{center}
\includegraphics[width=0.57\textwidth]{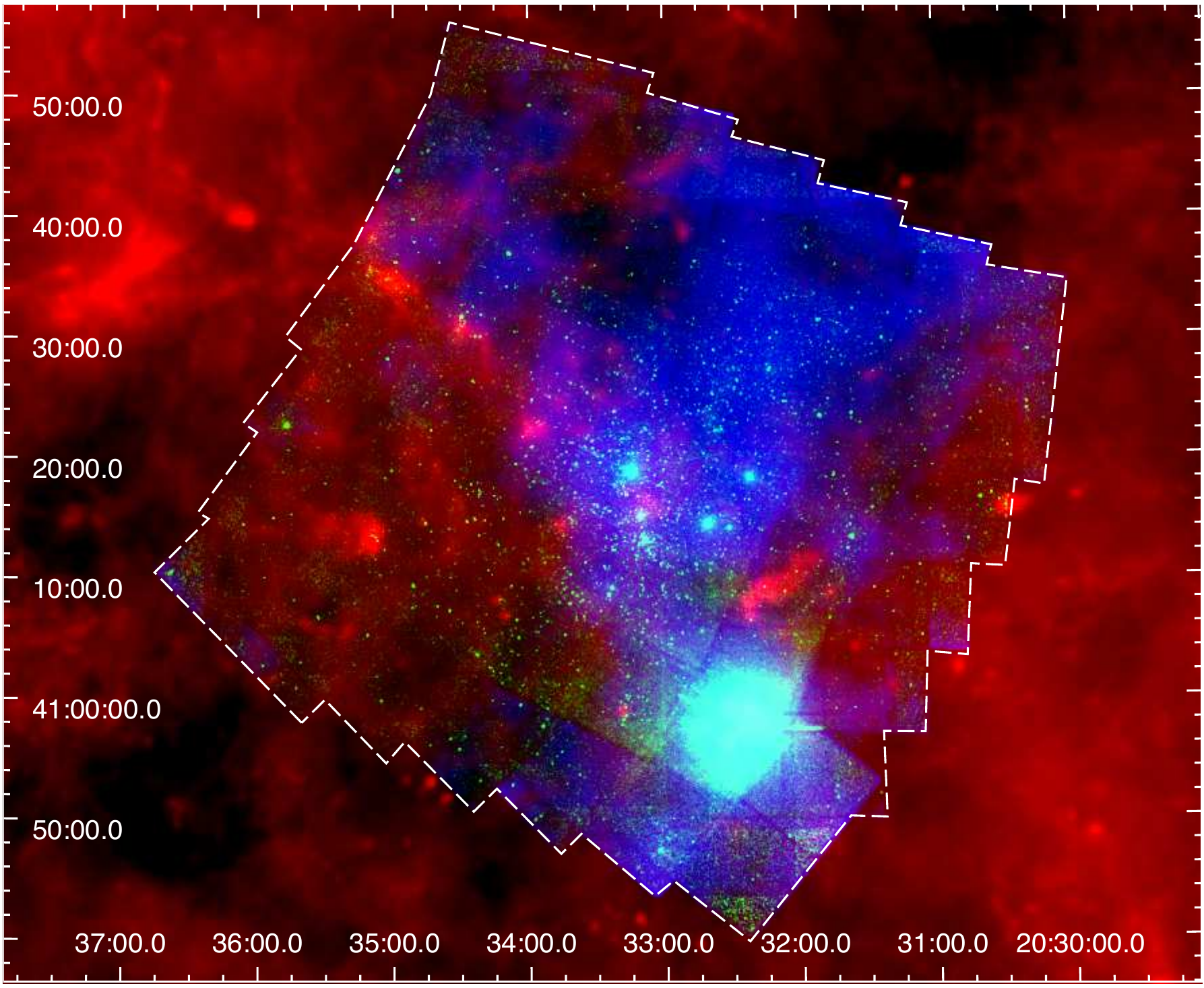}
\includegraphics[width=0.42\textwidth]{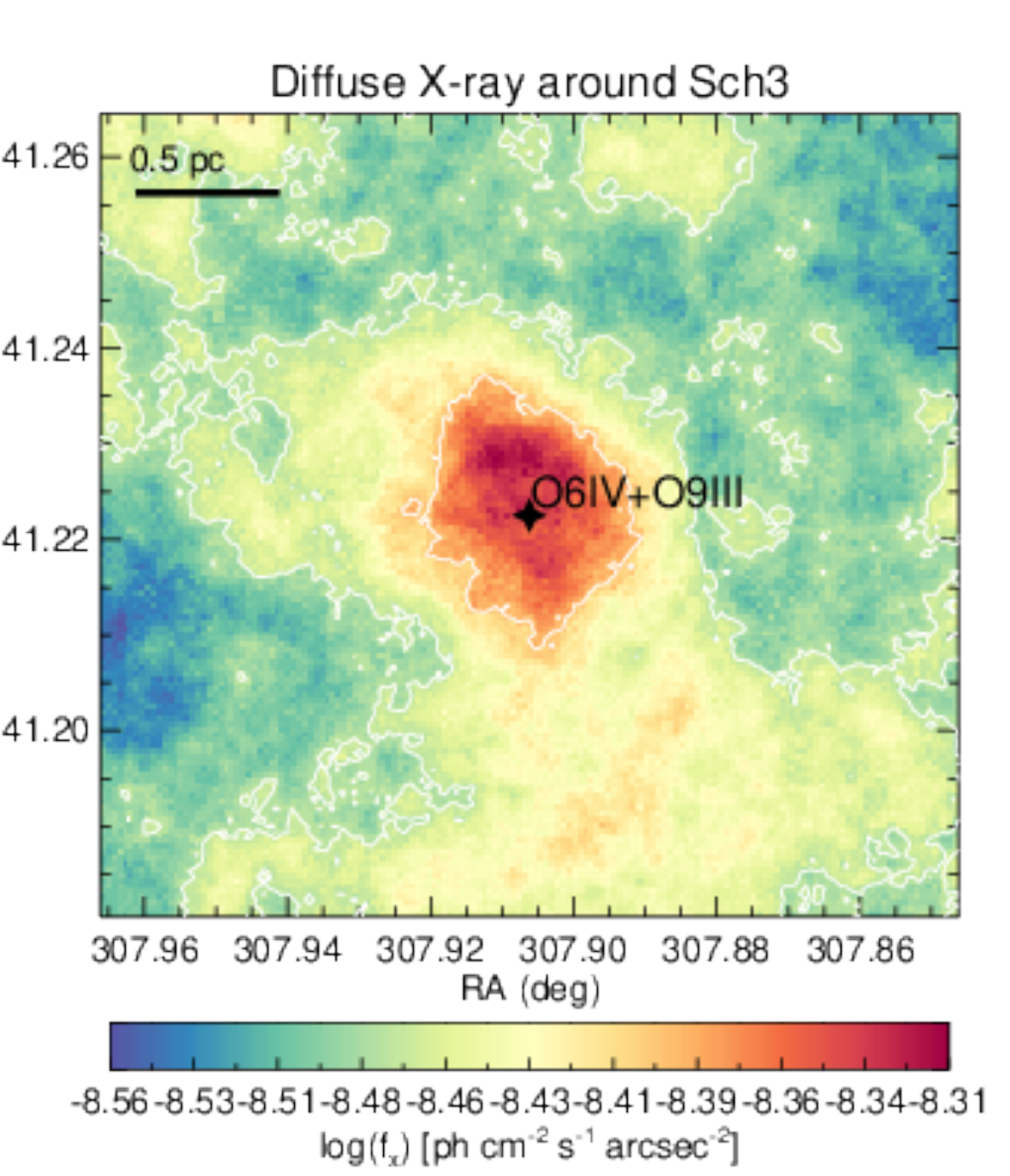}
\end{center}
\caption{The neighbourhood of Cygnus OB2. The ACIS-I mosaic of the Cygnus OB2 survey \citep{Drake.etal:19} is outlined in white. The false RGB color image is composed of the {\it Herschel} 500$\mu$m (T$\approx$10 K) cold gas emission in red, the mosaic of Chandra observations with the point sources in the 0.5--7.0 keV band in green, and the diffuse X-ray emission in the 0.5--2.5 keV energy range in blue.  The diffuse X-ray emission fills a cavity in the infrared {\it Herschel} data.
Right: Observed X-ray flux photon [ph cm$^{-2}$ s$^{-1}$ arcsec$^{-2}$] diffuse map computed with an adaptive smoothing algorithm to achieve a signal-to-noise ratio of 16. 
The black filled star symbol indicates the position of the massive O6V+O9III binary star Schulte~3 that possesses a bright X-ray emitting halo. From \citep{Drake.etal:19}.}
\label{f:diffuse}
\end{figure*}

\citet{Albacete_Colombo.etal:18} detected large-scale thermal X-ray diffuse emission in the Cygnus~OB2 association that follows the spatial distribution of massive stars and similarly fills a cavity that appears to have been excavated by the accumulated winds of the 169 catalogued OB stars of the region. The X-ray emission in the broad 0.5--7.0~keV energy band was reasonably well-matched by thermal plasma components with temperatures of 0.11, 0.40 and
1.2~keV and is illustrated together with {\it Herschel} 500$\mu$m (corresponding to a temperature T$\approx10$~K) cold gas emission in Figure~\ref{f:diffuse}.   The total luminosity of the diffuse emission was found to be  $4.2\times 10^{34}$ erg~s$^{-1}$, which is $10^4$ times less than the estimated total OB star wind kinetic energy injected into the region \citep{Drake.etal:19}.  Remarkably, this fraction is the same as that found in the hydrodynamic modelling study of the interaction of massive stellar winds with the ambient medium by \citet{Freyer.etal:06}.

The biggest surprise from Cygnus~OB2, however, was the detection of diffuse X-ray halos around some evolved massive stars.  Figure~\ref{f:diffuse} also shows an example around the massive O6V+O9III star Schulte~3 that appears to be a direct consequence of the collision and interaction of a fast, dense wind and surrounding gas.  

\citet{Townsley.etal:11b} also noted evidence for CX emission in all of the star forming regions they studied that exhibited diffuse X-ray emission which they interpreted to be a result of interaction of the hot plasma with gas at the interfaces of cold neutral pillars, ridges, and clumps.  Similar discrepancies were also found in Cyg~OB2 by \citet{Albacete_Colombo.etal:18}, reinforcing the conclusion that CX emission is a ubiquitous feature of massive star forming regions.

\section{X-ray Studies of Exoplanet Systems}

The explosion in exoplanet discoveries and science over the last two decades have sparked interest in the relevance of X-rays for exoplanetary systems.  There are at least four X-ray aspects to exoplanets: the effect of stellar X-rays on planets; the effects of planets on their host stars and possibly on their X-ray emission; the use of stellar X-rays to probe exoplanets; and the possibility of intrinsic exoplanetary X-ray emission analogous to the solar system processes discussed in \ref{s:xraysolsys}.  The latter remain well beyond the capabilities of {\it Chandra} and must await future missions to explore. 

\subsection{X-ray Induced Atmospheric Loss}

One important, and potentially critical, effect of stellar coronal emission on exoplanets is photoevaporation of the atmosphere.  
The rate at which gas is lost from an exoplanet's atmosphere is
critical for the survivability of surface water.  Atmospheric mass loss can
be driven by both thermal and non-thermal processes, all of which 
depend upon the radiation and winds of their host stars. The dominant
thermal process is hydrodynamical outflow energized by extreme ultraviolet
(EUV; 100--912~\AA) and X-radiation (0.1--100~\AA).  This short wavelength radiation is absorbed in the very top layers of the atmosphere, in the thermosphere, and the energy input can be sufficient to levitate gas against the exoplanetary gravitational
potential \citep[e.g.][]{Owen.Jackson:12}.

A simple ``energy limited" atmospheric mass loss rate for an X-ray flux $F_X$ received by a planet can be written 
\begin{equation}
\frac{dM}{dt}= \frac{4\pi R_{pl}^3 F_X}{G\mu M_{pl}},
\label{e:escape}
\end{equation}
where $\mu$ is the mean atmospheric particle mass, $M_{pl}$ and $R_{pl}$ are the planetary radius and mass, respectively, and G is the gravitational constant.
Equation~\ref{e:escape} assumes that the bulk of the energy from the X-ray (and in principle, also EUV) heating goes into the escape of the gas, and that the radius at which X-rays are absorbed is not significantly different to the planetary radius.  Using Equation~\ref{e:escape}, \citet{Penz.etal:08} showed that a significant fraction of the mass of a close-in gas giant can be lost to X-ray photoevaporation over several Gyr.

\begin{figure*}[ht!]
\begin{center}
\includegraphics[width=0.46\textwidth]{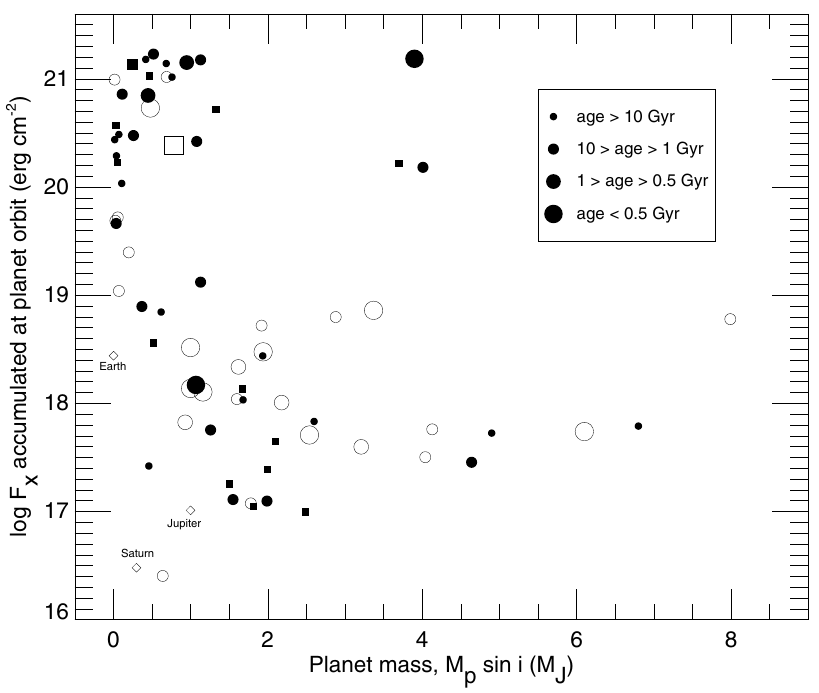}
\includegraphics[width=0.52\textwidth]{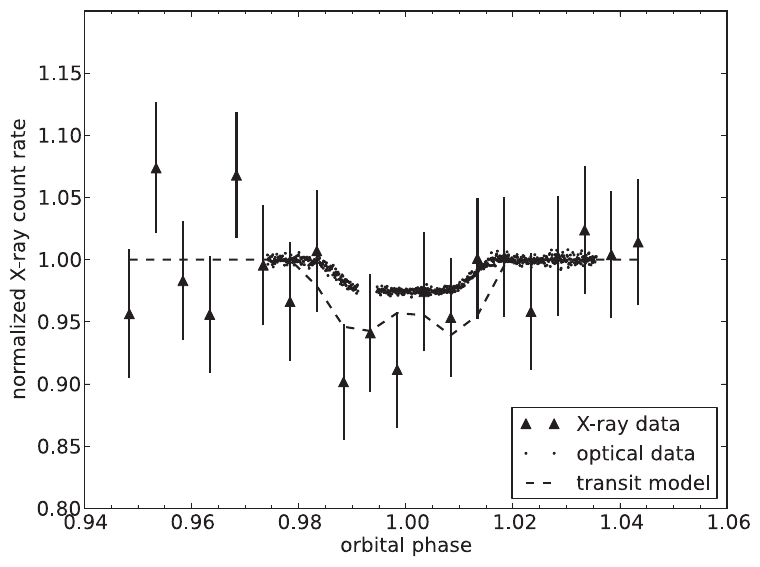}
\end{center}
\caption{Left: Distribution of planetary masses ($M_{pl} \sin i$) with accumulated X-ray dose from an age of 20 Myr to present.  Filled symbols (squares for subgiants, circles for dwarfs) represent  {\it Chandra} and {\it XMM-Newton} data, arrows denote upper limits, and open symbols are ROSAT data. From \citet{Sanz-Forcada.etal:10b}. 
Right:  Combined data for six {\it Chandra} and one {\it XMM-Newton} X-ray transits of the HD189733 hot Jupiter system in comparison with the optical transit. Dashed lines show the best fit to a limb-brightened transit model. From \citet{Poppenhaeger.etal:13}.}
\label{f:exoplanets}
\end{figure*}

Empirical evidence that X-ray photoevaporation is important was presented in a statistical study of exoplanet mass for stars observed by {\it Chandra} and {\it XMM-Newton}  
by \citet{Sanz-Forcada.etal:10b}. Tracing the accumulated X-ray dose over the lifetime of each system, they found  a distribution of planetary mass with X-ray dose (Figure~\ref{f:exoplanets}) consistent with a scenario in which the bulk of the mass of the lightest systems has been eroded away, with the most massive planets tending to have suffered the smallest X-ray doses.

\subsection{Star-Planet Interaction?}

Close-in planets can interact with their host stars either gravitationally or via their respective magnetic fields, with the latter expected to dominate any enhancement in the stellar X-ray signal.  Some weak evidence for such an enhancement based on a survey of exoplanet host X-ray fluxes was found by \citet{Kashyap.etal:08}, although subsequent studies with larger samples have found no positive statistical evidence that the effect is really attributable to planets \citep[e.g.][]{Poppenhaeger.Schmitt:11,Miller.etal:15}.  Firm evidence for significant star-planet interaction is still lacking. {\it XMM-Newton} observations of the hot Jupiter host HD189733 that revealed three flares within a fairly narrow range of planetary orbital phase prompted \citet{Pillitteri.etal:14} to suggest planetary magnetic interaction with stellar surface active regions as a possible mechanism. \citet{Maggio.etal:15} detected evidence of X-ray enhancement at periastron in the eccentric HD~17156 system hosting a Jupiter-mass planet. In this case, \citet{Maggio.etal:15} suggested accretion of material evaporated from the planet onto the star might be the cause.

\subsection{X-rays as Probes of Exoplanet Atmospheres}

X-rays provide potentially powerful diagnostics of planetary upper atmospheres through X-ray transit observations.  X-ray absorption through an atmosphere measures the absorbing column depth along the line of sight.  Since the absorption cross-section as a function of energy in the X-ray range depends on chemical composition, gas bulk chemical composition can also be determined.  This type of measurement is unique to the X-ray range. Since even the largest exoplanets---so-called ``hot Jupiters"---have cross-sectional areas of only a few percent of that of their stellar hosts and their transits across stellar disks typical last only a few hours, X-ray transit measurements are very challenging.

The transit of the hot Jupiter HD189733b in front of its K-type host star was detected through X-ray absorption by oxygen in {\it Chandra} ACIS-S observations by \citet{Poppenhaeger.etal:13}. The transit in soft X-rays, illustrated in Figure~\ref{f:exoplanets}, was significantly deeper than observed in the optical---removing 6\%--8\% or so of the stellar intensity versus a broadband optical transit depth of 2.41\%.  The deep transit was inferred to be caused by a thin outer planetary atmosphere that is transparent at optical wavelengths but sufficiently dense to be opaque in X-rays. The X-ray radius was also larger than observed at far-UV wavelengths, most likely due to the outer atmosphere being largely ionized, rendering the gas transparent in the UV but not in X-rays.

The exciting promise of using X-rays as a new probe of exoplanet atmospheres has been stymied by the build-up of a contamination layer on the ACIS instrument with a consequent loss of low-energy sensitivity that has prevented {\it Chandra} from pursuing further X-ray transit observations.

\section{X-rays from High Mass Stars}
\label{s:highmass}

X-rays from high mass stars were discovered serendipitously with the {\it Einstein} satellite shortly after launch during calibration observations using the powerful high mass X-ray binary ``microquasar" Cyg X-3 \citep{Harnden.etal:79}. {\it Einstein} had seen some of the stars in the nearest region of truly massive star formation, the Cygnus~OB2 Association. 
The early {\it Einstein} and later ROSAT observations established that the X-ray
luminosity of O and early B-type stars scales with their bolometric luminosity following approximately the relation $L_X/L_{bol} \sim 10^{-7}$ \citep{Long.White:80,Berghoefer.etal:97}.  

O-type to early B-type stars have masses of approximately $10M_\odot$ or more. They evolve rapidly, eventually becoming Wolf-Rayet stars and have total lifetimes of typically only 4--10 Myr. 
OB stars emit copious ultraviolet radiation, dominating the ionization of the interstellar medium in their vicinity. Their intense radiation fields also drive powerful and massive stellar winds through line opacity with terminal velocities exceeding 1000~km~s$^{-1}$.  Through their radiative and mechanical energy injection into the ISM during their lifetimes, together with their ultimate demise in colossal supernova explosions that enrich their surroundings in metals as well as in kinetic energy, they play a pivotal role in the evolution of their host galaxies. 

The ubiquity of rapidly expanding stellar winds from OB stars were one of the most unexpected and important discoveries of the early NASA space program \citep[e.g.\ ][]{Snow.Morton:76}.  The soft X-ray emission subsequently discovered by {\it Einstein} was originally understood in terms of shock instabilities in line-driven winds
\citep{Lucy.White:80}; however, understanding of the processes that lead to the shocks and X-ray emission are still incomplete.  The instabilities arise because of ``shadowing" in the continuum due to absorption in the photosphere such that a Doppler shift in the wind material out of the line results in an increase in driving force---the so-called ``line de-shadowing instability" \citep[see, e.g., the review by][]{Puls.etal:08}.  In cases where emission line driving is important, Doppler shifting out of the rest frame of the emitting source leads to a decrease in driving force.  In both cases, the variation of acceleration and velocity among streams and parcels of wind material lead to collisions and hydrodynamic shocks, and a highly inhomogeneous clumpy wind \citep[e.g.][]{Feldmeier.etal:97}.  

The X-ray luminosity of OB stars declines rapidly for stars later than B-type: a {\it ROSAT} OB star survey found detection rates of close to 100\% for O stars but less than 10\%\ for stars B stars later than B3 \citep{Berghoefer.etal:97}.  The weaker radiative driving power of later spectral types produces winds that are simply too tenuous to generate copious X-ray emitting shocked gas and X-ray emission is predicted to drop sharply (see Section~\ref{s:lxlbol} below).

As we shall see below, this picture has expanded to encompass the effects of stellar magnetic fields and the collisions of winds in high-mass binaries, as well as the notorious ``weak wind' problem uncovered for massive stars in which UV line diagnostics show clear wind signatures but derived mass-loss rates are lower by more than an order of magnitude compared with similar stars and with theoretical expectations \citep[e.g.][]{Puls.etal:96}.

\subsection{Universality of the $L_X$--$L_{bol}$ Relation}
\label{s:lxlbol}

\begin{figure*}[ht!]
\begin{center}
\includegraphics[width=0.49\textwidth]{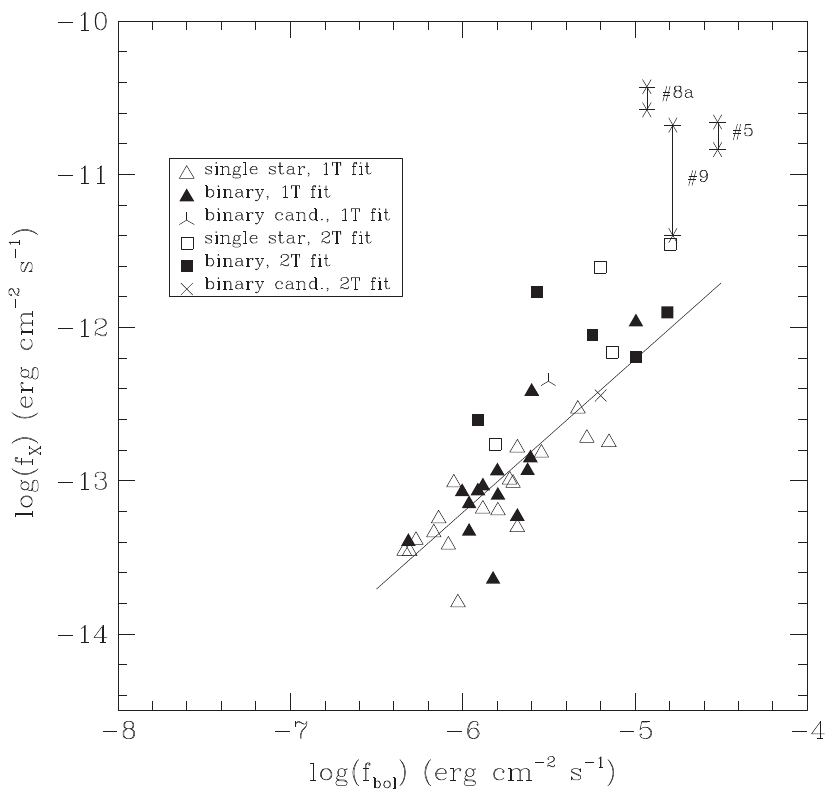}
\includegraphics[width=0.49\textwidth]{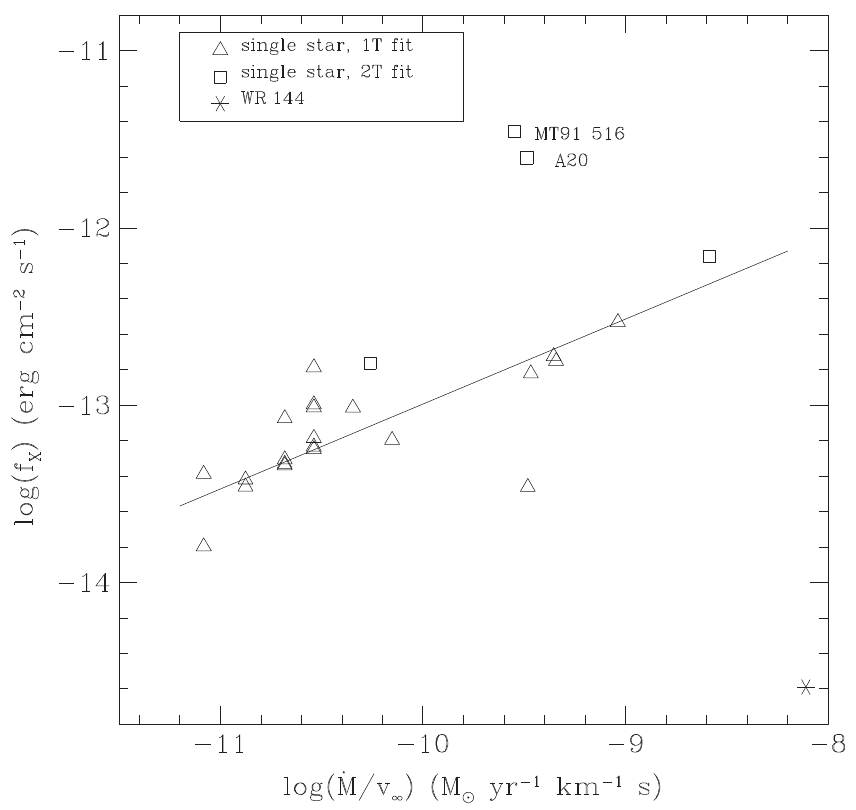}
\end{center}
\caption{Left: The relation between X-ray and bolometric luminosities for bright O stars with more than 30 counts in the {\it Chandra} Cygnus OB2 survey analysed by \citet{Rauw.etal:15}. X-ray luminosities are for the 0.5--10~keV band and have been corrected for interstellar absorption.  The straight line corresponds to unweighted best fit scaling relation $\log L_X/ L_{bol} = -7.21\pm 0.24$.
The variation of X-ray fluxes seen in {\it XMM-Newton} observations for the multiple systems Cyg OB2 \#5, \#8a, and \#9 that are likely colliding wind sources (see Section~\ref{s:colliding}) are indicated by asterisks. Right: The X-ray fluxes of presumably single stars in the Cygnus OB2 survey corrected for interstellar absorption as a function of the logarithm of the ratio of the wind mass loss rate and terminal velocity, $\log (\dot{M}/v_\infty)$.   Both figures from \citet{Rauw.etal:15}.}
\label{f:ostars_cygob2}
\end{figure*}

Massive stellar clusters containing numerous co-eval objects at a known distance are valuable targets for probing the high energy properties OB star winds.
The {\it Chandra} ACIS-I survey of the massive star formation region in the Carina OB1 Association, dubbed the {\it Chandra} Carina Complex Project (CCCP; \citealt{Townsley.etal:11}), garnered the largest sample of OB stars in a single survey to date, detecting 129 O and B stars \citep{Naze.etal:11}.  Of these, 78 had sufficient counts for spectral characterization and all of those were of spectral type B3 or earlier.  Based on the 60 of these that are O stars, \citet{Naze.etal:11} found a relation $\log L_X/ L_{bol} = -7.26\pm 0.21$, in good agreement with the rough canonical $\sim -7$ hewn out from early {\it Einstein} data.  The large number of stars enabled study of subsections of them. They derived similar $L_X/L_{bol}$ ratios for both bright and fainter objects, and for stars of different luminosity classes or spectral types. 

The universality of the $L_X/L_{bol}$ relation for O stars is further highlighted in the study of a large number of OB stars in the Cygnus~OB2 Association.  \citet{Rauw.etal:15} obtained  $\log L_X/ L_{bol} = -7.21\pm 0.24$ from 40 O stars bright enough for spectral characterization---essentially identical to the Carina~OB1 result (Figure~\ref{f:ostars_cygob2}).

Despite having essentially been established since the early 1980s, the physical origin of the $L_X$--$L_{bol}$ relation has proven difficult to reproduce from a theoretical perspective.   \citet{Owocki.etal:13} note that a wind subject to the line de-shadowing instability can be viewed as causing a small fraction of the wind material (say $10^{-3}$) to pass through an X-ray emitting shock. This rough picture has been verified in detail based on an analysis of line profiles in high resolution {\it Chandra} HETG X-ray spectra by \citet{Cohen.etal:14} who found that  the average wind mass element, as it advects through the wind, passes through roughly one shock that heats it to at least $10^6$~K. In this case, if the shocks cool purely radiatively, X-ray luminosity is limited by the kinetic energy flux and should scale approximately with the mass loss rate in the wind, $L_X\sim \dot{M}$.  The problem with this scaling is that the standard model for radiatively driven stellar winds predicts, to first order, a mass loss scaling with luminosity as $\dot{M}\sim L_{bol}^{1.7}$, implying  a much steeper $L_X\sim L_{bol}^{1.7}$ than is observed.  The situation is even worse if shocks are cooled adiabatically, with an $\dot{M}^2$ dependence expected, and therefore $L_X\sim L_{bol}^{3.4}$.

\citet{Owocki.etal:13} proposed a solution to the conundrum through mixing of cooler unshocked gas with the shocked gas through the thin-shell instability.  \citet{Kee.etal:14} show that the deformation of the shock compression by the instability leads to extended shear layers and fingers of cooled, more dense material.  Strong X-ray emission is limited to the tips and troughs of these fingers, and reduces the X-ray emission well below analytical expectations from radiative shocks without consideration of the instability.  Assuming the strength of this mixing-induced quenching of X-ray emission scales with the shock cooling length raised to a power $m$, termed the ``mixing exponent'' by \citet{Owocki.etal:13}, they show that if $m$ lies in the range 0.2--0.4 this leads to an X-ray luminosity scaling with mass loss as $L_X\sim \dot{M}^{0.6}$, thereby recovering $L_X\sim L_{bol}$.  

According to the \citet{Owocki.etal:13} the X-ray luminosity should also scale with $(\dot{M}/v_\infty)^{1-m})$ for O stars. Figure~\ref{f:ostars_cygob2} illustrates the relation between the observed X-ray flux and $\log (\dot{M}/v_\infty)$ for which \citet{Rauw.etal:15} obtained 
\begin{equation}
\log f_x=(0.48\pm 0.10)\log\frac{\dot{M}}{v_\infty} -8.19\pm0.97,
\end{equation}
which corresponds to $m=0.52\pm 0.1$.  Given the various uncertainties in the determination of $f_x$, this might be considered very marginally consistent with the  upper limit to the range proposed by \citet{Owocki.etal:13}.

While promising for explaining the observed X-ray to bolometric luminosity scaling for O star winds, the \citet{Owocki.etal:13} scenario does involve some {\it ad hoc} assumptions that can only be properly verified with detailed three-dimensional modeling.  Importantly, their model predicts different  behavior of the scaling of $L_X$ and $L_{bol}$ toward both low and high mass loss limits of O-star winds.  At the former, corresponding to low luminosities and later (B) spectral types, shocks should become adiabatic and the X-ray luminosity should drop steeply with $\dot{M}^2$.  Such a drop is indeed observed for stars of spectral type later than B3 \citep{Cohen.etal:97}. 

At the latter high luminosity end, winds should become optically thick and the X-ray luminosity should saturate and for very high mass-loss rate winds that should be optically-thick, even decrease with increasing luminosity and mass loss rate approximately as $L_X \sim 1/\dot{M}$. No such drop is apparent in the {\it Chandra} Cygnus~OB2 sample, although \citet{Rauw.etal:15} suggest that this is likely because the transition occurs at still higher luminosities than reached by those stars. 
\citet{Owocki.etal:13} suggest that the analysis of {\it Chandra} HETG X-ray data for the O2If star HD 93129A---one of the earliest and most massive
stars known in the Galaxy---by \citet{Cohen.etal:11} provides a potential example approaching the high mass loss rate limit.  While the X-ray luminosity was still well-approximated by the $L_X\sim 10^{-7}L_{bol}$ relation, absorption by bound-free opacity in the dense wind makes the spectrum harder.  Based on a catalog of O stars from {\it XMM-Newton} data, \citet{Nebot_Gomez-Moran.Oskinova:18} find instead that the $L_X$--$L_{bol}$ relation tends to break down for O supergiants, with a large scatter in X-ray luminosities  over three orders of magnitude for stars of rather similar bolometric luminosity.  Whether or not this is a symptom of the prediction of the \citet{Owocki.etal:13} model is as yet uncertain.

\subsubsection{Insights from High Resolution X-ray Spectroscopy}
\label{s:highres}

Massive stars present compelling cases for high-resolution spectroscopy with the {\it Chandra} diffraction gratings.  With wind outflow velocities in excess of 1000~km~$s^{-1}$, line profiles are expected to show significant Doppler broadening.  Intercombination to forbidden line ratio of of He-like ions, $z/(x+y)$, are sensitive to both plasma density and the intensity of the local radiation field (Section~\ref{s:helikes}) and are expected to provide powerful diagnostics of the location of the shocked plasma X-ray source.  It will come as no surprise that there are a myriad observations at high resolution with both {\it Chandra} and {\it XMM-Newton}; here we provide just some brief examples.  Extensive discussion of further results can be found in the review by \citet{Gudel.Naze:09}.

\begin{figure*}[ht!]
\begin{center}
\includegraphics[width=0.48\textwidth]{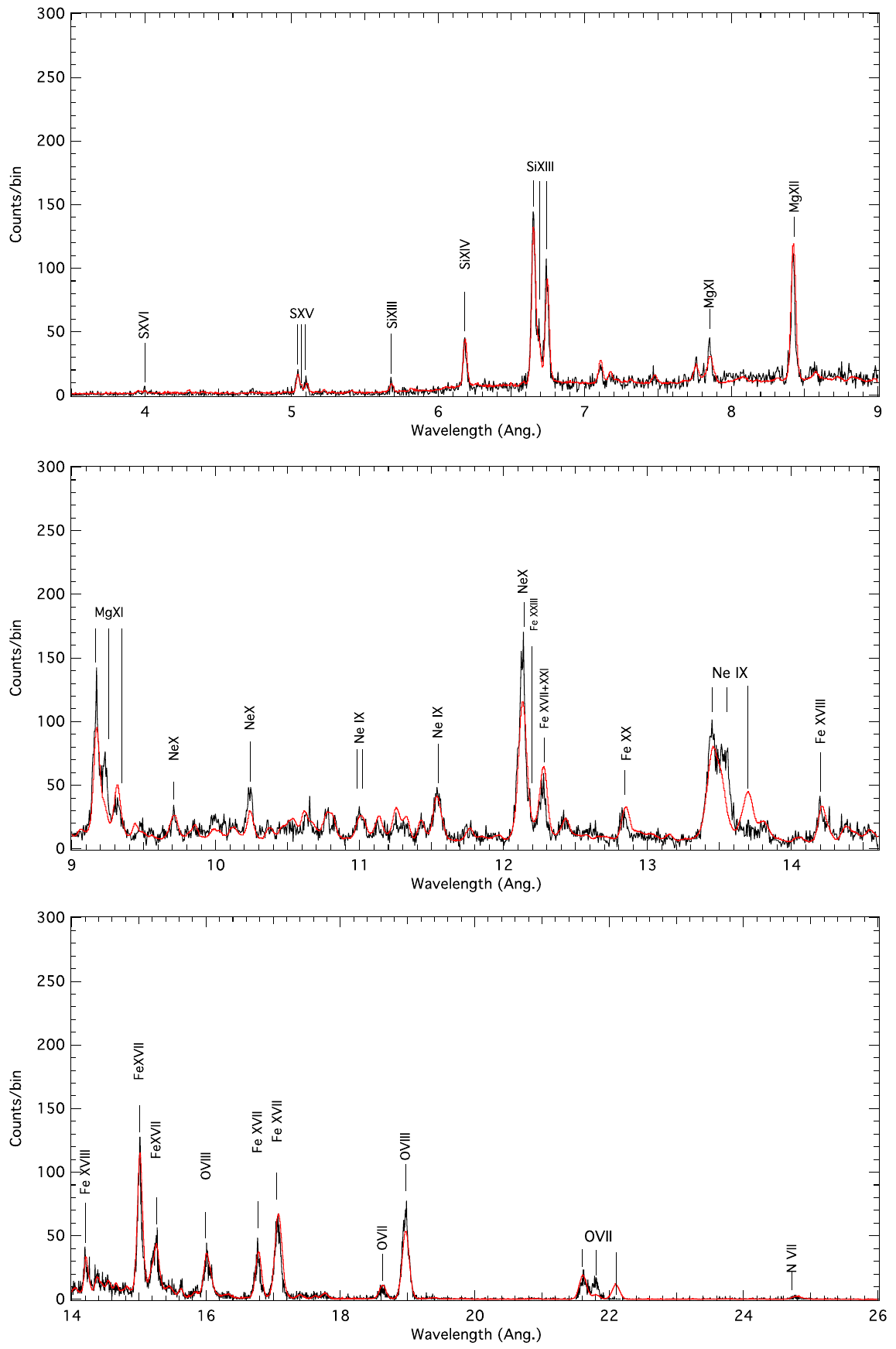}
\includegraphics[width=0.48\textwidth]{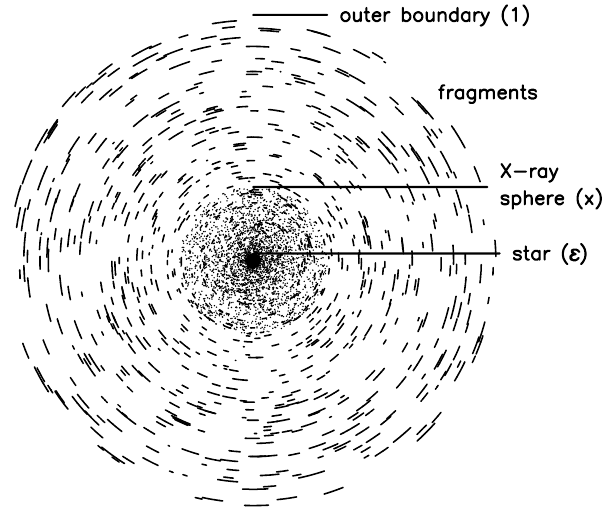}
\end{center}
\caption{Left: {\it Chandra} 
combined MEG+HEG spectrum of the binary $\delta$~Ori~Aa (black) from a 500~ks exposure covering nearly an entire orbit, demonstrating a line-dominated thermal spectrum with significant line broadening due to wind outflow velocities in the range 550--1000~km~s$^{-1}$.   
Model spectra shown in red assume low density and do not include effects of UV photoexcitation.  Consequently, the model overestimates the strength of the forbidden lines and underestimates the strengths of the intercombination lines in He-like ions.  From \citet{Corcoran.etal:15}.
Right: A sketch of the porous clumped wind model from \citet{Feldmeier.etal:03b} suggested to help explain observed line profiles for O star winds. The ``X-ray sphere'' corresponds to the region within which X-rays originate.}
\label{f:zetapup}
\end{figure*}

\begin{figure*}[ht!]
\begin{center}
\includegraphics[width=0.45\textwidth]{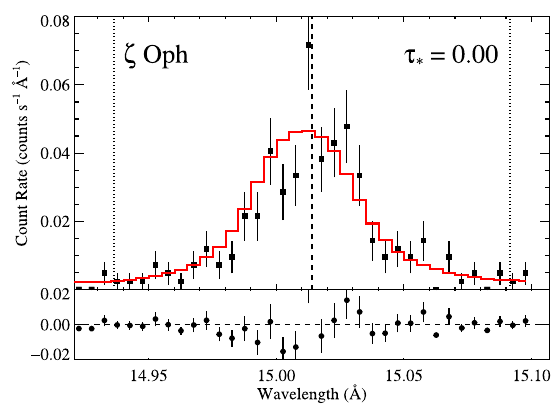}
\includegraphics[width=0.45\textwidth]{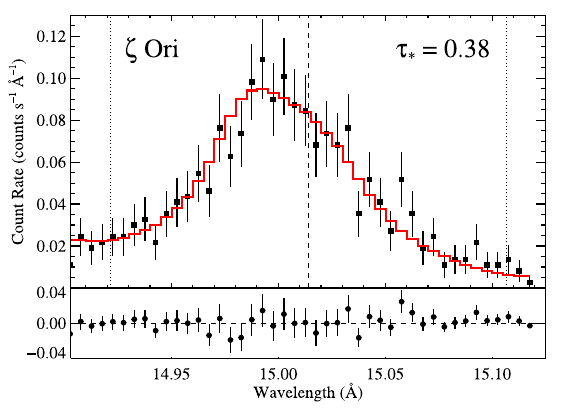}
\end{center}
\caption{Spectral line profiles for the Fe~XVII 15.02~\AA\ resonance line in $\zeta$~Oph and $\zeta$~Ori. Line center is denoted by the vertical dashed line and red histograms indicate the best-fit spherical shocked wind model. The line profile for $\zeta$~Oph appears symmetrical due to a weaker wind compared with the  $\zeta$~Ori profile that shows wind self absorption of the red-shifted, receding emission. From \citet{Cohen.etal:14b}.}
\label{f:zetas}
\end{figure*}

An early indication of the diagnostic power of high resolution X-ray spectra was provided by the HETG spectrum of $\theta^1$~Ori C, the most massive of the Orion ``Trapezium" stars and a close binary comprising O6Vp and B0V components.  \citet{Schulz.etal:00} found temperatures ranging from 0.5--$6\times 10^7$~K from a rich emission line spectrum. Lines were significantly broadened, with velocities ranging from 400 to 2000~km~s$^{-1}$, while the He triplets indicated electron densities $n_e>10^{12}$~cm$^{-3}$.  While some of these features confirmed the wind shock model for X-ray emission, the surprisingly high temperatures and densities were not expected. These features indeed were later to help confirm that $\theta^1$~Ori C is a case of a magnetically-confined wind shock (see Section~\ref{s:magconf} below).

Similar broad lines were obtained for the O 9.7Ib star $\zeta$~Ori by \citet{Waldron.Cassinelli:01}, and for the O4 f star $\zeta$~Pup by \citet{Cassinelli.etal:01}, in agreement with shocked wind models.
$\zeta$~Pup showed the classic line profile shapes expected from a shocked wind model: a broadened profile asymmetric and blue-shifted significantly due to absorption of the receding red-shifted part of the wind hidden behind the blue-shifted material.  This ``classic" case turned out to be a rarity, with the great majority of high resolution O star spectra showing more symmetrical profiles with little or no blueshift \citep[e.g.][]{Waldron.Cassinelli:07}!  

One of the best quality spectra obtained for a massive star is the {\it Chandra} HETG spectrum of the triple system $\delta$~Ori~A, which is dominated by $\delta$~Ori~Aa1 in X-rays (Figure~\ref{f:zetapup}). \citet{Corcoran.etal:15} also found largely symmetrical line profiles. He-like triplets were affected by UV excitation indicating formation within a few stellar radii, and increasing in distance with decreasing temperature of formation.  Lines were broadened by 550--1000~km~s$^{-1}$, which is 0.3--0.5 times the wind terminal velocity, indicating formation within the acceleration zone and again in broad agreement with the radiatively-driven wind shock scenario.

Two possible solutions to the lack of expected line asymmetries are wind clumping (see Figure~\ref{f:zetapup}) and lower mass loss rates, with both having the effect that winds are effectively less opaque and the receding red-shifted wind visible. 
The answer appears to be a combination of both. \citet{Cohen.etal:14b} revisited the HETG spectra of single O stars and applied a spherically symmetric wind model to derive mass-loss rates from the observed spectral line profiles in a way that is not sensitive optically-thin clumping. Two such profiles are illustrated in Figure~\ref{f:zetas}. They derived mass loss rates lower by an average factor of 3 than the theoretical models of \citet{Vink.etal:00}, and consistent with clumping-dependent H$\alpha$ measurements if clumping factors of $f_{cl}\sim 20$, where $f_{cl}=\langle \rho^2\rangle/\langle \rho \rangle^2$ for gas density $\rho$ are assumed. The same profile fitting technique found an onset radius for X-ray emission in the wind of typically at $r \approx 1.5R_\star$.

\subsection{Colliding Winds}
\label{s:colliding}

High mass stars tend to have a high multiplicity and most are expected to reside in binary systems.  Supersonic winds from two stars in relative proximity are expected to interact with each other leading to high Mach number shocks,  giving rise to a variety of observational signatures from radio  to $\gamma$~rays.  X-rays from ``colliding wind binaries'' were in fact predicted by \citet{Prilutskii.Usov:76} and \citet{Cherepashchuk:76} two years before the  detection of X-rays from massive stars \citep[see, e.g., the review by][]{Rauw.Naze:16}.

Early observations by {\it Einstein} and {\it ROSAT} came as a surprise in finding the X-ray luminosities of known binary system, both of Wolf-Rayet type and O-type stars, to be significantly lower than theoretical colliding wind calculations had suggested, and in many cases no larger than expected from single stars \citep{Pollock:87,Chlebowski.Garmany:91}.  However, some of the more luminous binaries did turn out to be significantly brighter in X-rays and this over-luminosity is now understood in terms of colliding winds.

The collision of two winds broadly results in two high Mach number shocks separated by an interface region whose shape depends on the relative ram pressures of the two winds. The general picture of the interaction has been investigated for different parameters using 3D hydrodynamical simulations \citet[e.g.][]{Pittard.Parkin:10}.  The shocks raise the gas involved to significantly higher temperatures than the line de-shadowing instability.  For winds in which X-ray optical depth effects are important, the observed X-ray emission can also be phase-dependent as the line-of-sight to the shock region changes.  In eccentric binaries the emission is also expected to vary with orbital phase as the conditions in the collision shock regions vary.  

\citet{Stevens.etal:92} introduced a cooling
parameter, $\chi=t_{cool}/t_{esc}$, for the ratio of the cooling time to the escape time for gas to flow from the system. In systems where $\chi > 1$, the gas does not have time to cool before it escapes and so the wind collision region behaves like an adiabatic shock and is large and extended.  Instead, in more dense interaction regions, when radiative cooling is important and $\chi < 1$, shocks are radiative and confined to relatively thin shells. In the case of adiabatic shocks, \citet{Stevens.etal:92} showed that the X-ray luminosity should scale with the inverse of the stellar separation, $L_X\sim 1/D_{sep}$.

While progress with theory and modelling of colliding wind binaries has been strong, the observations have tended to be difficult to understand within the theoretical framework.
Again, the large samples of stars collected in {\it Chandra} surveys have been pivotal in understanding the importance of colliding winds in explaining the X-ray emission of high mass stars.  

\citet{Naze.etal:11} found that high-mass binaries in the Carina OB1 survey that have colliding winds and therefore might be expected to appear X-ray bright were only marginally more X-ray luminous and harder than single stars to an extent that they declared not statistically significant. 
Similarly, \citet{Rauw.etal:15} also find no conspicuously large excess in X-ray luminosity in binaries as compared to single stars in the Cygnus~OB2 Association. However, they did find evidence for  modest X-ray over-luminosity as a function of wind kinetic power for the more massive known binaries. Examples are Cyg~OB2 \#5, \#8a and \#9 shown in Figure~\ref{f:ostars_cygob2}.

\citet{Naze.etal:11} noted that winds tend to collide at comparatively low speed in close binaries whereas the collision in wide systems is adiabatic and therefore not so X-ray bright. This leaves a  fairly small region of parameter space for X-ray-bright wind-wind collisions.  They also note that in cases where the wind momenta of the components are very different, modeling has shown that the stronger wind penetrates directly onto the companion \citet{Pittard.Parkin:10}.

\subsubsection{The Spectacular Case of $\eta$ Carinae} 
\label{s:etacar}

$\eta$ Carinae is one of the most remarkable stars in the Galaxy. It is the nearest example of a luminous blue variable---a supermassive, superluminous, unstable star, and likely the most luminous and massive object within the nearest 3~kpc or so.  Mass loss from $\eta$ Carinae shapes its circumstellar environment. While its current mass loss is in a slow $\sim 500$~km~s$^{-1}$, dense $\dot{M}\sim 10^{-3} M_\odot$~yr$^{-1}$ stellar wind it is also prone to eruptions during which large amounts of mass are expelled by processes that are not yet entirely understood (\citealt{Davidson.Humphreys:97}, \citealt{Corcoran.etal:04}). 

\begin{figure*}[ht!]
\begin{center}
\includegraphics[width=0.9\textwidth]{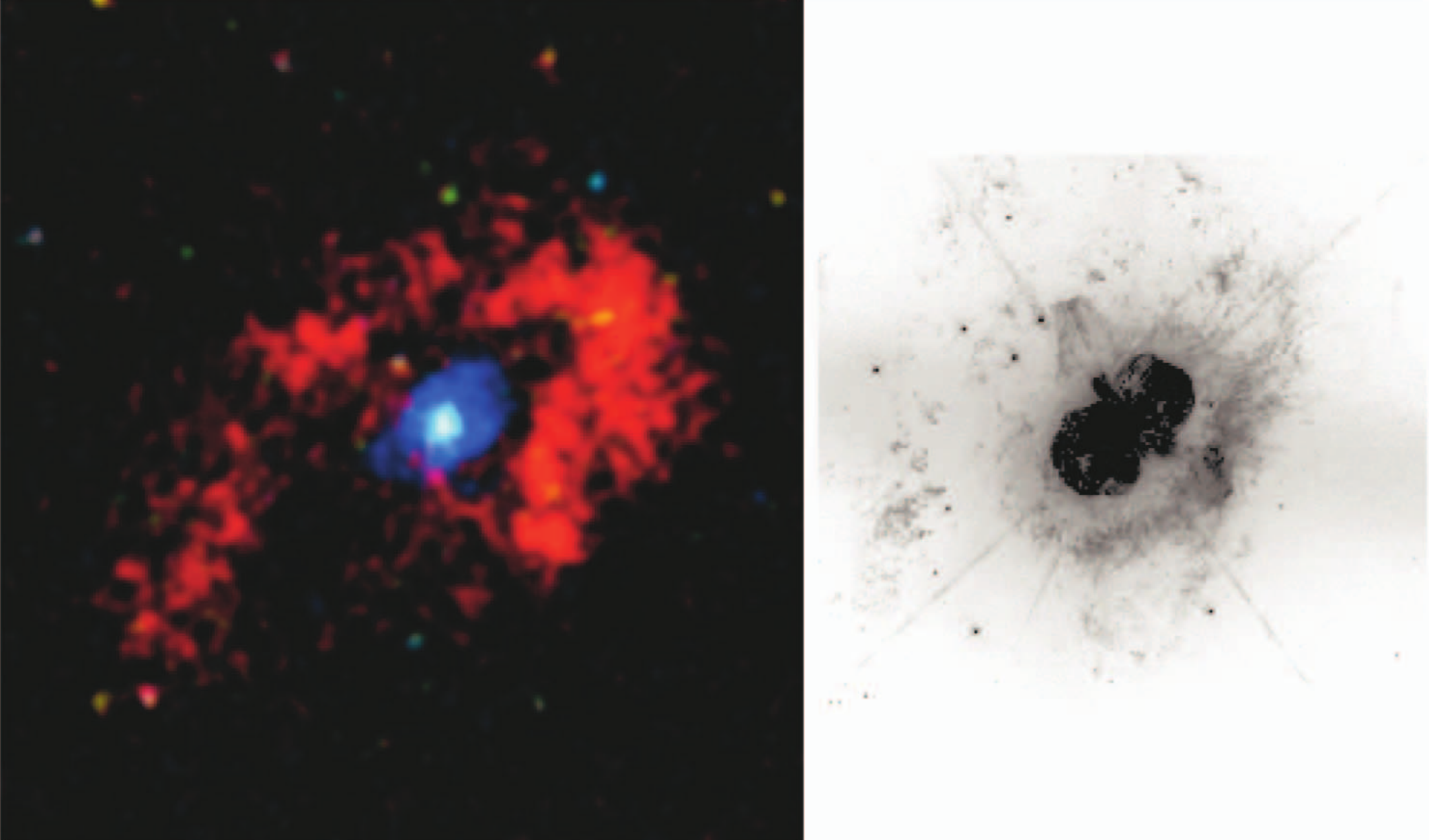}
\end{center}
\caption{Left: A false color {\it Chandra} ACIS X-ray image of the region centered on $\eta$~Car during X-ray minimum.  Colors correspond to different X-ray bands as follows: Red low-energy (0.2--1.5 keV), green medium-energy (1.5--3.0 keV), and blue high-energy (3.0--8.0 keV). Emission from $\eta$~Car itself corresponds to the central white point source.  A  broken elliptical ring of very soft X-ray (shown in red) emission from the shock arising from the ``Great Eruption'' lies in a partial ring surrounding the star. The bluish patch around the star inside this ring is reflected X-ray emission from the Homunculus nebula. Right: An HST WFPC2  [N II] 6583~\AA\ image of $\eta$~Car with the same plate scale and orientation as the {\it Chandra} image.
North is to the top, and east is to the left. From \citet{Corcoran.etal:04}.}
\label{f:etacar}
\end{figure*}

\begin{figure*}[ht!]
\begin{center}
\includegraphics[width=0.9\textwidth]{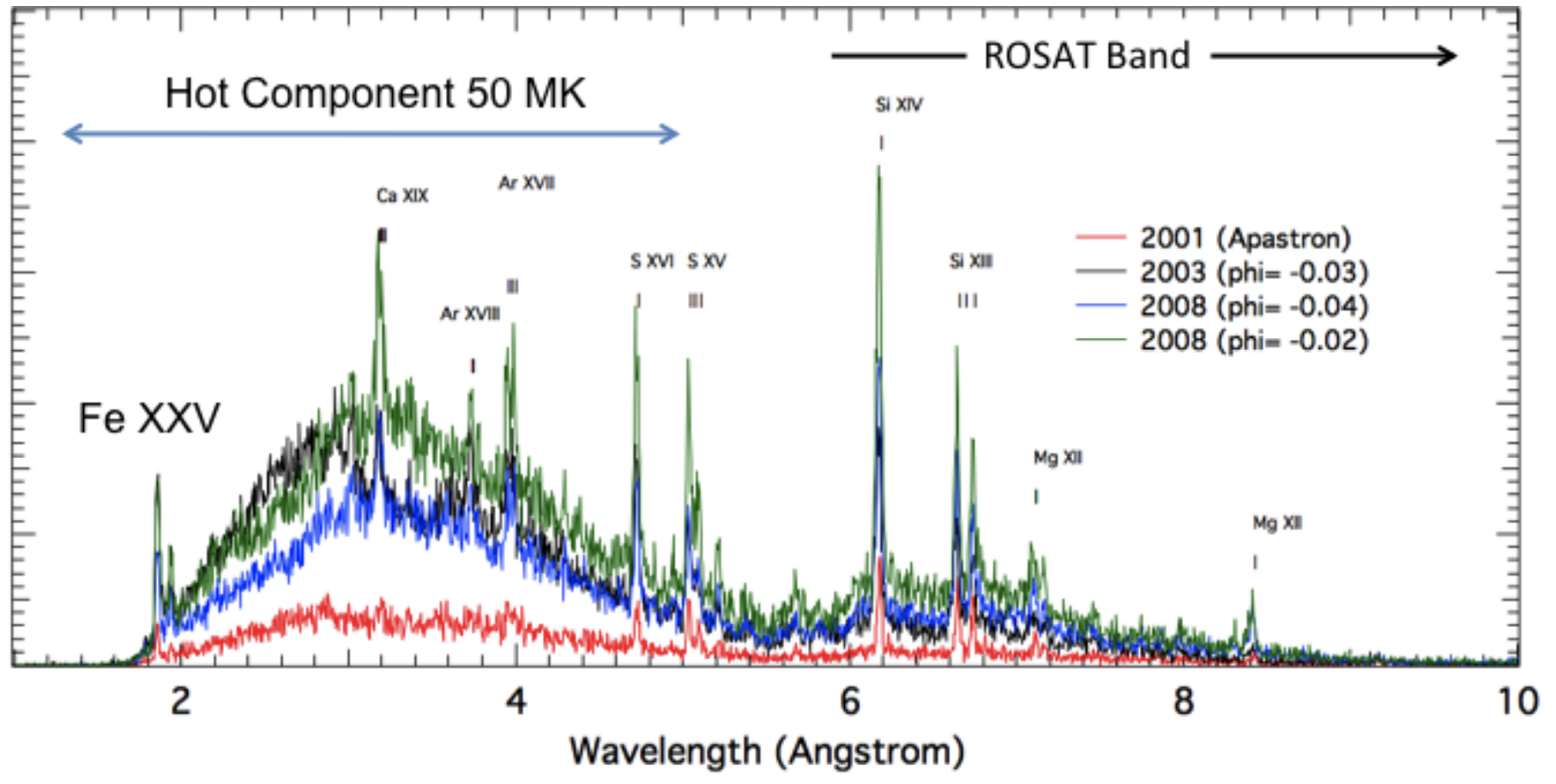}
\end{center}
\caption{{\it Chandra} HETG spectra of $\eta$~Car itself at different epochs showing strong X-ray spectral variations that appear to be reasonably well-explained by an adiabatic wind-wind shock with X-ray luminosity varying inversely with stellar separation, $L_X\sim 1/D_{sep}$.  The spectrum is largely thermal in nature,  with a temperature of about 4.5~keV that corresponds to a pre-shock stellar wind velocity of about 3000 km~s$^{-1}$.  From \citet{Corcoran.etal:15b}.}
\label{f:etacarspec}
\end{figure*}

$\eta$~Car itself is a colliding wind binary with a period of 2022~d, exhibiting strong variations in emission over a wide wavelength range including X-rays. These are driven by the collision of the dense wind of $\eta$~Car~A with the fast, less dense wind of $\eta$~Car~B  that otherwise remains hidden.  It undergoes a deep X-ray minimum every 5.53~yr resulting from its highly eccentric orbital motion and effects on the colliding wind X-ray source and its line-of-sight modulation.

 $\eta$~Car is also a source of non-thermal X-ray and $\gamma$-ray emission thought to result from Fermi acceleration at the shock interface of the wind-wind collision and subsequent inverse Compton upscattering of UV photons,  demonstrating that colliding wind binaries are a  source of energetic particles \citep{Hamaguchi.etal:18}.

A  {\it Chandra} ACIS image of the $\eta$~Carinae region centered on the Homunculus---the hollow, expanding bipolar nebula surrounding $\eta$ Car and corresponding to the shock front originating from ``Great Eruption" in the 1840s---was obtained in 2003 at a time when it was at X-ray minimum and is shown in Figure~\ref{f:etacar}.  This is the first time the X-ray nebula was detected, since it is normally swamped by the PSF wings of $\eta$ Car itself.  \citet{Corcoran.etal:04} found the emission to be characterized by an extremely high temperature in excess of 100~MK and consistent with scattering by the circumstellar environment of the time-delayed X-ray flux associated with the star. 
A strong Fe K fluorescent line at 6.4 keV was also detected. This line had a width of 4700~km~s$^{-1}$ which is much larger than the expansion velocity of the Homunculus itself.  \citet{Corcoran.etal:04} interpreted the line widths as likely resulting from reprocessing by fast flows in the lobes of the  Homunculus, perhaps with contributions from the companion stellar wind.

The X-ray signal from the central stars of $\eta$~Carinae undergoes dramatic variations on both long and short timescales---see the {\it Chandra} HETG spectra illustrated in Figure~\ref{f:etacarspec}.  On timescales of
years, the X-ray emission follows approximately the inverse of the stellar separation, $L_X\sim 1/D_{sep}$, as might be expected for adiabatic colliding wind conditions in a highly eccentric binary.
This long-term variability is interrupted by a deep, broad atmospheric eclipse near periastron when 
the dense inner wind of  $\eta$~Car A is in front of the bow shock region of the collision. Superimposed on the long-term trend are shorter ``flares", lasting from a couple to 100 days. The origin of these flares remains uncertain. \citet{Moffat.Corcoran:09} considered several mechanisms, including large-scale co-rotating interacting regions in the $\eta$~Car A wind sweeping across the wind collision zone and instabilities intrinsic to the collision zone, but 
argued that the most likely explanation lies in the largest of multi-scale stochastic wind
clumps from $\eta$~Car A entering hard X-ray-emitting
wind-wind collision zone.

\subsection{The Role of Magnetism}
\label{s:magconf}

A fraction amounting to approximately10\%\ of massive stars are observed to have X-ray emission characteristics that are difficult to explain in terms of pure shocked wind models.  These include variability, moderately hard X-ray emission, and in some cases higher plasma densities than expected \citep[e.g.][]{Naze.etal:14}.  It turned out that stars showing such characteristics also tended to have fairly strong magnetic fields thought to either be fossil remnants from star formation or else generated in convective cores.  Such magnetic fields channel the winds of massive stars towards the magnetic equator where they collide in a process now known as a ``magnetically confined wind".  This mechanism was originally proposed by \citet{Babel.Montmerle:97} to explain X-ray emission from the A0p star IQ Aur.  Since the magnetically-confined wind collides with itself at higher Mach number than the shocks produced by the line de-shadowing instability described in Section~\ref{s:lxlbol} the shocked plasma is expected to be hotter, as observed, and subject to variability due to instability in the flow \citep[e.g.][]{ud-Doula.etal:14}.

\begin{figure}
\begin{center}
\includegraphics[width=0.45\textwidth]{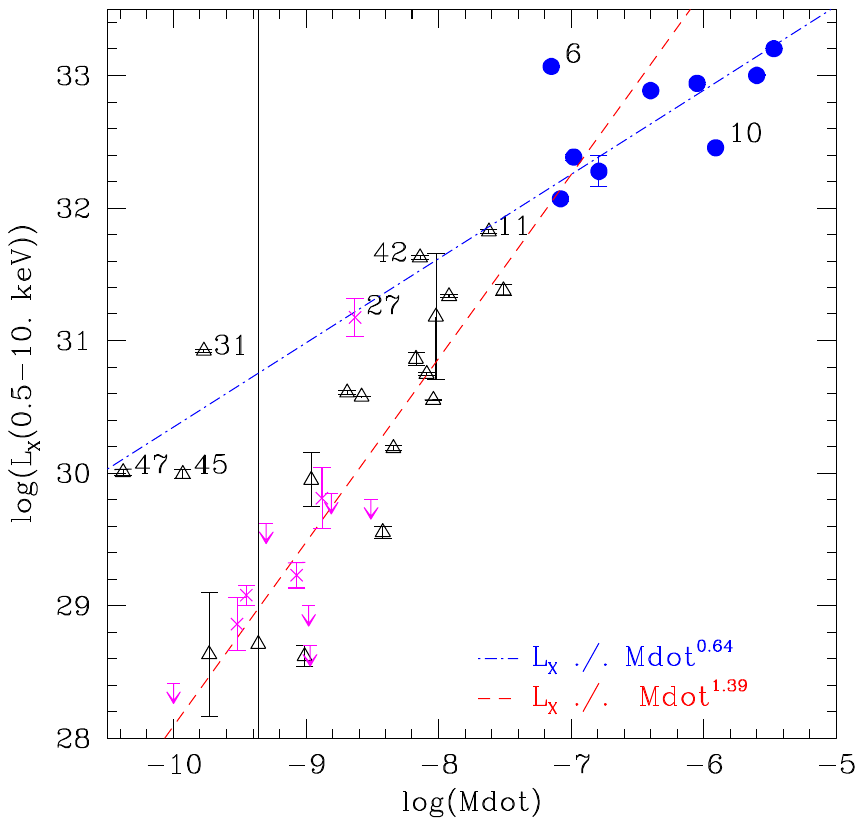}
\includegraphics[width=0.53\textwidth]{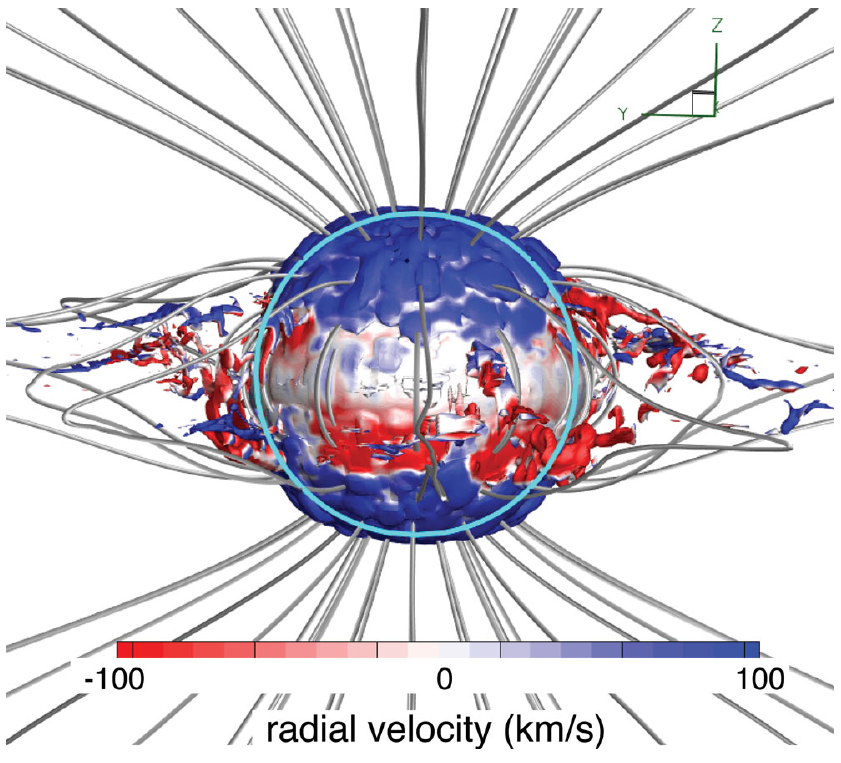}
\end{center}
\caption{Left: Absorption-corrected X-ray luminosities as a function of mass loss rate, $\dot{M}$, for magnetic OB stars analysed by \citet{Naze.etal:14}. Filled blue dots denote O stars, black empty triangles B stars, while  faint detections and X-ray luminosity upper limits are signified by magenta crosses and downward arrows, respectively. The  best-fit power-law relation are shown. Right: Illustration of a 3D MHD magnetically-confined wind shock model for $\theta^1$~Ori~C  by \cite{ud-Doula.etal:13} showing an iso-density surface at $\log \rho=12.5$~g~cm$^{-3}$, coloured with radial velocity.  Some of the gas  within the closed loop region falls back on to the stellar surface, while material in the open fields streams out radially. 
}
\label{f:magconf}
\end{figure}

Based on a sample of magnetic stars culled from {\it XMM-Newton} and {\it Chandra} archives, \citet{Naze.etal:14} have shown that the X-ray luminosity is strongly correlated with the stellar wind mass-loss rate and luminosity with relations illustrated in Figure~\ref{f:magconf}.  They determined a power-law dependency that is slightly steeper than linear for the less luminous B stars with lower mass loss rates, $\dot{M}$, and that flattens for the more luminous, higher $\dot{M}$ O stars.  The observed X-ray luminosities, and their trend with mass-loss rates, were found to be reasonably well reproduced by 2D and 3D magnetohydrodynamic  models \citep[][see Figure~\ref{f:magconf}]{ud-Doula.etal:13,ud-Doula.etal:14}.  \citet{ud-Doula.etal:14} find in general that for stars with lower mass loss rates, X-ray emission is reduced and softened by a ``shock retreat" that results from a large post-shock cooling length: within the fixed length of a closed magnetic loop the shock is forced back to lower pre-shock wind speeds.

This good agreement between 3D models and observations was confirmed in detail by models fitted directly to {\it Chandra} HETG spectra of the magnetically-confined wind Of?p star HD 191612 by \citet{Naze.etal:16}.  No significant line shifts were seen, with lines being relatively narrow and formed at a distance of about $2 R_\star$.

\citet{Naze.etal:14} still found puzzling exceptions to this picture of ``well-behaved" magnetically-confined wind stars that still bely explanation, such as the persistence of some X-ray overluminous stars not obviously attributable to colliding winds. Moreover, they found no relation between other X-ray plasma temperature or absorption and stellar or magnetic parameters as might have been expected, indicating that the the plasma properties are controlled by different processes to the X-ray luminosity itself.

\subsection{The Weak Winds Problem}

One of the outstanding problems in hot star winds is the so-called ``weak-wind" problem, in which mass-loss rates modeled from UV line diagnostics for stars that show clear classical P Cygni line profiles can be discrepant by more than an order of magnitude from expected values based on O star statistical trends and based on theoretical models.  The problem then has two aspects: lower 
mass-loss rates than other stars of similar spectral type; on the other hand, other weak-wind stars have lower mass-loss rates than what atmospheric models predict. High X-ray luminosities have sometimes been invoked to explain the weakness of the winds, as they can modify the wind ionization and hence the efficiency of wind acceleration \citep[e.g.][]{Martins.etal:05}.

\citet{Huenemoerder.etal:12} combined {\it Chandra} LETG+ACIS-S spectroscopy and {\it Suzaku} observations of the notorious weak wind star $\mu$~Col and found an X-ray emission measure corresponding to an outflow an order of magnitude greater than suggested by UV lines, and comparable with a standard wind luminosity relationship for O stars. The spectrum of $\mu$~Col is soft and lines are broadened, with radiative excitation of the He-like triplets indicating that the bulk of the X-ray emission is formed within five stellar radii, all in agreement with expectations of a typical shocked wind. 

The ``weak-wind'' problem" identified from cool wind UV and optical spectra is then largely resolved, at least in $\mu$~Col, by accounting for the hot wind only seen in X-rays.

\subsection{The Mysterious X-rays from Cepheid Variables}
\label{s:cepheids}

Classical Cepheid variables are pulsating yellow supergiant stars with masses in the range 4--$20~M_\odot$ and pulsation periods typically ranging from 2 to 60 days. They undergo periodic changes in size, temperature, and brightness as a result of the ``$\kappa$-mechanism",  so called because the instability is driven by the sensitivity of the stellar opacity to changes in the stellar structure.  In Cepheids, the instability is driven by He ionization. The cooling of gas in an expanding atmosphere leads to lower He ionization and a lower opacity that causes the outer envelope to contract under gravity.  Under collapse, the gas is heated and He eventually becomes more ionized and opaque, leading to another expansion and contraction cycle.  Cepheids lay the foundations of the cosmic distance scale through the Leavitt Law that describes the correlation between the period of the Cepheid and its luminosity.

\begin{figure*}[ht!]
\begin{center}
\includegraphics[width=0.5\textwidth]{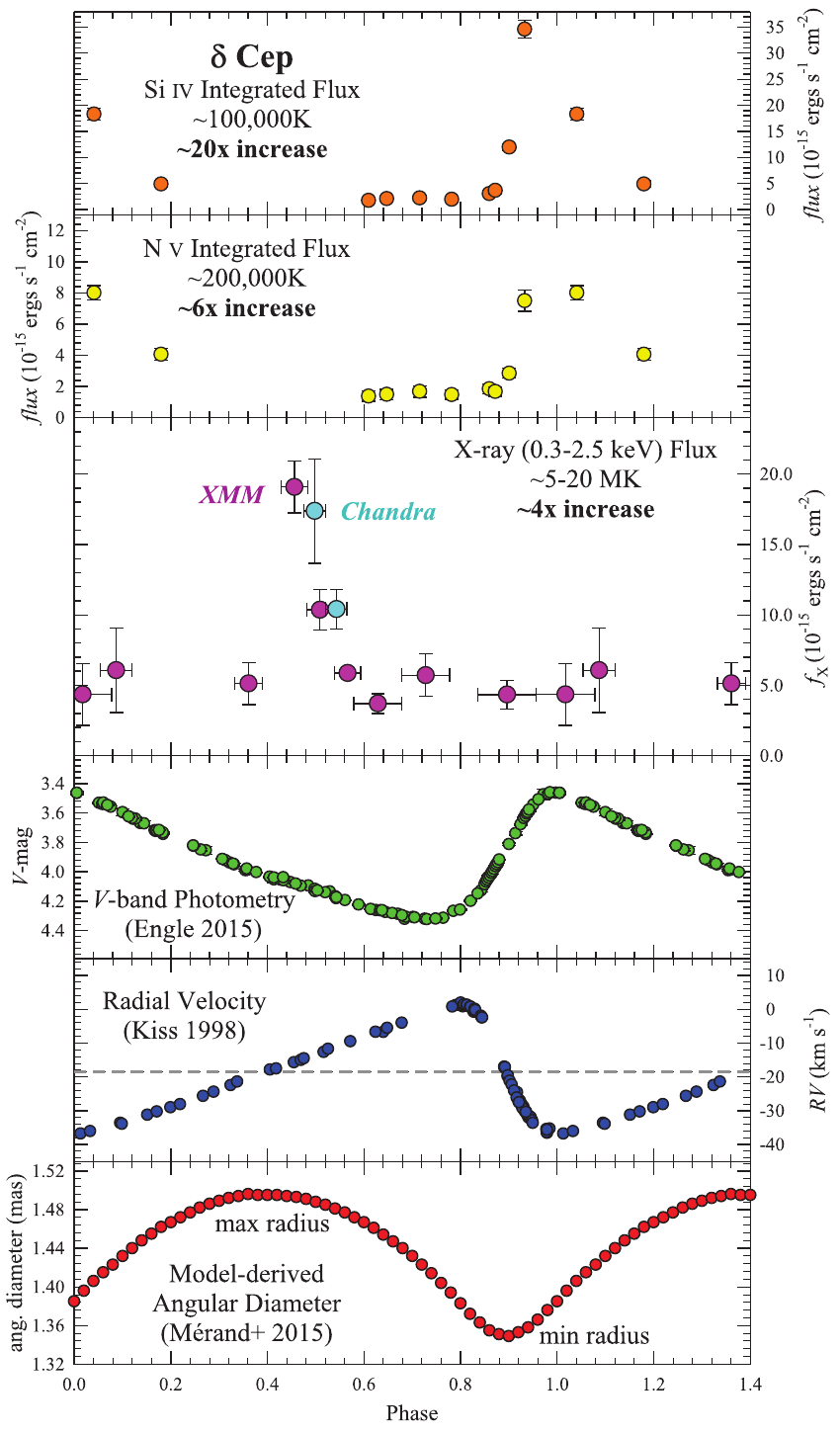}
\end{center}
\caption{UV (upper two panels), X-ray (third panel) and photospheric (lower three panels) variations of $\delta$~Cep as a function of pulsation phase. Pink and cyan X-ray data points are from {\it XMM-Newton}, and 2015 {\it Chandra} observations, respectively. From \citet{Engle.etal:17}.
}
\label{f:delcep}
\end{figure*}

X-rays were surprising detected from the prototypical Classical Cepheid $\delta$~Cep in {\it XMM-Newton} observations \citep{Engle.etal:14}.  Subsequent {\it Chandra} observations confirmed the hint from the discovery observations that the X-ray flux is correlated with the pulsation phase \citep{Engle.etal:17}. Observations obtained in far ultraviolet lines with the {\it Hubble Space Telescope} also demonstrated strong phase-dependent emission.  The data for $\delta$~Cep are illustrated in Figure~\ref{f:delcep}. 

For the majority of the five day pulsation cycle of $\delta$~Cep, the X-ray luminosity is comparatively low at $\log L_X = 28.5$--29~erg~s$^{-1}$, similar to that of non-pulsating yellow supergiants and  corresponding to an X-ray surface flux an order of magnitude or more below that of the typical Sun. However, near
maximum radius close to phase 0.5 the X-ray flux was observed to rise 
rapidly by a factor of 4 or so, and then to fall equally rapidly only 0.10 later in phase. Surprisingly, the X-ray peak occurs at the phase of maximum radius, which is quite different to the the beginning of the rise phase at which the peak in the far UV fluxes is observed.

There are at least two likely mechanisms explaining the X-ray emission.
Firstly, a shock wave propagating through the atmosphere might be expected to develop from the pulsation wave gradually propagating into a medium of lower and lower density.  X-ray emission from Cepheids was indeed predicted from such a mechanism by \citet{Sasselov.Lester:94}.  Secondly, if $\delta$~Cep possessed a sufficiently strong magnetic field, the periodic stressing of the field due to pulsations could drive a periodic type of flaring. Magnetic fields of the order of 1~G have indeed been reported in the Cepheid $\eta$~Aql, as well as in the supergiant $\alpha$~Per \citep{Grunhut.etal:10}.

There are now X-ray detection of several other Cepheids \citep{Engle.etal:17}, although only one of them, $\beta$~Dor, shows strong evidence for similar phase-dependent X-ray emission so far.  Progress is hampered because these stars are just generally too distant and too faint in X-rays for detailed study with {\it Chandra} and {\it XMM-Newton}.   Thorough examination of the potential of the Cepheids to be a new class of X-ray pulsating sources will likely have to await future missions with larger collecting areas.

\section{Intermediate Mass Stars}
\label{s:intermediate}

We noted previously in Section~\ref{s:coronae} 
that early observations by the {\it Einstein Observatory} showed that
X-rays from stars are common throughout the Hertzsprung-Russell
diagram \citep{Vaiana.etal:81}.  Notable exceptions were late-type
giants and main-sequence stars from late B-type through middle
A-type.  While the bright early A-type stars Vega and Sirius seemed to have been detected by the {\it Einstein} High Resolution Imager these were subsequently attributed to UV
leaks \cite[][a similar situation noted by \citealt{Zombeck.etal:97} also arose with the later {\it ROSAT} HRI]{Golub.etal:84}.  

The preceding text has detailed two ways in which single, non-degenerate stars produce X-rays:  1) through shock-heated regions in radiatively-driven hot stellar winds and 2) in hot coronae
sustained by dynamo-driven magnetic activity that arises in stars with
convective envelopes.  The lack of X-rays from early A-type stars
is then naturally explained by their location in a transition regime between stars that are able to
produce X-rays by these mechanisms.  Their temperatures are too low and radiation fields too weak to drive strong supersonic winds, but high enough that hydrogen---the dominant source of opacity giving rise to outer convection zones in stars---is largely ionized rendering their envelopes radiative.  The effective temperatures defining the transition
regime fall in the range $\sim 8500$-12000~K, with $B-V \sim -0.1-0.2$.

One problem arises with the above reasoning: at least some intermediate mass main-sequence stars are found to be X-ray emitters. Pre-main
sequence Herbig Ae/Be (HAeBe) stars---the intermediate mass nearly fully radiative counterparts to
classical T~Tauri stars---have routinely been found coincident with X-ray sources with 
luminosities of a few $10^{31}$~erg~s$^{-1}$ down to $\sim
10^{29}$~erg~s$^{-1}$
\citep{Damiani.etal:94,Zinnecker.Preibisch:94,Stelzer.etal:06}.  While unresolved lower mass companions might still be responsible for  some of the detections, \citet{Stelzer.etal:06} found an
overall detection fraction of 76\%\ for a sample of 17 HAeBes, and  only half of these have known unresolved companions.

A plausible scenario raised by the Herbig Ae/Be stars is that an early active magnetic dynamo gives rise to X-ray emission, feeding from the natal shear and differential rotation within the star.  This shear is gradually dissipated by the torque exerted by Lorentz forces as a result of magnetic fields generated by the shear operating in concert with the Taylor magnetic instability \citep{Tayler:73}, which is analogous to the pinch instability in plasma physics.   Such a dynamo scenario was outlined by \citet{Tout.Pringle:95} and further considered by \citet{Spruit:02} and \citet{Braithwaite:06}.  The key for testing such a scenario is to follow early A-type stars at very young ages to map out their X-ray emission.

\citet{Drake.etal:14b} observed HR 4796A, a nearby ($\sim 70$~pc) 8 Myr old main sequence A0 star using the {\it Chandra} HRC-I, which is more sensitive than the ACIS detector to very soft X-ray emission that might arise from a cooler corona than exhibited by more active stars.  HR 4796A possesses a remnant dusty disk that has been scrutinized intensively from both the ground and in space using {\it Hubble} that has essentially ruled out the possibility of a stellar mass companion. A 21~ks exposure failed to detect a single photon from HR 4796A!  The implied X-ray luminosity upper limit was $L_X\leq 1.3\times 10^{27}$~erg~s$^{-1}$. 

One other key star of spectral type A0 is Vega, even closer at $\sim 7.8$~pc star and also with a dusty remnant disk and thought to be single.  Vega is older than HR4796A, with a rather uncertain age in the range 100-400~Myr. {\it Chandra} uses Vega for routine monitoring of the detectors for possible UV leaks, and \citet{Pease.etal:06} used accumulated observations to search for an X-ray signal. None was found that could not be attributed to background or a very small residual UV leak and an associated upper limit to the flux of $L_X < 3\times 10^{25}$~erg~s$^{-1}$ was derived, corresponding to a bolometric fraction limit of $L_X/L_{bol}<9\times 10^{-11}$ \citep[see also][]{Ayres:08}, or about a million times lower than the Sun near solar minimum.  The surface X-ray flux of Vega is in fact at least 2 orders of magnitude below the a ``universal minimum" surface flux for F-M main-sequence stars found by \citet{Schmitt:97} based on {\it ROSAT} data, which corresponds to the surface flux of solar coronal holes.

\begin{figure}
\begin{center}
\includegraphics[width=0.6\textwidth]{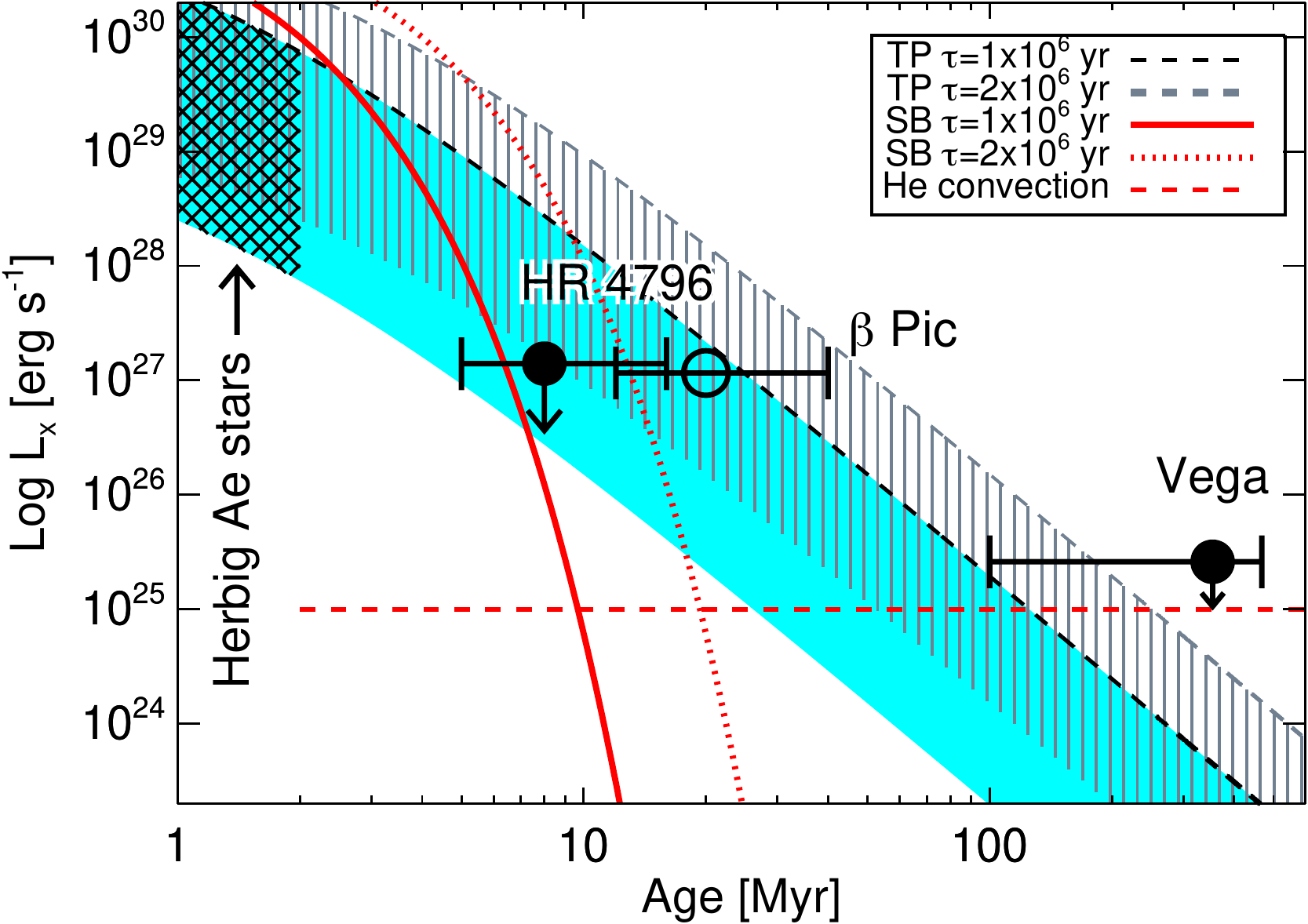}
\end{center}
\caption{X-ray luminosity vs.\ time for shear dynamo models compared with observations of single A-type stars from \citet{Drake.etal:14c}.   The hashed region to the left represents the range of
 X-ray luminosities of Herbig~Ae stars.  The dashed curves correspond to a natal X-ray luminosity of $L_{X_0}=2\times 10^{31}$~erg~s$^{-1}$ and two different values for the rotation shear decay timescale for the  \citet[][TP]{Tout.Pringle:95} model.  The solid and dotted curves correspond to an exponential X-ray luminosity decay corresponding to the rotational shear dissipation found by \citet[][SB; see also \citealt{Braithwaite:06}]{Spruit:02} assuming the \citet{Tout.Pringle:95} prescription for magnetic flux rising due to buoyancy.  
  The  horizontal dashed curve corresponds to the  \citet{Drake.etal:14b} estimate of a base level X-ray luminosity due to a weak sub-surface convection zone.  The $\beta$~Pic detection of \citet{Gunther.etal:12} is shown with a hollow symbol to denote the likely origin of its X-ray emission from a thin surface convection zone.
 \label{f:intermediatem_lx}
}
\end{figure}

The implications of the non-detections of HR 4796A and Vega are illustrated in Figure~\ref{f:intermediatem_lx}, and are compared with typical fluxes observed for Herbig Ae/Be stars and with predictions of the \citet{Tout.Pringle:95} and the \citet{Spruit:02} and  \citet{Braithwaite:06} shear dynamo formalisms.  The former predicts a decay of an initial X-ray luminosity, $L_{X_0}$, with time according to 
\begin{equation}
L_X(t)=\frac{L_{X_0}}{(1+t/\tau)^{3}},
\label{e:lxtp}
\end{equation}
where $\tau$ is the decay timescale.  Based on reasonable guesses of relevant parameters, \cite{Tout.Pringle:95} estimated $\tau\sim 10^6$~yr,  which also corresponds to the timescale over which HAeBe stars appear X-ray bright \citep[e.g.][]{Hamaguchi.etal:05}.  The analogous formula derived by \citet{Drake.etal:14b} based on \citet{Spruit:02} and \citet{Braithwaite:06} gives
\begin{equation}
L_X(t)=L_{X_0}\exp \left( - \frac{3t}{2 \tau}\right)
\label{e:lxsp}
\end{equation}
although with a considerably shorter timescale possibly as short as  few hundred years.  The decay timescale for both formalisms is, however, extremely uncertain, and by orders of magnitude.  Figure~\ref{f:intermediatem_lx} shows the X-ray decay for timescales of 1 and $2\times 10^6$~yr.  Another difficulty is in the rise of field from deeper layers through diffusive buoyancy which acts very slowly.

\citet{Drake.etal:14b} also discuss two other mechanisms by which early A-type and late B-type stars might produce X-rays: decay of primordial magnetic field, and a thin, weak subsurface convection zone driven by opacity bumps from the ionization of iron and helium.  They dismissed the former as insignificant, and for the latter derived 
\begin{equation}
\label{e:lx_conv}
L_{\rm X} \sim 10^{25} \left(\frac{P}{\rm 12 hr}\right)^{-2} {\rm erg\, s}^{-1}
\end{equation}
for a $2M_\odot$ star with $L=25L_\odot$, where the approximate power of $-2$ on the period originates from the same $Ro$ relation in Equation~\ref{e:dynnum} and $\beta=2$.  

The X-ray luminosity for HR4796A marginally conflicts with the \citet{Tout.Pringle:95}-based model, whereas the X-rays based on \citet{Spruit:02} and \citet{Braithwaite:06} decay too quickly to be seen.   Subsurface convection appears only able to produce X-ray luminosities of order $10^{25}$ erg s$^{-1}$, which lies only moderately below the current detection limit for Vega.  Based on these difficulties, \cite{Drake.etal:14b} instead favoured an accretion or jet-based mechanism to explain the X-ray activity of Herbig Ae/Be stars, an idea suggested by \citet{Hamaguchi.etal:05}.  A lack of photoexcitation of the intercombination lines of H-like O in an {\it XMM-Newton} observation of the Herbig Ae star HD163296 \citet{Gunther.Schmitt:09} suggested the cooler X-ray emission originates away from the star at the base of a jet, whereas hotter emission might be coronal in origin.  

Regardless of the exact mechanisms at play Following the Herbig Ae/Be pre-main sequence phase, early A stars then generally decline in X-ray luminosity at least 100,000 fold in only a few million years. 

The spectral type limit earlier than which A-stars begin to be plausibly X-ray dark appears to be about A5 based on a detection of $\beta$~Pictoris in a 20~ks {\it Chandra} HRC-I observation reported by  \citet{Gunther.etal:12}. For later A-type stars, weak X-ray emission can be supported by a thin convection zone.

\section{White Dwarfs and White Dwarf Binary Systems}
\label{s:wds}

White dwarfs are the final evolutionary states of low-mass and intermediate-mass stars---stars of less than about $8 M_\odot$ that are not massive enough to become supernovae and neutron stars and make up over 97\% of stars in the Milky Way \citep{Fontaine.etal:01}. After main-sequence evolution, such a star expands to a red giant in which core He burning through the triple-alpha process builds up a carbon and oxygen core.  Helium shell burning proceeds on the asymptotic giant branch (AGB) and the star loses mass at a much higher rate, eventually shedding its envelope. At the end of the AGB phase, the remnant envelope material forms a planetary nebula inside which the remnant CO core becomes a nascent white dwarf.  

Much more rare higher mass progenitors in the 8--$10~M_\odot$ range are expected to have had higher core temperatures on the red giant branch and to have produced cores composed of oxygen, neon and magnesium \citep[e.g.][]{Iben.etal:97}.  However, the dividing line between production of a white dwarf and a core-collapse supernova is thought to be quite fine and evidence for the remnant ONeMg white dwarfs is fairly scarce.

Stars of less than $0.5 M_\odot$ have insufficient core temperatures for helium burning, and are thought to end their lives as He white dwarfs.  Since their evolutionary time scales are longer than the age of the Universe, their only appearance is expected to be in binary systems in which their envelopes have been stripped away gravitationally. 

A white dwarf is very hot when it forms, with a surface temperature that can easily exceed 100,000~K and therefore having a photosphere whose Wien Tail is visible into the soft X-ray range.  Having no internal source of energy, white dwarfs gradually cool with a timescale of millions to hundreds of Myr as this energy is radiated away.  The temperature of a single white dwarf is then an age diagnostic, rendering white dwarfs of considerable interest for cosmochronology \citep[e.g.][]{Fontaine.etal:01}. 

White dwarfs forming in binary systems can also become X-ray sources through accreting matter captured from their companions, becoming cataclysmic variables and novae.  The average surface gravity of white dwarf stars is  $\log g \sim 8$, compared with 4.4 for the Sun. Unlike the accreting protostars discussed in Section~\ref{s:accretion}, whose X-ray spectra can show traces of accretion energy, the strong gravitational well of accreting white dwarfs can generate bright, hot, hard X-ray spectra.  

{\it Chandra} has observed white dwarfs stars in all their phases of activity, from nascent stars in planetary nebulae (PNe) to single white dwarfs on the cooling track, and binary white dwarfs in cataclysmic variables and nova explosions.

\subsection{White Dwarf Birth in Planetary Nebulae}
\label{s:pne}

PNe are commonly detected as X-ray sources. The X-ray emission is typically in two forms: from the compact point-like sources associated with the central stars; and extended diffuse X-ray
emission from shocks and hot bubbles in the nebula itself resulting from the outflow and excitation by the hot ionizing central source. 

The most comprehensive study of X-rays from PNe to date has been the {\it Chandra Planetary Nebula Survey} (ChanPlaNS; \citealt{Kastner.etal:12}, \citealt{Freeman.etal:14}) that targeted a total of 59 PNe within $\sim 1.5$~kpc.  The diffuse X-ray detection rate of the sample was 27\% and the point source detection rate 36\%. 

\begin{figure}
\begin{center}
\includegraphics[width=1.0\textwidth]{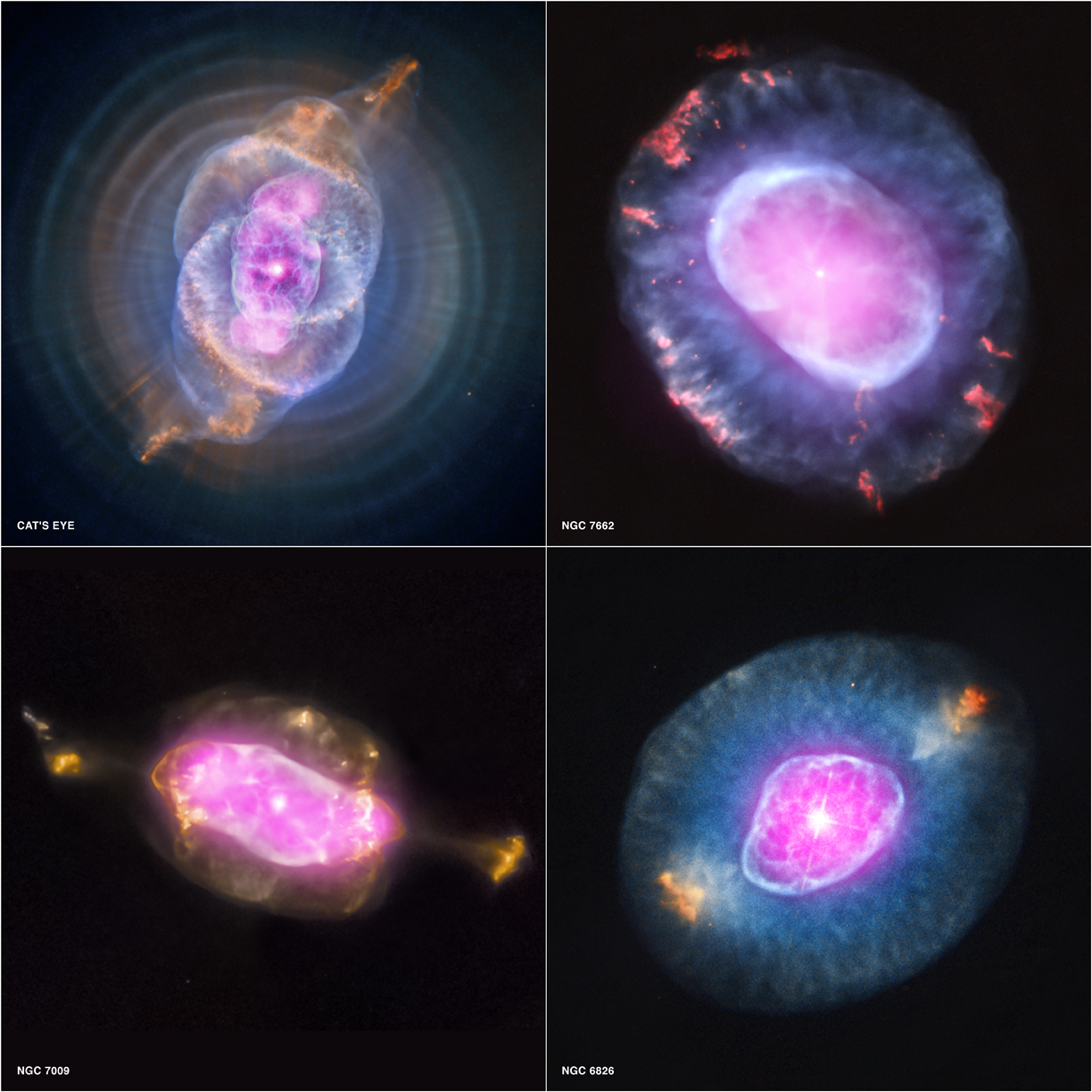}
\end{center}
\caption{Four objects from the first systematic X-ray survey of planetary nebulae in the solar neighborhood exhibiting diffuse X-ray emission.  Shown here are NGC 6543, also known as the Cat's Eye, NGC 7662, NGC 7009 and NGC 6826. In each case, X-ray emission from Chandra is colored purple and optical emission from the Hubble Space Telescope is colored red, green and blue.  X-ray emission is produced by the interaction of the fast, radiatively-driven outflow from the central nascent pre-white dwarf with itself and with the ejected red giant envelope. From http://chandra.harvard.edu/photo/2012/pne/  and \citet{Kastner.etal:12}.
 \label{f:pne}
}
\end{figure}

Diffuse X-ray emission is expected on the basis of the fast (500--1500~km~s$^{-1}$) radiatively-driven winds from the pre-WD that collide with the rejected red giant envelope, sweeping the ejecta into a thin shell. This interaction results in shocks that can heat the wind gas to temperatures exceeding $10^6$~K, leading to the formation of a hot bubble of soft X-ray emitting gas. \citealt{Freeman.etal:14} found that diffuse X-ray emission is associated with young (\lax $5 \times 10^3$~yr), and correspondingly compact (\lax 0.15~pc in radius) PNe with closed morphologies (Figure~\ref{f:pne}).   

All five of the PNe in the sample with central stars of the so-called ``Wolf-Rayet-type'' or [WR]-type---so named because of their H deficiencies, high mass loss rates up to $10^{-6} M_\odot$~yr$^{-1}$ and general superficial resemblance to true massive Wolf-Rayet stars---were detected in diffuse X-rays. In these cases \citealt{Freeman.etal:14} noted that the diffuse X-ray emission resembles the limb-brightened, wind-blown bubbles blown by massive WR stars.

All of the PNe with diffuse emission detections have relatively high central electron densities of $n_e $\gax$ 1000$~cm$^{-3}$, again reflecting their relatively compact size and the vigor of the outflow of central star during its early stages. PNe typically last of the order of 20,000~yr, and the {\it Chandra} diffuse emission detections  indicate that beyond the first 5000 years the systems are either too large and/or the central outflow too weak to maintain the wind interactions necessary to produce detectable X-ray emission.

X-ray detections of the central stars themselves have proved a bit more difficult to interpret.  Twenty of the sample of 59 were detected by {\it Chandra}.  \citet{Montez.etal:15} found the majority of them to be associated with luminous central stars within the relatively young and compact nebulae of the sample. While very soft X-ray emission from the continua of the hot nascent photospheres of these emerging white dwarfs might be expected, the great majority of the detections displayed comparatively hard X-ray emission that \citet{Montez.etal:15} discovered were consistent with optically-thin thermal plasma emission.  Based on the fitted plasma properties, they 
 identified two classes of central star X-ray emission. One has relatively cool plasma emission characterized by temperatures of \lax~$3\times 10^6$~K and with X-ray to bolometric luminosity ratios $L_X/L_{bol} \sim 10^{-7}$, reminiscent of the radiatively-driven wind shocks seen in high mass stars described in Section~\ref{s:highmass}. The other has high-temperature plasma at $10^7$~K or so and with X-ray luminosities apparently uncorrelated with the stellar bolometric luminosities.  \citet{Montez.etal:15} identified this latter category with magnetically active binary companions. The binary companion can be spun up in the common envelope phase and by accreting matter from the envelope of its evolving companion, invigorating its magnetic dynamo (see Section~\ref{s:rotation}).
 
\subsection{Photospheric Emission}
\label{s:wdphot}

One of the surprises from the {\it ROSAT} PSPC X-ray all sky survey was the unexpectedly small number of white dwarf stars detected compared with pre-launch expectations \citet{Fleming.etal:96}. The great majority of the 175 detections (161)  are of DA spectral type---essentially photospheric emission from a pure hydrogen envelope. The remainder comprised three helium-dominated DO types, three DAO types (H+He envelopes), and eight PG1159 stars with He-C-O dominated photospheres.   This sample amounted to only 10\%\ of the white dwarfs thought hot enough to emit in soft X-rays ($T_{eff} > 20,000$~K and be detected by {\it ROSAT}. 

The reason for the dearth of detections lies in the radiative levitation of metals in the atmospheres of hot white dwarfs that would otherwise sink due to the strong gravitational field.  The degree of levitation depends on the white dwarf gravity and on the effective temperature and consequent strength of the photospheric radiation field.  The picture can also be complicated by weak mass loss that alters the chemical equilibrium, as as the star gradually cools the photospheric chemical composition is expected to change. For stars with effective temperatures $T_{eff} > $40,000~K, the atmospheric opacity of metals is very large in the EUV and soft X-ray range, effectively blocking the radiation field and redistributing much of it toward longer wavelengths outside of the X-ray band \citep[e.g.][]{Barstow.etal:97}. 

As noted in the introduction to this section, the Wien Tail of the photospheric emission from hot white dwarfs extends from the UV and far UV into the extreme ultraviolet and soft X-ray range---and the longer wavelengths covered by the {\it Chandra} LETG+HRC-S (1.2--170~\AA). This has enabled {\it Chandra} to study the detailed spectra of a small handful of hot white dwarfs and attempt to unravel the nature of their short wavelength photospheric spectra.  

The well-known DA white dwarfs Sirius~B and HZ~43 are observed to have pure H atmospheres in the LETG range, devoid of metal lines. Their smooth continua have been used quite extensively as calibration targets for the {\it Chandra} LETGS, acting as standard candles providing sources of absolute flux as a function of wavelength through normalization in the UV and application of accurate model atmosphere spectral energy distribution predictions \citep[e.g.][]{Pease.etal:03}. 

\begin{figure}
\begin{center}
\includegraphics[width=0.56\textwidth]{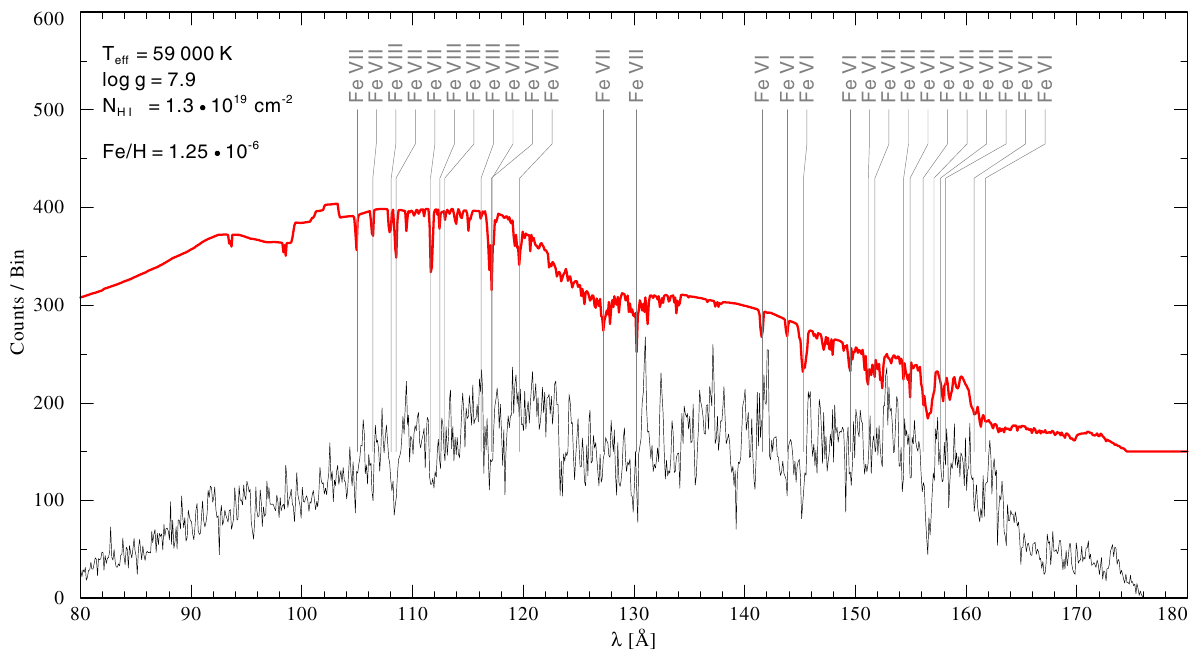}
\includegraphics[width=0.43\textwidth]{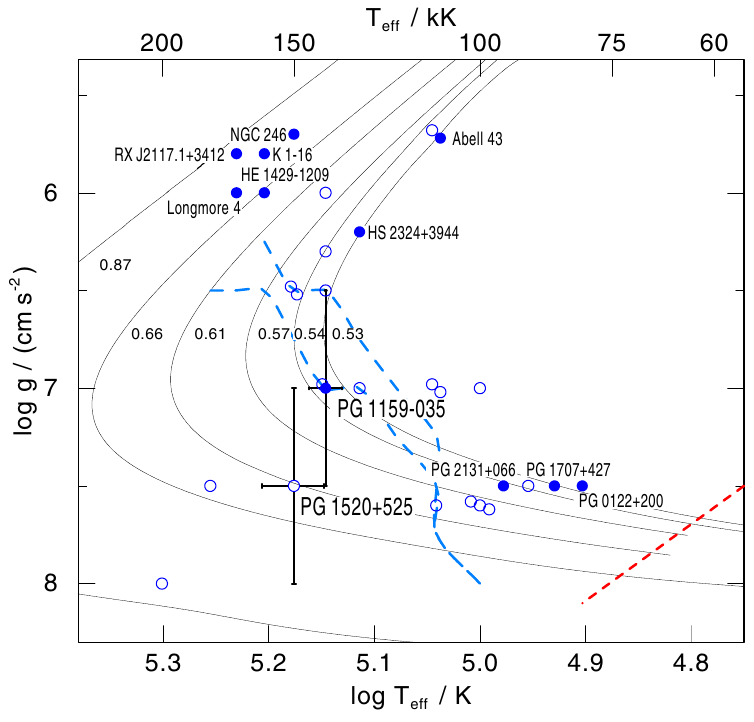}
\end{center}
\caption{Left: The {\it Chandra} LETG+HRC-S spectrum of the DA white dwarf GD246 (black line)
and model spectrum (red; shifted upward by 150 counts for clarity) with some lines due to highly ionized Fe identified.  
Right: The $\log Teff$-$\log g$ diagram illustrating the GW~Vir instability strip constrained by the measured effective temperatures of the 
pulsating star PG 1159?035 and the non-pulsator PG~1520+525. Other 
pulsating (filled circles) and non-pulsating (empty circles) stars are illustrated.  Evolutionary tracks are labeled according to stellar mass. The red edge of the instability region (short dashed line) and two blue edges (long dashed lines) from the models of \citet{Quirion.etal:07} are shown, the latter corresponding to metallicities of $z = 0$ (upper) and 0.007 (lower), respectively.
Both figures from \citet{Adamczak.etal:12}.
\label{f:wd} }
\end{figure}

{\it Chandra} LETGS observations of the DA white dwarf GD~246, illustrated in Figure~\ref{f:wd},  enabled the first unambiguous identification of lines from highly ionized iron, revealing a slew of features from Fe~VI, Fe~VII and Fe~VIII in the 100-170~\AA\ range \citep{Vennes.Dupuis:02,Adamczak.etal:12}.  It remains the only white dwarf observed that shows identifiable individual iron lines in the soft X-ray range. A detailed model atmosphere analysis by \citet{Adamczak.etal:12} failed to provide a good match to the observed spectrum with either homogeneous chemical composition or chemically stratified models.  The latter models compute the element abundance pattern at each depth point in the atmosphere  assuming equilibrium between gravitational and radiative forces.  As a result, the atmospheres are no longer chemically homogeneous but vertically stratified in composition.   \citet{Adamczak.etal:12} found that the only way to match the GD~246  spectrum with a stratified model was to artificially remove Ni from the atmosphere.

Similar problems were encountered with the spectrum of the DA star LB~1919.  This object has a significantly lower metallicity than other DA white dwarfs with otherwise similar atmospheric parameters, and indeed no metal lines were detected in the LETG spectra.  The problem here is that chemically stratified models including radiative levitation predict that significant amounts of light and heavy metals should be accumulated in the atmosphere and readily detectable metal lines should be present.  HZ~43 illustrates similar problems, although the situation of LB~1919 is arguably more severe owing to its hotter effective temperature (56,000~K vs 51,000~K). Like HZ~43,  \citet{Adamczak.etal:12} found the LETG spectrum of LB~1919 to be well-matched with a pure H atmospheric model.

The failure of chemically-stratified models to match the soft X-ray spectra of stars like GD~246 and LB~1919 points to missing physics somewhere \citep[see also][]{Barstow.etal:03b}. Mass loss, or even accretion of circumstellar material, are not included, while model atoms employed in the models are likely incomplete to some significant extent, though as \citet{Adamczak.etal:12} note, it is difficult to understand how any of these can explain the model over-prediction of metals.

Instead, \citet{Adamczak.etal:12} were able to obtain a good match between observed and model spectra for the pre-white dwarf PG~1159-type star PG~1520+525. The PG~1159 stars are hot, post-AGB stars thought to be H-deficient because of having their H-envelope burned during a late He-shell flash \citep[e.g.,][]{Werner.Herwig:06}.  Their surface abundances are not expected to be affected by gravitational settling or radiative leviation because a weak radiatively-driven wind drives them to a homogenous equilibrium.  

By measuring the effective temperature of PG 1520+525, \citet{Adamczak.etal:12} were able to constrain the blue edge of the GW Vir instability region in the HR diagram illustrated in Figure~\ref{f:wd}.  The GW~Vir instability results in pulsations with periods of 300-5000~s and is thought to be caused by the $\kappa$ mechanism, similar to the Cepheids discussed in Section~\ref{s:cepheids} although driven by C and O instead of He \citep{Quirion.etal:07}. PG 1520+525 is not pulsating, whereas the very similar but slightly cooler PG 1159$-$035 is. The position of the edge constrained is this way is consistent with predictions of the pulsation models of \citet{Quirion.etal:07}, providing strong verification for the predictive power of these models and their use in asteroseismological analyses.

{\it Chandra} LETGS spectroscopy of another PG~1159 star, H~1504+65 revealed potential evidence for a rare ONeMg core.  H~1504+65 is extremely hot, with an effective temperature of approximately 200,000 K.  The LETG spectra analysed by \citet{Werner.etal:04} confirmed that it is not only H-deficient, but also He-deficient, being primarily composed of carbon and oxygen but also exhibiting lines from highly ionized Ne and Mg.  \citet{Werner.etal:04} note that the spectroscopic evidence supports H~1504+65 being a naked CO stellar core, but could also be produced by a CO envelope on top of an ONeMg white dwarf. 

\subsection{Cataclysmic Variables and Nova Explosions}
\label{s:cvs}

With of the order of a stellar mass in an object of a similar size to the Earth, the gravitational fields of white dwarfs lead to a considerable zoo of systems and behavior when combined with other stars in binary systems.  Close binaries involving white dwarfs are typically formed due to frictional drag on the secondary companion when the white dwarf progenitor AGB star expands to engulf it. In the most common case of a late-type dwarf star companion, the same angular momentum loss mechanism noted in Section~\ref{s:rotation} that leads to gradual stellar spin-down operates.  When the stars are close enough that tidal forces provide spin-orbit coupling, the net loss of angular momentum leads to an inward spiral of the two stars, and eventually to Roche Lobe overflow of the companion (see, e.g., \citealt{Knigge.etal:11} for a detailed review, and \citealt{Garraffo.etal:18} for recent theoretical refinements to CV angular momentum evolution).  This typically occurs at separations of a few secondary stellar radii and orbital periods of a few hours. Strong surface gravity then leads to the liberation of considerable amounts of energy from accreting material falling onto the white dwarf.   The term ``cataclysmic variable" (CV) refers to the rapid and strong variability exhibited by the class as a result of this accretion activity.

A confusing array of sub-classes of CV has arisen that divides them based on their different types of behavior.  Novae undergo major outbursts of many magnitudes, appearing as a ``new star" in the case of classical novae, and being observed to repeat such outbursts in the case of recurrent novae.  Dwarf novae of which there are several sub-types, undergo frequent outbursts of much small amplitude than novae and with typical cycle times of tens of days. Nova-likes have spectra resembling novae spectroscopically but have not been observed to erupt as classical novae. The remaining important subdivision is the magnetic CVs making up 10-20\%\ of the population whose fields are strong enough to influence accretion.  They include two sub-types based on the strength of the magnetic field of the white dwarf: the polars, in which the field prevents formation of an accretion disk, dominating the accretion process within the Roche Lobe such that accreting matter follows along magnetic field lines near the inner Lagrangian point and down the magnetic poles of the white dwarf; and intermediate polars in which an accretion disk forms but is truncated at some distance from the white dwarf surface by the magnetic field that again channels accretion onto the magnetic poles.  

SS~Cygni, a type of dwarf nova, was the first cataclysmic variable to be detected in X-rays, albeit tentatively, based on a rocket flight \citep{Rappaport.etal:74}. The potential for X-ray detection of CVs as a class was subsequently explored by \citet{Warner:74}.  We shall barely scratch the surface of this now very active subfield of X-ray astronomy and only touch upon novae, dwarf novae and magnetic CVs, here, in addition to the symbiotic binaries that we introduce below.  The interested reader is referred to the book by \citet{Warner:95} for a thorough treatise on these fascinating binary systems, and to the review by \citet{Mukai:17} for a recent summary of X-ray properties of CVs.

\subsubsection{Symbiotic Binaries in X-rays}
\label{s:symbiotics}

Symbiotic binaries are a particular type of cataclysmic variable, sometimes referred to as Z Andromedae type, comprising a white dwarf star orbiting within the extended envelope or wind of a companion red giant.  The white dwarf accretes stellar or stellar wind material via Roche lobe overflow, usually through an accretion disk.  Some symbiotics have been observed to display collimated, bi-polar outflows, or jets, extending from the white dwarf.  The accretion of matter onto the white dwarf is expected to result in X-ray emission, and indeed this has been observed in several systems.

\begin{figure}
\begin{center}
\includegraphics[width=0.50\textwidth]{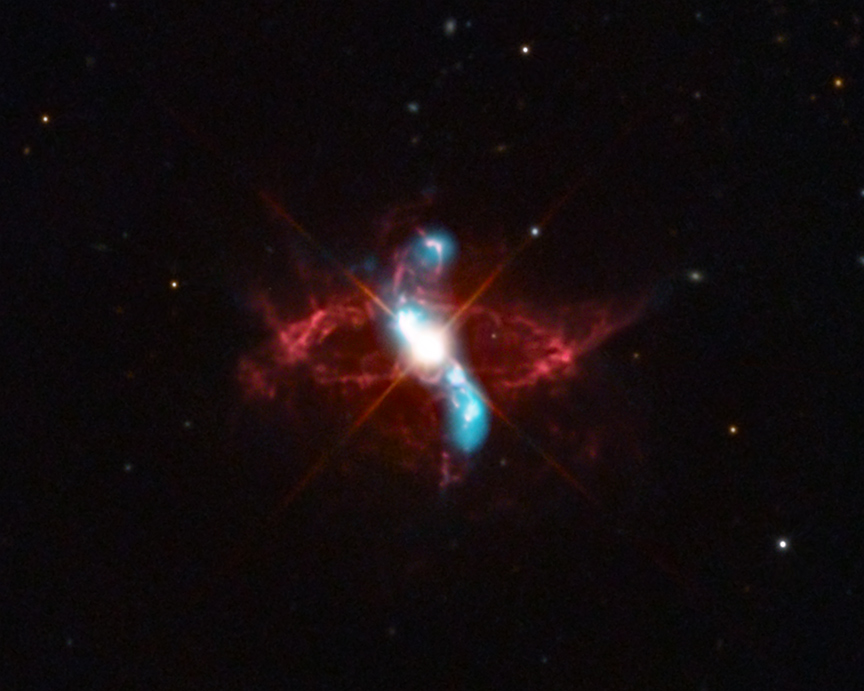}
\includegraphics[width=0.48\textwidth]{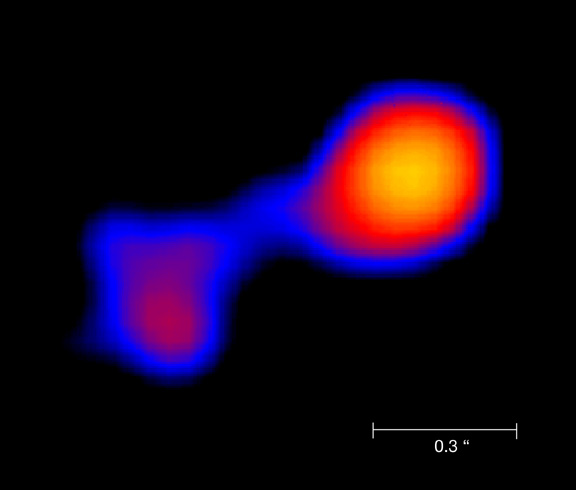}
\end{center}
\caption{Left: A composite {\it Chandra} ACIS (blue) and optical (red) image of the symbiotic system R~Aqr.  {\it Chandra} data were obtained in three separate pointings between  2001 September and 2005 October.  The ring of optical emission shown in red is thought to have arisen from an earlier nova explosion resulting from thermonuclear runaway of material accreted onto the white dwarf (see Section~\ref{s:novae}). 
Image credit: http://chandra.harvard.edu/photo/2017/raqr/.  Right: The Mira system observed and resolved using the {\it Chandra} ACIS-S detector.  The wind-accreting white dwarf is to the left, while cool AGB star Mira A is to the right.  \citet{Karovska.etal:05} speculate that the X-rays from Mira A result from a magnetic reconnection flare. Image from http://chandra.harvard.edu/photo/2005/mira/ 
\label{f:symbiotics}
}
\end{figure}

{\it Chandra} made the first observation of a two-sided X-ray jet in a symbiotic system in an ACIS-S observation of R~Aquarii in 2000 September \citep{Kellogg.etal:01} (Figure~\ref{f:symbiotics}). R~Aqr is a well-studied symbiotic system comprising a mass-losing Mira-like long-period variable with a 387 day period and a $\sim 1 M_\odot$ compact star thought to be a white dwarf.  The object is thought to have undergone a nova explosion in the past (see Section~\ref{s:novae} below). The X-ray jets found by {\it Chandra} were soft, with emission observed only in the energy range $< 1$~keV and are likely shock-heated by interaction with the ambient wind from the red giant star.  In addition to the jets, the central, likely white dwarf, source was also detected and found to be consistent with a blackbody spectrum with a temperature of $2\times 10^6$~K. However, evidence was also present for a cold Fe K$\alpha$ and K$\beta$ fluorescence lines at 6.4 and 7.0 keV, together with other weak K lines corresponding to Cr, V, Ca, Ar, and S.  Since production of Fe fluorescence requires an significant ionizing flux at energies of $\sim 7$~keV and higher, \citet{Kellogg.etal:01} suggested that a hidden hard source might be associated with an accretion disk.

Observations taken 3.3 years later showed significant changes in X-ray spectral and morphological characteristics. One jet was observed to have moved outward with an apparent projected velocity of $\sim 580$~km~s$^{-1}$, while the other faded, likely due to adiabatic expansion and cooling \citep{Kellogg.etal:07}.  More centrally, evidence for formation of a new jet within the inner 500~AU of the system was uncovered, and the central object itself hardened considerably, revealing a component consistent with a $T\sim 10^8$~K thermal plasma \citep{Nichols.etal:07}.   Similar X-ray emission associated with both a central source and jet activity was seen in the symbiotic binary CH~Cyg by \citet{Karovska.etal:07}, with later observations showing the jets to have precessed \citep{Karovska.etal:10}.

Perhaps one of the most surprising discoveries in a symbiotic system came from a {\it Chandra} observation of Mira---the eponymous star of the pulsating AGB class of luminous variables.   Mira is losing mass to a slow wind at a rate of about $10^7 M_\odot$~yr$^{-1}$, while a companion white dwarf (Mira B) approximately 70~AU away (corresponding to 0.6 arcsec) accretes from this wind.  An ACIS-S observation on 2003 December  found not only X-rays from Mira A, but also from B.  X-rays had never before been detected from an AGB star and \citet{Karovska.etal:05} speculated that the emission arose from a magnetic reconnection flare similar to those discussed in Section~\ref{s:flares} (Figure~\ref{s:symbiotics}).

\subsubsection{Two Types of Cataclysmic Variable X-ray Spectra}
\label{s:cvspec}

It might be expected that the X-ray spectra of different types of CVs can be different based on the characteristics of the accretion flow.  In the case of magnetic systems, accretion streams are expected to impact the white dwarf surface vertically, whereas in non-magnetic systems gas accretes unimpeded via a disk, forming a boundary layer as it connects to the stellar surface.  \citet{Mukai.etal:03} noted that in both cases the emergent X-ray spectrum is expected to be from plasmas with a continuous range of temperatures, from the temperature of the shock that forms at the white dwarf surface down to its photospheric temperature.  For strong shock conditions, the shock temperature, $T_s$, is related to the gravitational potential as follows:
\begin{equation}
T_s=\frac{3}{8}\frac{\mu m_H}{k}\frac{GM}{R}
\label{e:strongshock}
\end{equation}
where $\mu$ is the mean molecular weight, $m_H$ the mass of the hydrogren atom, $k$ is the Boltzmann constant, $G$ the gravitational constant, and $M$ and $R$ the white dwarf mass and radius, respectively.  Since white dwarf radii {\it decrease} for increasing mass, the shock temperature increases significantly with increasing mass, from $\sim 10$~keV for a $M=0.5M_\odot$ to $\sim 200$~keV for $M=1.4M_\odot$ \citep[e.g.][]{Mukai:17}.  Indeed, X-ray shock temperatures have been used quite extensively to infer CV white dwarf masses \citep[e.g.][and references therein]{Yu.etal:18}.

\begin{figure}
\begin{center}
\includegraphics[width=0.49\textwidth]{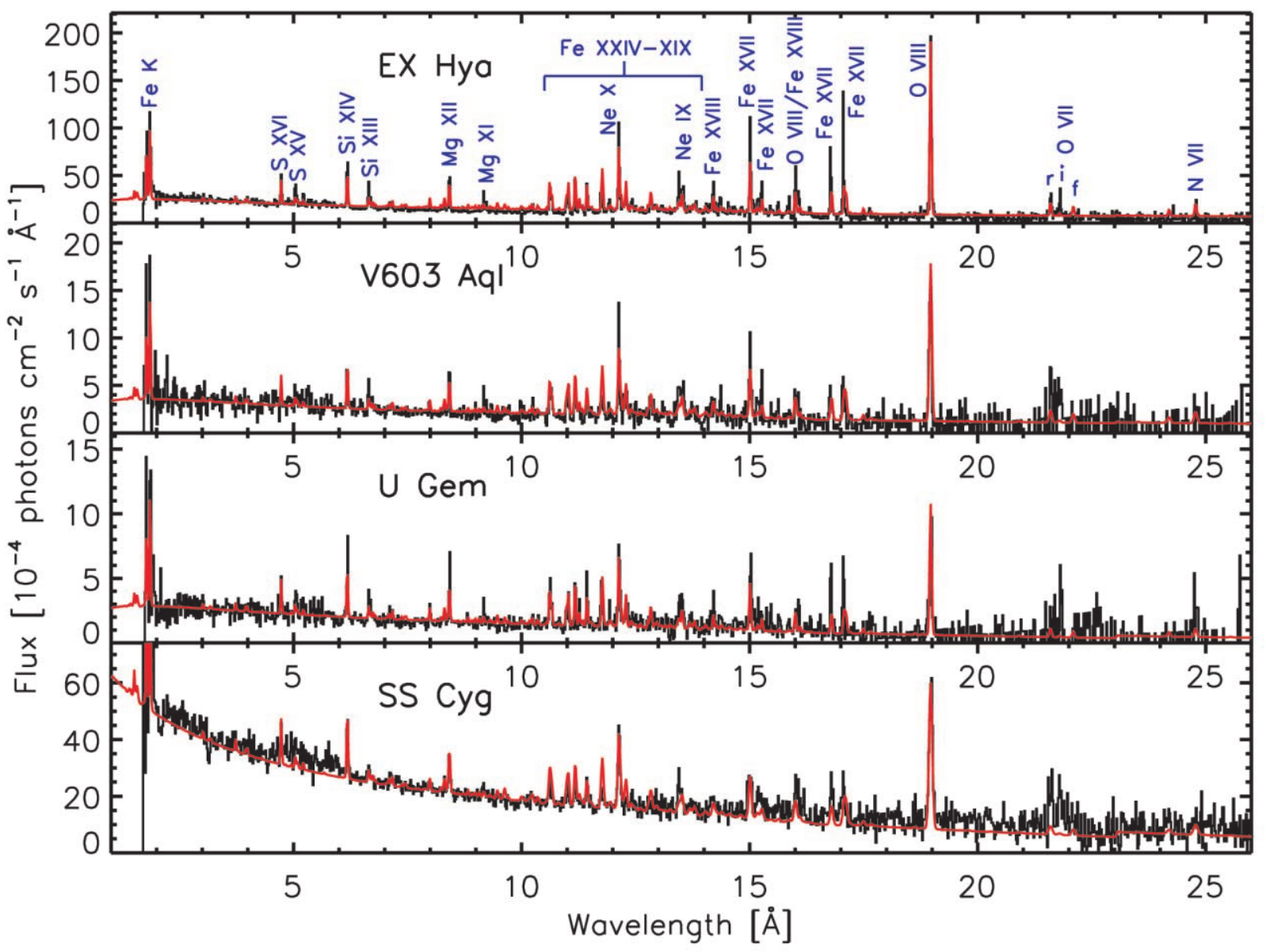}
\includegraphics[width=0.49\textwidth]{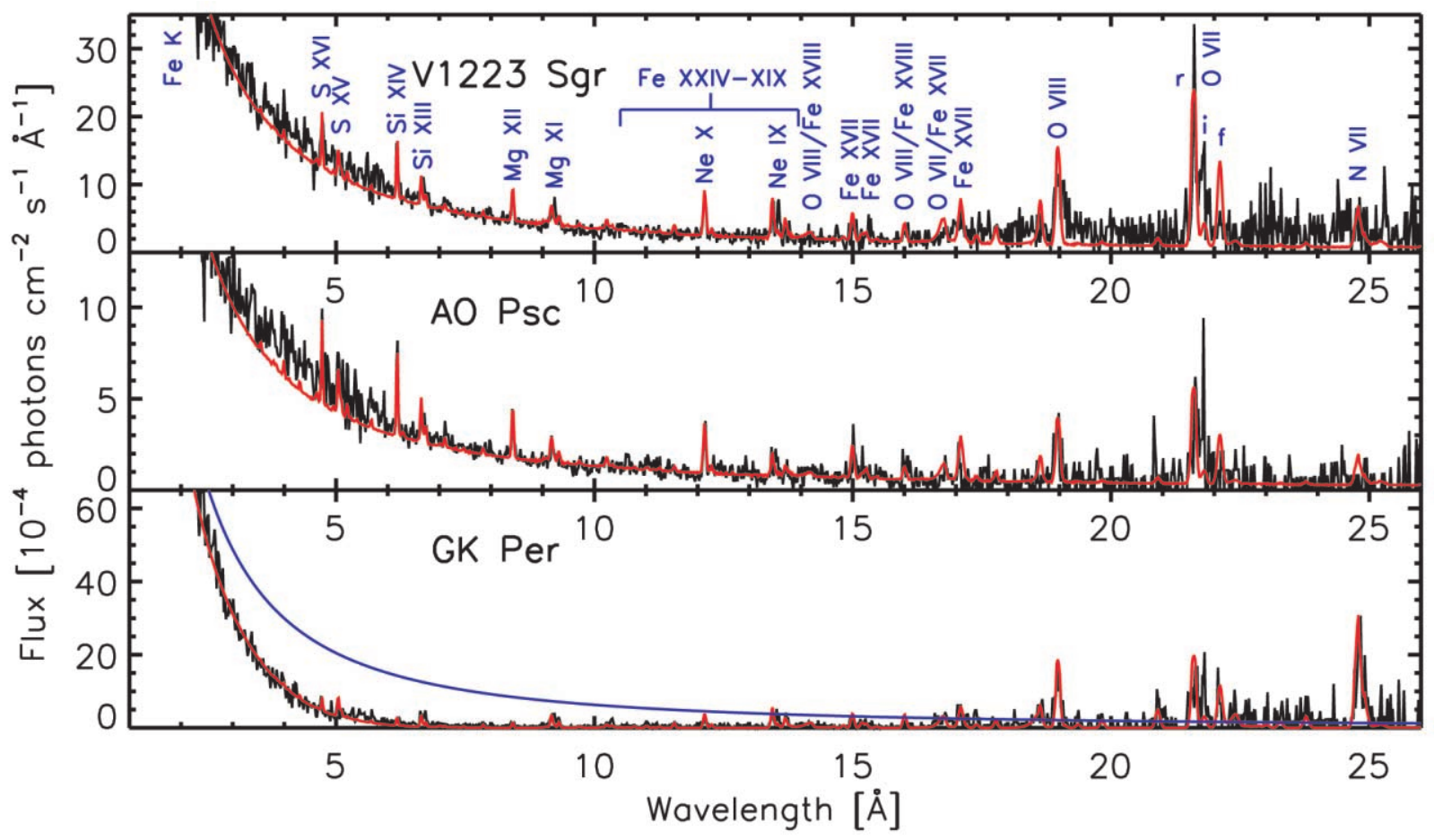}
\end{center}
\caption{Chandra combined MEG $\pm$1st order data of four CVs that exhibit cooling flow spectra (left) and three showing photoionized spectra (right). Data are shown in black with the model spectra in red. From \citet{Mukai.etal:03}.
\label{f:cvs}
}
\end{figure}

One of the first breakthroughs enabled by {\it Chandra} observations of CVs was spearheaded by a study of high-resolution HETG spectra of seven different systems by \citet{Mukai.etal:03}, four of which were magnetic systems and three non-magnetic. They found that the spectra divide into two distinct types.
The non-magnetic systems are remarkably well-fitted by a simple ``cooling flow'' model that assumes steady-state isobaric radiative cooling of an optically-thin plasma, such as those forming stellar coronae (Section~\ref{s:coronae}), between maximum and minimum temperatures.  Here, the maximum corresponds to the shock temperature of the impacting flow, and the minimum to the white dwarf photospheric temperature.  Cooling flow models were originally developed to describe the X-ray emitting gas in the centers of clusters of galaxies, although the similarities to flows settling onto the surfaces of accreting white dwarfs was noted by \citet{Fabian.Nulsen:77}.  \citet{Mukai.etal:03} note with irony that these models can be used to fit CV spectra, but are now known to be a poor description of galaxy clusters emission!
 
The spectra of the magnetic systems, all of which are of intermediate polar (IP) type, instead were found to be grossly inconsistent with cooling flow models, and exhibited hard continua superimposed by strong H-like and He-like ion emission of abundant elements but little Fe ``L-shell" (ions with $n=2$ ground states) emission.  \citet{Mukai.etal:03} found these spectra could be matched with a photoionized plasma model, and in some ways were similar to spectra of black hole systems and active galactic nuclei.

The fly in the ointment is that one of the cooling flow-like spectra, that of EX Hya, is also an IP system. What then is the common physics between the different types of spectra? \citet{Mukai.etal:03} suggest that it is the specific accretion rate---the accretion rate per unit area.  Accretion on  non-magnetic CVs occurs over a much larger area than in magnetic systems, and so the specific accretion rate is generally expected to be lower than for magnetic systems where accretion is restricted to a small fraction of the stellar surface.  The compact ``pillbox" geometry of a high specific accretion rate shock renders photon escape from the sides of the postshock region more difficult.   EX~Hya lies below the CV ``period gap" of 2--3~hr where few CVs are found.  Below the gap, gravitational radiation tends to dominate the angular momentum loss which proceeds at quite a low rate, an accretion rates are commensurately lower.  EX~Hya therefore has a lower accretion rate than most IP systems above the gap, and thus it has a taller shock geometry more similar to non-magnetic systems, and more amenable to ready escape of X-radiation.

The cooling flow model for EX~Hya was qualitatively verified by \citet{Luna.etal:15} based on a much longer 496~ks HETG+ACIS-S observation obtained in 2007 May,  but they noted that in detail the model failed to reproduce certain spectral features such as the ratios of resonance lines of He-like and H-like ions, the former of which were generally stronger than the model prediction.  The spectrum suggests that extra heating must be deposited at the base of the accretion column, where cooler ions form. Simple modifications to the cooling flow model were unable to match the observed spectrum and \citet{Luna.etal:15} concluded that additional physics, such as thermally unstable cooling that would modify the temperature distribution of the emitting gas might be required.  A similar problem in which cooler lines were underpredicted was found earlier for the dwarf nova WX~Hydri by \citet{Perna.etal:03} and the likely IP system V426 Oph by \citet{Homer.etal:04}.

The exquisite quality of the EX~Hya {\it Chandra} spectra has enabled several other important measurements. Based on earlier data obtained in, \citet{Mauche.etal:01} and \citet{Mauche.etal:03} were able to infer a plasma density of $\sim 10^{14}$~cm$^{-3}$ based on lines of Fe~XVII and Fe~XXII, respectively---orders of magnitude larger than observed in stellar coronae.  By co-adding spectral lines, \citet{Hoogerwerf.etal:04} were able to achieve a velocity precision of 15~km~s$^{-1}$ and made the first detection of orbital motion using X-ray spectroscopy.  The white dwarf mass inferred, $0.49 \pm 0.13 M_\odot$, is consistent with optical and UV measurements. 
\citet{Luna.etal:10} identified narrow ($\sim 150$~km~s$^{-1}$) and broad ($\sim 1600$~km~s$^{-1}$) components in several emission lines and interpreted the broad component as due to photoionization of the pre-shock flow by radiation from the post-shock flow.  Since the photoionized region has to be close to the radiation source in order to produce strong photoionized emission lines the model was able to constrain the height of the standing shock above the white dwarf surface to with in the approximate range 0.2--0.6 white dwarf radii. Thus, EX~Hya, while dominated by a cooling flow-type spectrum, shows the link between this thermal emission and the photoionization-dominated sources.

\subsubsection{Nova Explosions}
\label{s:novae}

The term ``nova" originates from its use by Tycho Brahe in his book ``De Nova Stella''---or ``On the New Star" as we would write today---which featured a description of the supernova SN 1572 now commonly referred to as Tycho's supernova. Unable to see the progenitor of the explosion, from his perspective it appeared like a bright new star where before there was none.  Following the accumulation of observational evidence that suggested novae and supernovae are different phenomena,  novae were subsequently classified as ``classical novae" (CNe).  Novae have since been further divided, with the term ``recurrent novae'' (RNe) being used for objects that have been observed to have undergone more than one outburst.  Of course, since we now understand novae to originate in white dwarf binaries, it is interesting to note that ``new stars" are actually a manifestation of processes in very old stellar systems!

Novae originate from the explosive detonation of H-rich material that accumulates on the surface of a white dwarf by mass transfer from the stellar companion in cataclysmic variable systems.  The bottom layer of the accreted H-rich gas is compressed under the strong white dwarf gravity and becomes hot and degenerate. Once the temperature at the bottom of the accreted layer has reached several million degrees, H fusion can occur through the $p$-$p$ chain, heating the accreted degenerate material even further.   At a temperature of  about $7\times 10^7$~K, thermal energy is comparable to the Fermi energy of degeneracy and the gas can begin to expand. At the same time the energy liberated by nuclear reactions, including the CNO cycle at later times, is  increasing rapidly: a thermonuclear runaway occurs and explosively ejects the accreted envelope. 

Novae are of special interest not just because of the broad array of astrophysics they encompass, but because they, along with CVs in general, are also the likely progenitors for Type~1a supernovae. For a detailed description of the nova phenomenon, the reader is referred to the reviews of \citet{Starrfield.etal:16}  and \citet{Williams:92}.

The amount of accreted material required to initiate thermonuclear runaway depends on the temperature and mass of the underlying white dwarf.  Heat flow from the white dwarf interior, or from prior outbursts, can be important for heating the accreted layers, while the mass controls the compressional heating and pressure within the layer.   A first order estimate for the ignition mass, $M_{ig}$, is given by the critical pressure for ignition, $P_{crit}$, and white dwarf mass and radius $M_{WD}$ and $R_{WD}$, respectively, by 
\begin{equation}
M_{ig} = \frac{4\pi R^4_{WD} P_{crit} }{G M_{WD}},
\label{e:mig}
\end{equation}
where $P_{crit}\sim 10^{20}$~dyne~cm$^{-2}$ \citep[see, e.g.,][]{Starrfield.etal:16}.  Values of $M_{ig}$ based on Equation~\ref{e:mig} range from $10^{-3}$--$10^{-6} M_\odot$ \citep[e.g.][]{Drake.etal:16}.   More massive white dwarfs, requiring less accreted material to initiate thermonuclear runaway, tend to be ``fast" novae---evolving and fading rapidly.

X-ray emission is important for understanding many of the different aspects of nova explosions and {\it Chandra} has made key contributions in all of these areas.  Relatively hard X-rays, and even $\gamma$-rays, can be produced in the early stages of the explosion itself when the ejected envelope material interacts violently with circumbinary material inducing particle accleration in the resulting shocks \citet{Cheung.etal:15}. Following the initial blast, nuclear burning continues on the surface of the white dwarf, often leading to a super-Eddington luminosity in which radiation pressure drives a an outflow.  As the ejecta and the radiatively-driven outflow have thinned, the photosphere gradually shrinks, so that the observed effective temperature increases until reaching several hundred thousand degrees.  At this point, analogous to the hot white dwarfs discussed in Section~\ref{s:wdphot}, the Wien Tail of the photospheric emission is well into the soft X-ray regime and, provided the intervening interstellar column density is not too high, the nova becomes a ``super-soft source'' (SSS).   At certain accretion rates thought to be close to a few $10^{-7} M_\odot$~yr$^{-1}$ steady nuclear burning can also result without thermonuclear runaway, producing a persistent supersoft source, such as CAL 83 and CAL 87 in the Large Magellanic Cloud first seen by {\it Einstein} \citep[][]{Long.etal:81,van_den_Heuvel.etal:92}.

Novae can remain X-ray bright for timescales from days to years.  The most fruitful observation strategy has been to employ the {\it Swift} X-ray satellite \citep{Gehrels.etal:04} for making short observations to follow X-ray evolution over these long timescales, and {\it Chandra} and {\it XMM-Newton} to target key episodes and provide detailed high-resolution X-ray spectra.

\paragraph{Blast Waves}
\label{s:blast}

Arguably the most spectacular novae in X-rays are the rare events that occur in the types of symbiotic binaries described in Section~\ref{s:symbiotics}.  In these systems, the explosion occurs inside the wind of the red giant companion and generates an outwardly propagating blast wave that is similar to that of a Type~II supernova in which a massive star explodes within the medium of its massive wind.  The difference is in the amount of energy involved.  Symbiotic nova explosions typically involve energies of the order of $10^{43}$--$10^{44}$~erg, compared with $\sim 10^{51}$~erg liberated in a supernova.  As a consequence, symbiotic novae evolve on timescales of weeks instead of millennia, allowing detailed observation of the whole event. 

\begin{figure}
\begin{center}
\includegraphics[width=0.99\textwidth]{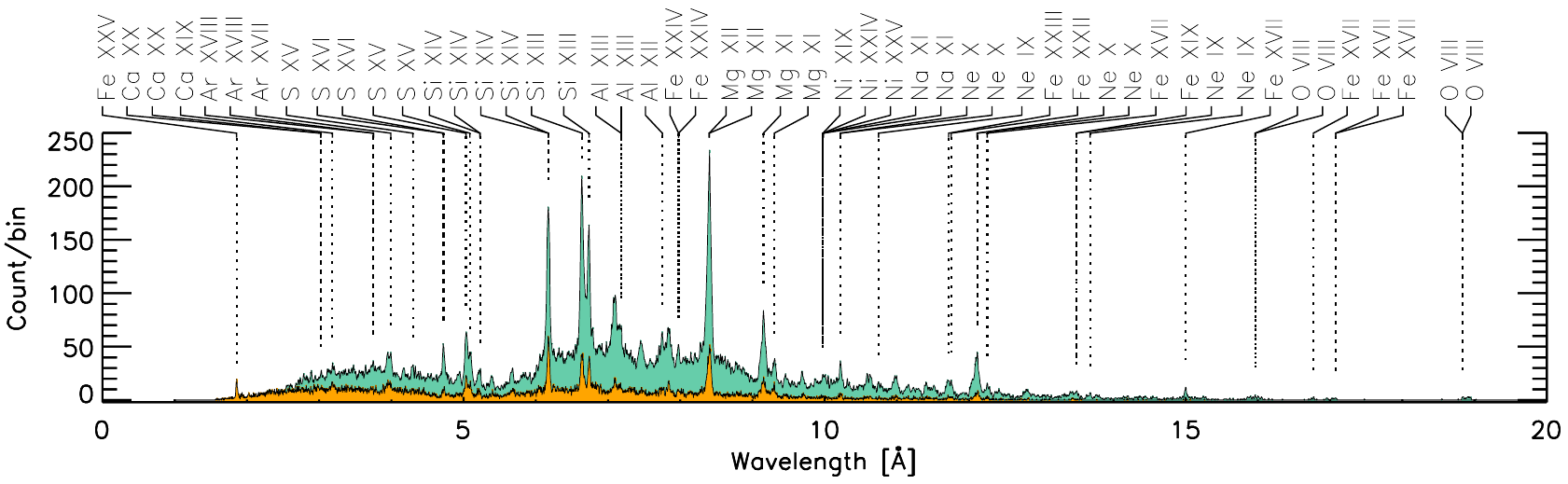}
\includegraphics[width=0.54\textwidth]{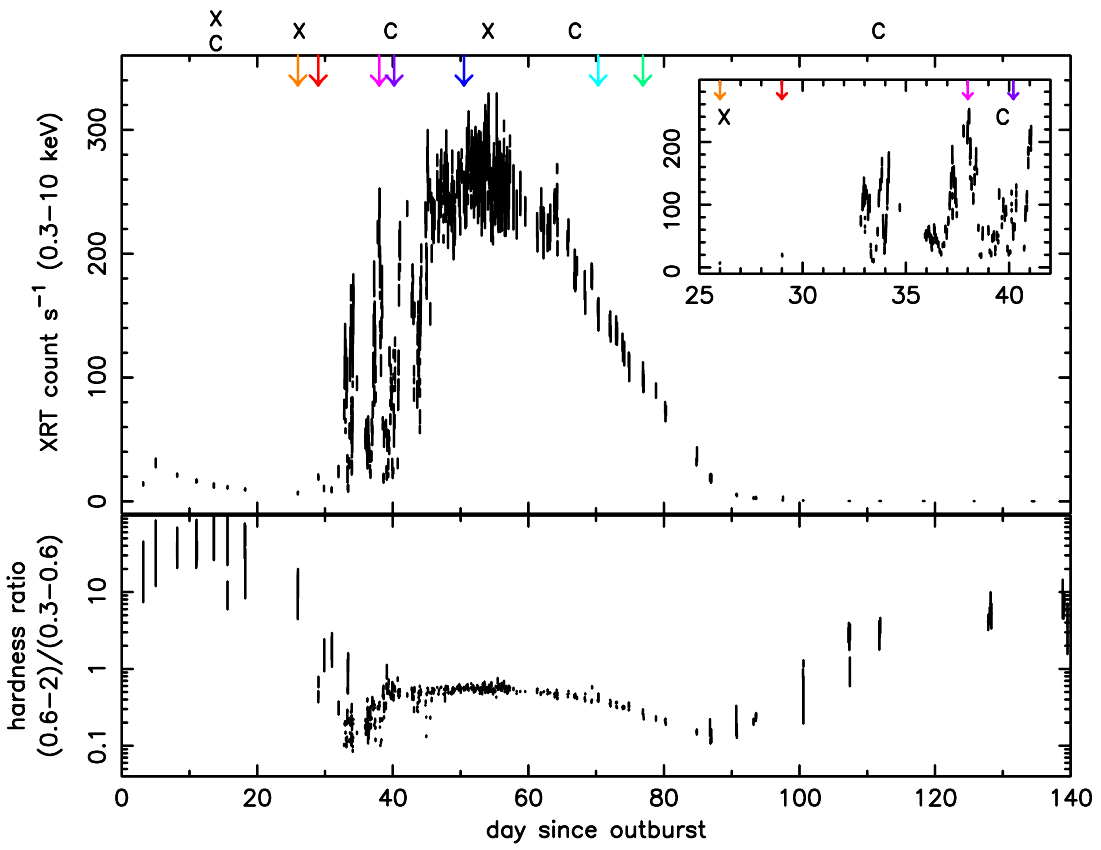}
\includegraphics[width=0.44\textwidth]{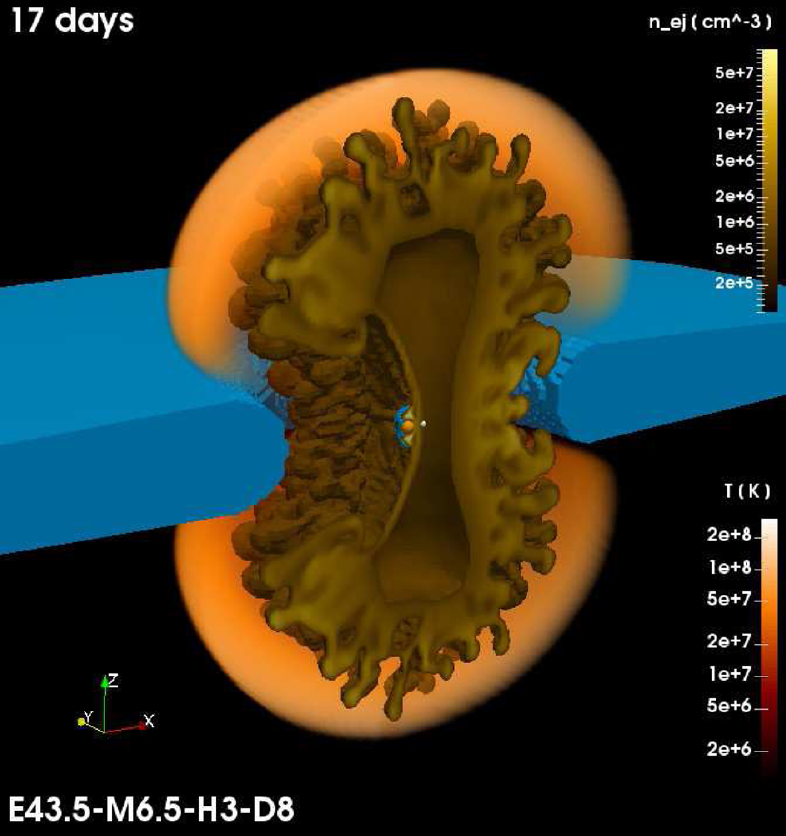}
\end{center}
\caption{Top:  {\it Chandra} MEG (background) and HEG (foreground) spectra of RS Oph obtained on day 13.9 of the 2006 outburst showing a hot thermal spectrum from the blast wave.  Prominent lines are identified. From \citet{Drake.etal:09}.
Bottom left: {\it Swift} XRT light curve and the (0.6--2~keV)/(0.3--0.6~keV) hardness ratio for RS Oph from day 3 to day 135 after outburst discovery.  The blast wave emission dominates before day 30, after which violent variability is seen in at the onset of SSS emergence.  Observations by {\it Chandra} (C) and {\it XMM-Newton} (X) are marked. From \citet{Osborne.etal:11b}.  Bottom right: A 3D rendering of the hydrodynamic model of the 2014 February 6 explosion of the symbiotic nova V745~Sco at 17 days after outburst showing the density distribution of the ejecta and the plasma temperature 17 days after the outburst. The blue region is the unshocked equatorial density enhancement isosurface  for ejecta densities larger than $10^7$~cm$^{-3}$. The small orange and white spheres at the center represents the red giant and white dwarf, separated by 1.7~AU.  The explosion is strongly collimated and shaped by the circumbinary material. From \citet{Orlando.etal:17}.
\label{f:rsoph}
}
\end{figure}

The first symbiotic nova to be observed in detail in the {\it Chandra} and {\it XMM-Newton} era occurred on RS~Ophiuchi on 2006 February 12, triggering an extensive international multiwavelength campaign.  RS~Oph is a recurrent nova at a distance of about 1.6~kpc and had been seen in outburst on several previous occasions, with an outburst frequency of approximately once every 20 years.  The {\it Swift} XRT light curve of the 2006 outburst analysed by \citet{Osborne.etal:11b} is illustrated in Figure~\ref{f:rsoph} and indicates the times of {\it Chandra} and {\it XMM-Newton} observations.  The first 30 days are dominated by the blast wave that peaked at a temperature of 10~keV before it gradually faded and cooled as energy was dissipated by the expanding shock in the surrounding medium.  The SSS then began to appear during a period of chaotic variability that turned out to be quite common at the beginning of the SSS phase (see Section~\ref{s:sss} below), peaking at an amazing 300 counts/s.

{\it Chandra} HETG spectra taken on day 13.9 of the outburst (Figure~\ref{f:rsoph}) revealed a rich thermal plasma spectrum exhibiting Doppler-broadened emission lines with a half-width at zero intensity of $\sim 2400$~km~s$^{-1}$ and formed over a wide temperature range from 3--60 million K \citep{Nelson.etal:08,Drake.etal:11}.  Lines were shaped by differential absorption by the remnant itself and indicated a non-spherical, collimated blast expanding largely in the plane of the sky, consistent with the picture provided by radio, and later, {\it Hubble} observations \citep{O'Brien.etal:06,Bode.etal:07}.   Extended X-ray emission from the blast was detected by {\it Chandra} one and a half years later, confirming the greatest expansion was close to the plane of the sky and indicating an average expansion velocity of 6000~km~s$^{-1}$ \citep{Luna.etal:09}.

Using hydrodynamic simulations \citet{Walder.etal:08} and \citet{Orlando.etal:09} showed that the blast collimation was due to circumstellar material and a density enhancement in the equatorial plane, a situation that is probably common to all novae (see also the hydrodynamic models of the symbiotic novae V407 Cyg and V745~Sco;  \citealt{Orlando.Drake:12}, \citealt{Drake.etal:16}, \citealt{Orlando.etal:17}; and of the recurrent CNe system with an accretion disk, U~Sco, \citealt{Drake.Orlando:10}).  The {\it Chandra} gratings also caught the symbiotic nova V745~Sco that exploded on 2014 February 6. Spectra 17 days after the outburst revealed a similar blast wave spectrum to RS~Oph, with line profiles indicating a collimated asymmetric outflow.  Hydrodynamic models tuned to match the X-ray observations of symbiotic novae enable estimation of the explosion energy and ejecta mass.  \citet{Orlando.etal:09} found the explosion energy and ejected mass for RS~Oph were $E_{ex}\approx10^{44}$~erg and $M_{ej}\approx 10^{-6} M_\odot$.  

Understanding the ejecta mass in comparison with the inter-outburst accretion rate is important for determining whether the white dwarf is gradually increasing in mass, or whether each outburst reduces its mass. V745~Sco was a ``faster" nova than RS~Oph, declining by 3 magnitudes in only 9 days, and expectations of a lower outburst energy and ejected mass were borne out by both 1D and 3D hydrodynamic simulations that found  $E_{ex}\approx3\times 10^{43}$~erg and $M_{ej}\approx 1$--$3\times10^{-7} M_\odot$ \citep{Drake.etal:16,Orlando.etal:17}. The ejected mass is an order of magnitude smaller than required for thermonuclear runaway for a massive white dwarf based on Equation~\ref{e:mig}, indicating the white dwarf is a mass gainer and likely supernova 1a progenitor.

\paragraph{The Super-Soft Source Phase}
\label{s:sss}

High resolution {\it Chandra} and {\it XMM-Newton} observations of the SSS phase of nova outbursts have been revolutionary in revealing the complex line profiles and chemical compositions of what are often apparently super-Eddington sources.  The steady nuclear burning phase of nova outbursts that produces the characteristic X-ray SSS is thought to begin at the time of outburst, and its appearance simply depends on when the effective photosphere shrinks to yield X-ray emitting effective temperatures of a few hundred thousand K.  The SSS is thought to last as long as the H-rich material, such that its duration is a function of white dwarf mass that controls both the surface pressure and the accreted mass.  This picture has been largely confirmed by a mostly {\it Swift}-based study by \citet{Schwarz.etal:11} of 52 Galactic and Magellanic Cloud novae of which SSS emission was detected in 26, and by the {\it Chandra} study of novae in M31 by \citet{Henze.etal:14}.  \citet{Schwarz.etal:11} did, however, conclude that white dwarf mass is likely not the only controlling factor in SSS turn-on and turn-off times.

\begin{figure}
\begin{center}
\includegraphics[width=0.99\textwidth]{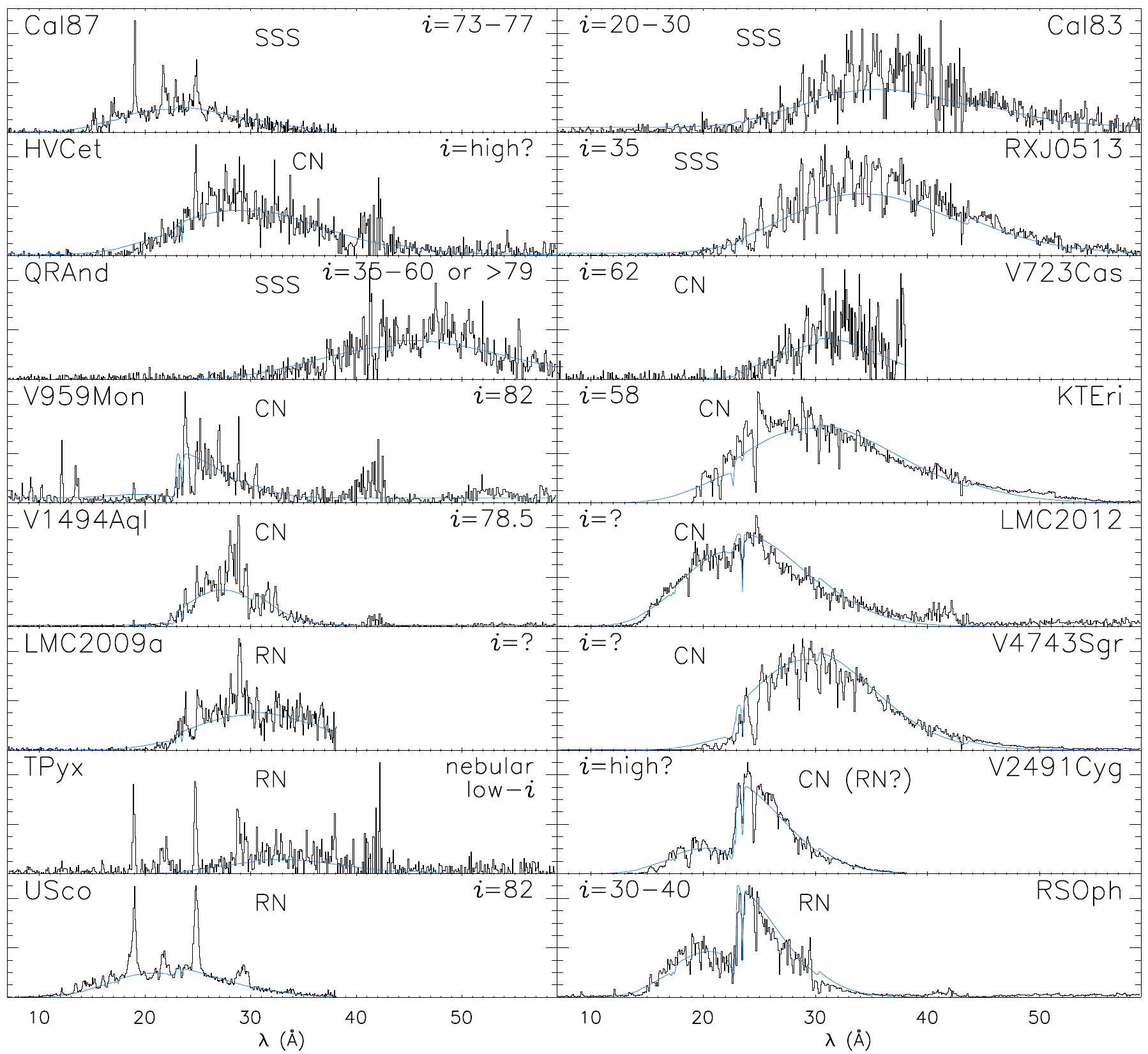}
\end{center}
\caption{High-resolution {\it Chandra} and {\it XMM-Newton} spectra in arbitrary flux units of SSS exhibiting emission lines without absorption features (SSe, left column) and those that show just  continua and absorption lines (SSa, right column), both labeled with inclination angles where known. Thin blue lines represent absorbed blackbody curves.  Labels SSS, CN, and RN denote persistent SSS, CNe and RNe. Note that features near 40~\AA\ in some {\it Chandra} spectra are due  to division by low effective areas near the C~K edge. From \citet{Ness.etal:13}.
\label{f:sss}
}
\end{figure}

The most startling aspect of SSS emission has turned out to be the extreme variability at onset, in which the X-ray count rates are seen to change by factors of 100 or more (e.g., RS~Oph in Figure~\ref{f:rsoph}).  This remains largely unexplained, although possible explanations include temporary obscuration by clumpy material, or photospheric radius changes, although neither of these can explain both the X-ray hardness changes and the variability observed. Other dramatic variations include oscillations with a period of the order of 20 minutes thought to be non-radial $g+$-modes on V1494~Aql and V4743~Sgr, the former of which also exhibited a sudden burst of X-rays in which the count rate increased by a factor of 10 for about 1000s \citep{Drake.etal:03}, an the latter of which showed a very sudden drop in X-rays \citep{Ness.etal:03}. These sudden variations remain unexplained. 

SSS X-ray spectra sometimes exhibit P Cygni-like line profiles, with emission features on the redward side of an absorption line, and absorption features redshifted by up to 2000~km~s$^{-1}$ or so (e.g., V4743~Sgr, \citealt{Rauch.etal:10}; Nova SMC 2016, \citealt{Orio.etal:18}), indicting a photosphere formed in rapidly outflowing gas.  \citet{Ness.etal:13} found that SSS with emission lines (``SSe" in their notation) were less luminous than those with just absorption features (``SSa'') and appear to be in mostly high inclination systems (Figure~\ref{f:sss}).  This suggests influence of a disk on the spectrum, perhaps as a source of obscuration and the emission line producing gas.

The problem facing research into the SSS of nova outbursts is that the theorical modelling side is now lagging behind observational developments.  While non-local thermondynamic equilibrium atmospheric models appear to work well for white dwarfs, and even for SSS spectra to some extent \cite[e.g.][]{Rauch.etal:10}, readily available models do yet incorporate the outflow and associated radiative transfer within the dynamic atmosphere.

\section{Epilogue}

Part of the aim of this primer is to give the reader a sense of the remarkably broad scope of the field of X-rays from stars and planetary systems.  Fascinating in its own right, this field encompasses almost all of the physical processes at work in the more distant Universe and in systems much less well understood.  In this sense, our solar system, the Sun and stars can be considered true laboratories for X-ray astronomy.  

I have given only a brief taste of the astrophysics to be learned from {\it Chandra} observations of stellar and planetary systems here.  By nature, what has been presented is {\it Chandra}-centric and incomplete, glaringly so in some topics that have been unnaturally compressed by the limited space available,  and the reader is encouraged to follow the articles referenced in order to get a deeper picture of the various subjects touched upon. 

{\it Chandra} has provided us with exquisite mirrors and diffraction gratings to resolve objects both spatially and spectrally for the first time. However, many compelling and important targets remain beyond {\it Chandra's} reach and must await a next generation mission with a larger collecting area. {\it Chandra} has lit the path that should be followed to make these next generation breakthroughs.

\section{Acknowledgements}

JJD thanks  M.~Corcoran, P.~Edmonds, W.~Dunn, V.~Kashyap, B.~Snios, W.~Tucker and B.~Wargelin for helpful comments, corrections and suggestions, and the CXC Director, Belinda Wilkes, for continuing advice and support.

\backgroundsetup{contents=\includegraphics{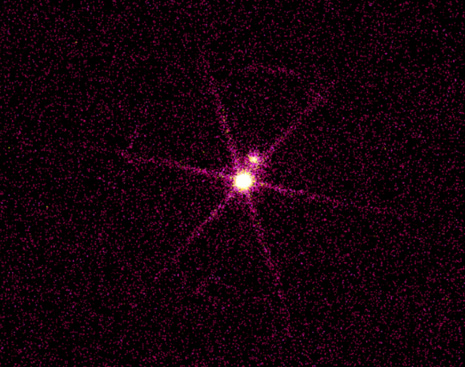},scale=0.6,opacity=1.0,vshift=-370pt,angle=0}
\BgThispage

\newpage

\small
\bibliographystyle{aasjournal}
\bibliography{solsys,cx,atomic,chianti,dem,coronae_general,cycles,corabun,bds,planetevap,star_planet,highmassx,cepheids,rotation,collmagwinds,herbigae,twhya,vega,jjdrake,coup,proplyd,cygob2,cygob2_mario_ettore_facundo,xphotevap,hr4796,pne,wds,cvs,nova_misc}



\backgroundsetup{contents=\includegraphics{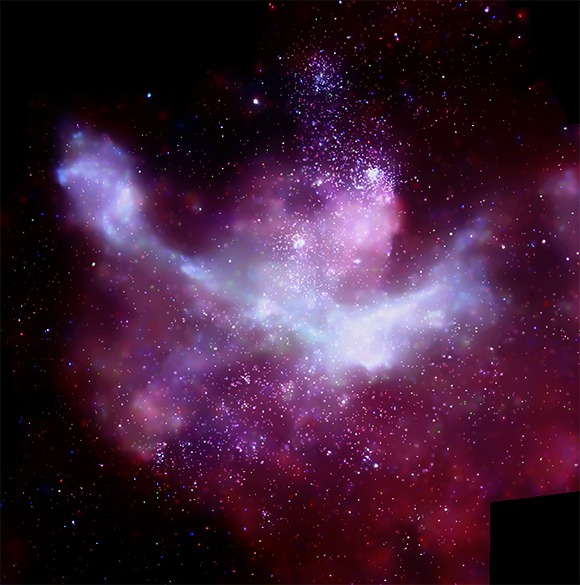},scale=0.75,opacity=1.0,vshift=-110pt,angle=0}
\BgThispage

\end{document}